\documentclass[journal,draftclsnofoot,onecolumn,12pt,oneside]{IEEEtran} % for single column no footer
\usepackage{cite}
\ifCLASSINFOpdf
  \usepackage[pdftex]{graphicx}
  \usepackage{epstopdf} % remove this
\else
\fi
\usepackage{amsmath}
\usepackage{algorithm}
\usepackage{algorithmic}

  \usepackage[caption=false,font=footnotesize]{subfig}
\usepackage{url}
\usepackage{amsthm}

\theoremstyle{definition}

\newtheoremstyle{myremark}% name of the style to be used
{}% measure of space to leave above the theorem. E.g.: 3pt
{}% measure of space to leave below the theorem. E.g.: 3pt
{}% name of font to use in the body of the theorem
{0pt}% measure of space to indent
{\bfseries}% name of head font
{.}% punctuation between head and body
{ }% space after theorem head; " " = normal interword space
{\thmname{#1}\thmnumber{ #2}: \thmnote{#3}}

\theoremstyle{theoremdd}

\theoremstyle{myremark}
\newcounter{rem}
\newtheorem{remark}[rem]{Remark}

\newcommand{\comment}[1]{{}}

\usepackage{amssymb}
\usepackage{xspace}
\usepackage{bm}
\usepackage{bbm}

\newcommand{\set}[1]{\ensuremath{\mathcal{#1}}\xspace} % caligraphic set notation
\newcommand{\mat}[1]{\ensuremath{\mathbf{#1}}\xspace} % matrices
\renewcommand{\vec}[1]{\ensuremath{\mathbf{#1}}\xspace} % vectors
\newcommand{\parens}[1]{{\left(#1\right)}\xspace}
\newcommand{\brackets}[1]{{\left[#1\right]}\xspace}
\newcommand{\braces}[1]{{\left\{#1\right\}}\xspace}
\newcommand{\bars}[1]{{\left\vert#1\right\vert}\xspace}
\newcommand{\doublebars}[1]{{\left\Vert#1\right\Vert}\xspace}

\newcommand{\complex}{\ensuremath{\mathbb{C}}\xspace}

\newcommand{\floor}[1]{\ensuremath{\left\lfloor{#1}\right\rfloor}\xspace}

\newcommand{\card}[1]{\bars{#1}}

\newcommand{\setcomplex}{\ensuremath{\complex}}

\newcommand{\setvector}[2]{\ensuremath{#1^{#2 \times 1}}\xspace}

\newcommand{\setvectorcomplex}[1]{\setvector{\setcomplex}{#1}}

\newcommand{\setmatrix}[3]{\ensuremath{#1^{#2 \times #3}}\xspace}

\newcommand{\setmatrixcomplex}[2]{\setmatrix{\setcomplex}{#1}{#2}}

\newcommand{\inv}{\ensuremath{^{-1}}\xspace}

\newcommand{\ctrans}{\ensuremath{^{{*}}}\xspace}
\newcommand{\entry}[2]{\ensuremath{\brackets{#1}_{#2}}\xspace}

\newcommand{\logtwo}[1]{\ensuremath{\mathrm{log}_{2}\parens{#1}}}

\newcommand{\pnorm}[2]{\ensuremath{\doublebars{#2}_{#1}}\xspace}

\newcommand{\normtwo}[1]{\pnorm{2}{#1}}

\newcommand{\normfro}[1]{\pnorm{\mathrm{F}}{#1}}

\newcommand{\ind}[1]{\ensuremath{\mathbbm{1}\braces{#1}}\xspace}

\DeclareMathOperator*{\argmin}{\mathrm{argmin}}
\DeclareMathOperator*{\argmax}{\mathrm{argmax}}

\newcommand{\maxop}[1]{\ensuremath{\mathrm{max}\parens{#1}}\xspace}

\newcommand{\minop}[1]{\ensuremath{\mathrm{min}\parens{#1}}\xspace}

\newcommand{\st}{\ensuremath{\mathrm{s.t.~}}\xspace}
\newcommand{\opt}{\ensuremath{^{\star}}\xspace}

\newcommand{\todB}[1]{\ensuremath{\brackets{#1}_{\mathrm{dB}}}}

\newcommand{\powernoise}{\ensuremath{P_{\mathrm{noise}}}\xspace}

\newcommand{\powertx}{\ensuremath{P_{\mathrm{tx}}}\xspace}

\newcommand{\powertxue}{\ensuremath{\powertx^{\mathrm{UE}}}\xspace}

\newcommand{\powertxdonor}{\ensuremath{\powertx^{\mathrm{Don}}}\xspace}
\newcommand{\powertxiab}{\ensuremath{\powertx^{\mathrm{IAB}}}\xspace}

\newcommand{\snr}{\ensuremath{\mathsf{SNR}}\xspace}

\newcommand{\sinr}{\ensuremath{\mathsf{SINR}}\xspace}

\newcommand{\inr}{\ensuremath{\mathsf{INR}}\xspace}

\newcommand{\inrul}{\ensuremath{\inr_{\mathrm{UL}}}\xspace}

\newcommand{\inrulthresh}{\ensuremath{\inrul^{\mathrm{thresh}}}\xspace}

\newcommand{\atx}[1]{\ensuremath{\vec{a}_{\mathrm{tx}}\parens{#1}}\xspace}
\newcommand{\arx}[1]{\ensuremath{\vec{a}_{\mathrm{rx}}\parens{#1}}\xspace}

\newcommand{\Na}{\ensuremath{N_{\mathrm{a}}}\xspace} % number of antennas

\newcommand{\precb}{\ensuremath{\mathcal{F}}\xspace}

\newcommand{\comcb}{\ensuremath{\mathcal{W}}\xspace}

\newcommand{\nbr}{\ensuremath{\parens{\Delta\theta,\Delta\phi}}\xspace}
\newcommand{\nbrv}{\ensuremath{\parens{\Deltavartheta,\Deltavarphi}}\xspace}
\newcommand{\nbrd}{\ensuremath{\parens{\deltatheta,\deltaphi}}\xspace}
\newcommand{\nbrnbrd}{\ensuremath{\parens{\Deltatheta,\Deltaphi,\deltatheta,\deltaphi}}\xspace}
\newcommand{\nbrnbrdth}{\ensuremath{\parens{\Deltatheta,\deltatheta}}\xspace}
\newcommand{\nbrnbrdph}{\ensuremath{\parens{\Deltaphi,\deltaphi}}\xspace}
\newcommand{\nbrvnbrd}{\ensuremath{\parens{\Deltavartheta,\Deltavarphi,\deltatheta,\deltaphi}}\xspace}

\newcommand{\nbroneone}{\ensuremath{\parens{1^\circ,1^\circ}}\xspace}

\newcommand{\nbrtwotwo}{\ensuremath{\parens{2^\circ,2^\circ}}\xspace}

\newcommand{\thph}{\ensuremath{\parens{\theta,\phi}}\xspace}
\newcommand{\thphtx}{\ensuremath{\parens{\thetatx,\phitx}}\xspace}
\newcommand{\thphrx}{\ensuremath{\parens{\thetarx,\phirx}}\xspace}
\newcommand{\thphtxrx}{\ensuremath{\parens{\thetatx,\phitx,\thetarx,\phirx}}\xspace}
\newcommand{\thphtxi}{\ensuremath{\parens{\thetatx\idx{i},\phitx\idx{i}}}\xspace}
\newcommand{\thphrxj}{\ensuremath{\parens{\thetarx\idx{j},\phirx\idx{j}}}\xspace}
\newcommand{\thphtxiopt}{\ensuremath{\parens{\thetatx\idx{i\opt},\phitx\idx{i\opt}}}\xspace}
\newcommand{\thphrxjopt}{\ensuremath{\parens{\thetarx\idx{j\opt},\phirx\idx{j\opt}}}\xspace}
\newcommand{\thphtxbar}{\ensuremath{\parens{\thetatxbar,\phitxbar}}\xspace}
\newcommand{\thphrxbar}{\ensuremath{\parens{\thetarxbar,\phirxbar}}\xspace}
\newcommand{\thphtxrxbar}{\ensuremath{\parens{\thetatxbar,\phitxbar,\thetarxbar,\phirxbar}}\xspace}
\newcommand{\thphtxopt}{\ensuremath{\parens{\thetatx\opt,\phitx\opt}}\xspace}
\newcommand{\thphrxopt}{\ensuremath{\parens{\thetarx\opt,\phirx\opt}}\xspace}
\newcommand{\thphtxrxopt}{\ensuremath{\parens{\thetatx\opt,\phitx\opt,\thetarx\opt,\phirx\opt}}\xspace}
\newcommand{\thphtxrxijopt}{\ensuremath{\parens{\thetatx\idx{i\opt},\phitx\idx{i\opt},\thetarx\idx{j\opt},\phirx\idx{j\opt}}}\xspace}

\newcommand{\Kth}{\ensuremath{K_{\theta}}\xspace}
\newcommand{\Kph}{\ensuremath{K_{\phi}}\xspace}

\newcommand{\idx}[1]{\ensuremath{^{\parens{#1}}}\xspace}

\newcommand{\txdirsetmeas}{\ensuremath{\set{T}}\xspace}
\newcommand{\rxdirsetmeas}{\ensuremath{\set{R}}\xspace}

\newcommand{\txdirsetcb}{\ensuremath{\set{A}_{\mathrm{tx}}}\xspace}
\newcommand{\rxdirsetcb}{\ensuremath{\set{A}_{\mathrm{rx}}}\xspace}

\newcommand{\thetatx}{\ensuremath{\theta_{\mathrm{tx}}}\xspace}
\newcommand{\phitx}{\ensuremath{\phi_{\mathrm{tx}}}\xspace}

\newcommand{\thetarx}{\ensuremath{\theta_{\mathrm{rx}}}\xspace}
\newcommand{\phirx}{\ensuremath{\phi_{\mathrm{rx}}}\xspace}

\newcommand{\thetatxbar}{\ensuremath{\bar{\theta}_{\mathrm{tx}}}\xspace}
\newcommand{\phitxbar}{\ensuremath{\bar{\phi}_{\mathrm{tx}}}\xspace}
\newcommand{\thetarxbar}{\ensuremath{\bar{\theta}_{\mathrm{rx}}}\xspace}
\newcommand{\phirxbar}{\ensuremath{\bar{\phi}_{\mathrm{rx}}}\xspace}

\newcommand{\Ntx}{\ensuremath{N_{\mathrm{tx}}}\xspace}
\newcommand{\Nrx}{\ensuremath{N_{\mathrm{rx}}}\xspace}

\newcommand{\labelbackhaul}{\mathrm{BH}}
\newcommand{\labelaccess}{\mathrm{AC}}
\newcommand{\labelsi}{\mathrm{SI}}
\newcommand{\labeliui}{\mathrm{CL}}
\newcommand{\labeldonor}{\mathrm{Don}}
\newcommand{\labeliab}{\mathrm{IAB}}
\newcommand{\labelue}{\mathrm{UE}}
\newcommand{\labeltx}{\mathrm{tx}}
\newcommand{\labelrx}{\mathrm{rx}}
\newcommand{\labelsum}{\mathrm{sum}}
\newcommand{\labeldl}{\mathrm{DL}}
\newcommand{\labelul}{\mathrm{UL}}
\newcommand{\labeltdd}{\mathrm{TDD}}
\newcommand{\labelgtdd}{\mathrm{TDD}\textrm{-}\mathrm{PC}}
\newcommand{\labelcb}{\mathrm{cb}}
\newcommand{\labelnom}{\mathrm{nom}}
\newcommand{\labelours}{\mathrm{ours}}

\newcommand{\snrtx}{\ensuremath{\snr_{\labeltx}}\xspace}
\newcommand{\snrrx}{\ensuremath{\snr_{\labelrx}}\xspace}
\newcommand{\snrtxnom}{\ensuremath{\snr_{\labeltx}^{\labelnom}}\xspace}
\newcommand{\snrrxnom}{\ensuremath{\snr_{\labelrx}^{\labelnom}}\xspace}
\newcommand{\snrtxours}{\ensuremath{\snr_{\labeltx}^{\labelours}}\xspace}
\newcommand{\snrrxours}{\ensuremath{\snr_{\labelrx}^{\labelours}}\xspace}

\newcommand{\sinrtx}{\ensuremath{\sinr_{\labeltx}}\xspace}
\newcommand{\sinrrx}{\ensuremath{\sinr_{\labelrx}}\xspace}
\newcommand{\sinrtxnom}{\ensuremath{\sinr_{\labeltx}^{\labelnom}}\xspace}
\newcommand{\sinrrxnom}{\ensuremath{\sinr_{\labelrx}^{\labelnom}}\xspace}
\newcommand{\sinrtxours}{\ensuremath{\sinr_{\labeltx}^{\labelours}}\xspace}
\newcommand{\sinrrxours}{\ensuremath{\sinr_{\labelrx}^{\labelours}}\xspace}

\newcommand{\inrtx}{\ensuremath{\inr_{\labeltx}}\xspace}
\newcommand{\inrrx}{\ensuremath{\inr_{\labelrx}}\xspace}
\newcommand{\inrrxthresh}{\ensuremath{\inr_{\labelrx}^{\mathrm{tgt}}}\xspace}
\newcommand{\inrrxnom}{\ensuremath{\inr_{\labelrx}^{\labelnom}}\xspace}
\newcommand{\inrrxours}{\ensuremath{\inr_{\labelrx}^{\labelours}}\xspace}
\newcommand{\inrrxmin}{\ensuremath{\inr_{\labelrx}^{\mathrm{min}}}\xspace}

\newcommand{\snrbhdl}{\ensuremath{\snr_{\labelbackhaul}^{\labeldl}}\xspace}
\newcommand{\snracdl}{\ensuremath{\snr_{\labelaccess}^{\labeldl}}\xspace}
\newcommand{\snrbhul}{\ensuremath{\snr_{\labelbackhaul}^{\labelul}}\xspace}
\newcommand{\snracul}{\ensuremath{\snr_{\labelaccess}^{\labelul}}\xspace}

\newcommand{\snrbhdlbar}{\ensuremath{\overline{\snr}_{\labelbackhaul}^{\labeldl}}\xspace}
\newcommand{\snracdlbar}{\ensuremath{\overline{\snr}_{\labelaccess}^{\labeldl}}\xspace}
\newcommand{\snrbhulbar}{\ensuremath{\overline{\snr}_{\labelbackhaul}^{\labelul}}\xspace}
\newcommand{\snraculbar}{\ensuremath{\overline{\snr}_{\labelaccess}^{\labelul}}\xspace}

\newcommand{\inrbhdl}{\ensuremath{\inr_{\labelbackhaul}^{\labeldl}}\xspace}
\newcommand{\inracdl}{\ensuremath{\inr_{\labelaccess}^{\labeldl}}\xspace}
\newcommand{\inrbhul}{\ensuremath{\inr_{\labelbackhaul}^{\labelul}}\xspace}
\newcommand{\inracul}{\ensuremath{\inr_{\labelaccess}^{\labelul}}\xspace}

\newcommand{\snrtxbar}{\ensuremath{\overline{\snr}_{\labeltx}}\xspace}
\newcommand{\snrrxbar}{\ensuremath{\overline{\snr}_{\labelrx}}\xspace}

\newcommand{\captx}{\ensuremath{C_{\labeltx}}\xspace}
\newcommand{\caprx}{\ensuremath{C_{\labelrx}}\xspace}
\newcommand{\capsum}{C_{\labelsum}\xspace}

\newcommand{\captxcb}{C_{\labeltx}^{\labelcb}\xspace}
\newcommand{\caprxcb}{C_{\labelrx}^{\labelcb}\xspace}

\newcommand{\capsumcb}{C_{\labelsum}^{\labelcb}\xspace}

\newcommand{\setx}{\ensuremath{R_{\labeltx}}\xspace}
\newcommand{\serx}{\ensuremath{R_{\labelrx}}\xspace}

\newcommand{\setxtdd}{\ensuremath{R_{\labeltx}^{\labeltdd}}\xspace}
\newcommand{\serxtdd}{\ensuremath{R_{\labelrx}^{\labeltdd}}\xspace}
\newcommand{\sesumtdd}{\ensuremath{R_{\labelsum}^{\labeltdd}}\xspace}

\newcommand{\setxgtdd}{\ensuremath{R_{\labeltx}^{\labelgtdd}}\xspace}
\newcommand{\serxgtdd}{\ensuremath{R_{\labelrx}^{\labelgtdd}}\xspace}
\newcommand{\sesumgtdd}{\ensuremath{R_{\labelsum}^{\labelgtdd}}\xspace}

\newcommand{\setxnom}{\ensuremath{R_{\labeltx}^{\labelnom}}\xspace}
\newcommand{\serxnom}{\ensuremath{R_{\labelrx}^{\labelnom}}\xspace}
\newcommand{\sesumnom}{\ensuremath{R_{\labelsum}^{\labelnom}}\xspace}

\newcommand{\setxours}{\ensuremath{R_{\labeltx}^{\labelours}}\xspace}
\newcommand{\serxours}{\ensuremath{R_{\labelrx}^{\labelours}}\xspace}
\newcommand{\sesumours}{\ensuremath{R_{\labelsum}^{\labelours}}\xspace}

\newcommand{\segain}{\ensuremath{\eta}\xspace}
\newcommand{\segainsum}{\ensuremath{\segain_{\labelsum}}\xspace}

\newcommand{\segaintxtdd}{\ensuremath{\segain_{\labeltx}^{\labeltdd}}\xspace}
\newcommand{\segainrxtdd}{\ensuremath{\segain_{\labelrx}^{\labeltdd}}\xspace}
\newcommand{\segainsumtdd}{\ensuremath{\segain_{\labelsum}^{\labeltdd}}\xspace}

\newcommand{\segaintxgtdd}{\ensuremath{\segain_{\labeltx}^{\labelgtdd}}\xspace}
\newcommand{\segainrxgtdd}{\ensuremath{\segain_{\labelrx}^{\labelgtdd}}\xspace}
\newcommand{\segainsumgtdd}{\ensuremath{\segain_{\labelsum}^{\labelgtdd}}\xspace}

\newcommand{\segaintxnom}{\ensuremath{\segain_{\labeltx}^{\labelnom}}\xspace}
\newcommand{\segainrxnom}{\ensuremath{\segain_{\labelrx}^{\labelnom}}\xspace}
\newcommand{\segainsumnom}{\ensuremath{\segain_{\labelsum}^{\labelnom}}\xspace}

\newcommand{\capfrac}{\ensuremath{\gamma}\xspace}
\newcommand{\capfracsum}{\ensuremath{\capfrac_{\labelsum}}\xspace}

\newcommand{\capfractxtdd}{\ensuremath{\capfrac_{\labeltx}^{\labeltdd}}\xspace}
\newcommand{\capfracrxtdd}{\ensuremath{\capfrac_{\labelrx}^{\labeltdd}}\xspace}
\newcommand{\capfracsumtdd}{\ensuremath{\capfrac_{\labelsum}^{\labeltdd}}\xspace}

\newcommand{\capfractxgtdd}{\ensuremath{\capfrac_{\labeltx}^{\labelgtdd}}\xspace}
\newcommand{\capfracrxgtdd}{\ensuremath{\capfrac_{\labelrx}^{\labelgtdd}}\xspace}
\newcommand{\capfracsumgtdd}{\ensuremath{\capfrac_{\labelsum}^{\labelgtdd}}\xspace}

\newcommand{\capfractxnom}{\ensuremath{\capfrac_{\labeltx}^{\labelnom}}\xspace}
\newcommand{\capfracrxnom}{\ensuremath{\capfrac_{\labelrx}^{\labelnom}}\xspace}
\newcommand{\capfracsumnom}{\ensuremath{\capfrac_{\labelsum}^{\labelnom}}\xspace}

\newcommand{\capfractxours}{\ensuremath{\capfrac_{\labeltx}^{\labelours}}\xspace}
\newcommand{\capfracrxours}{\ensuremath{\capfrac_{\labelrx}^{\labelours}}\xspace}
\newcommand{\capfracsumours}{\ensuremath{\capfrac_{\labelsum}^{\labelours}}\xspace}

\newcommand{\mHsi}{\mH_{\labelsi}\xspace}
\newcommand{\mHbh}{\mH_{\labelbackhaul}\xspace}
\newcommand{\vhbh}{\vh_{\labelbackhaul}\xspace}
\newcommand{\vhac}{\vh_{\labelaccess}\xspace}
\newcommand{\vhiui}{\vh_{\labeliui}\xspace}

\newcommand{\hiui}{h_{\labeliui}\xspace}

\newcommand{\vhtx}{\vh_{\labeltx}\xspace}
\newcommand{\vhrx}{\vh_{\labelrx}\xspace}

\newcommand{\Gtx}{\ensuremath{G_{\labeltx}}\xspace}

\newcommand{\Gbh}{\ensuremath{G_{\labelbackhaul}}\xspace}
\newcommand{\Gac}{\ensuremath{G_{\labelaccess}}\xspace}

\newcommand{\Gsi}{\ensuremath{G_{\mathrm{SI}}}\xspace}

\newcommand{\Giui}{\ensuremath{G_{\labeliui}}\xspace}

\newcommand{\powernoiseue}{\ensuremath{\powernoise^{\labelue}}\xspace}

\newcommand{\powernoiseiab}{\ensuremath{\powernoise^{\labeliab}}\xspace}
\newcommand{\powernoisedonor}{\ensuremath{\powernoise^{\labeldonor}}\xspace}

\newcommand{\casedl}{\mathsf{DL}\textsf{-}\mathsf{DL}}
\newcommand{\caseul}{\mathsf{UL}\textsf{-}\mathsf{UL}}

\newcommand{\deltatheta}{\ensuremath{\delta\theta}\xspace}
\newcommand{\deltaphi}{\ensuremath{\delta\phi}\xspace}
\newcommand{\Deltatheta}{\ensuremath{\Delta\theta}\xspace}
\newcommand{\Deltaphi}{\ensuremath{\Delta\phi}\xspace}

\newcommand{\Deltavartheta}{\ensuremath{\Delta\vartheta}\xspace}
\newcommand{\Deltavarphi}{\ensuremath{\Delta\varphi}\xspace}

\newcommand{\setinrmeas}{\ensuremath{\mathcal{I}_{\labelrx}}\xspace}
\newcommand{\setnbr}{\ensuremath{\mathcal{N}}\xspace}
\newcommand{\setnbrth}{\ensuremath{\mathcal{N}_{\theta}}\xspace}
\newcommand{\setnbrph}{\ensuremath{\mathcal{N}_{\phi}}\xspace}

\def\vf{{\vec{f}}}

\def\vh{{\vec{h}}}

\def\vv{{\vec{v}}}
\def\vw{{\vec{w}}}
\def\vx{{\vec{x}}}

\def\mH{{\mat{H}}}

\usepackage{xspace}
\usepackage[acronym,nogroupskip,nonumberlist,nopostdot]{glossaries}
\makeglossaries

\usepackage{xcolor}
\newcommand{\edit}[1]{\textcolor{black}{#1}}

\newacronym{snr}{SNR}{signal-to-noise ratio}
\newacronym{sinr}{SINR}{signal-to-interference-plus-noise ratio}
\newacronym{inr}{INR}{interference-to-noise ratio}
\newacronym{sir}{SIR}{signal-to-interference ratio}
\newacronym{sqr}{SQR}{signal-to-quantization-noise ratio}
\newacronym{sqnr}{SQNR}{signal-to-quantization-plus-noise ratio}
\newacronym{ian}{IAN}{interference as noise}
\newacronym{ber}{BER}{bit error rate}
\newacronym{pn}{PN}{pseudorandom noise}
\newacronym{bfsk}{BFSK}{binary frequency shift keying}
\newacronym{fh}{FH}{frequency-hopped}
\newacronym{fh-bfsk}{FH-BFSK}{frequency-hopped binary frequency shift keying}
\newacronym{crc}{CRC}{cyclic redundancy check}
\newacronym{isi}{ISI}{intersymbol interference}
\newacronym{dsss}{DSSS}{direct-sequence spread spectrum}
\newacronym{ofdm}{OFDM}{orthogonal frequency-division multiplexing}
\newacronym{ofdma}{OFDMA}{orthogonal frequency-division multiple access}
\newacronym{sdr}{SDR}{software-defined radio}
\newacronym{tx}{TX}{transmitter}
\newacronym{rx}{RX}{receiver}
\newacronym{fdd}{FDD}{frequency-division duplexing}
\newacronym{tdd}{TDD}{time-division duplexing}
\newacronym{fdma}{FDMA}{frequency-division multiple access}
\newacronym{tdma}{TDMA}{time-division multiple access}
\newacronym{sdma}{SDMA}{space-division multiple access}
\newacronym[plural=MPCs]{mpc}{MPC}{multipath component}
\newacronym{mui}{MUI}{multi-user interference}

\newacronym{qam}{QAM}{quadrature amplitude modulation}
\newacronym{mqam}{MQAM}{M-ary quadrature amplitude modulation}

\newacronym{ls}{LS}{least-squares}
\newacronym{lms}{LMS}{least mean squares}
\newacronym{rls}{RLS}{recursive least-squares}
\newacronym{rzf}{RZF}{regularized zero-forcing}
\newacronym{mmse}{MMSE}{minimum mean square error}
\newacronym{lmmse}{LMMSE}{linear minimum mean square error}
\newacronym{mse}{MSE}{mean square error}
\newacronym{fft}{FFT}{fast Fourier transform}
\newacronym{dft}{DFT}{discrete Fourier transform}
\newacronym{dtft}{DTFT}{discrete-time Fourier transform}
\newacronym{ctft}{CTFT}{continuous-time Fourier transform}
\newacronym{ml}{ML}{machine learning}
\newacronym[plural=NNs]{nn}{NN}{neural network}
\newacronym[plural=RNNs]{rnn}{RNN}{recurrent neural network}
\newacronym[plural=ADCs]{adc}{ADC}{analog-to-digital converter}
\newacronym[plural=DACs]{dac}{DAC}{digital-to-analog converter}
\newacronym[plural=FPGAs]{fpga}{FPGA}{field-programmable gate array}
\newacronym{evm}{EVM}{error vector magnitude}
\newacronym{enob}{ENOB}{effective number of bits}
\newacronym{zf}{ZF}{zero-forcing}
\newacronym{rv}{r.v.}{random variable}
\newacronym{omp}{OMP}{orthogonal matching pursuit}
\newacronym{svd}{SVD}{singular value decomposition}

\newacronym{sdp}{SDP}{semidefinite programming}
\newacronym{psd}{PSD}{positive semidefinite}
\newacronym{nsd}{NSD}{negative semidefinite}

\newacronym{ks}{K-S}{Kolmogorov-Smirnov}

\newacronym{mad}{MAD}{median absolute deviation around the median}

\newacronym{agc}{AGC}{automatic gain control}
\newacronym{rf}{RF}{radio frequency}
\newacronym{if}{IF}{intermediate frequency}
\newacronym{los}{LOS}{line-of-sight}
\newacronym{nlos}{NLOS}{non-line-of-sight}
\newacronym{ple}{PLE}{path loss exponent}
\newacronym[plural=dB,firstplural=decibels (dB)]{db}{dB}{decibel}
\newacronym[plural=dBm,firstplural=decibel milliwatts (dBm)]{dbm}{dBm}{decibel milliwatts}
\newacronym{pa}{PA}{power amplifier}
\newacronym{lna}{LNA}{low noise amplifier}
\newacronym{cw}{CW}{continuous wave}
\newacronym{papr}{PAPR}{peak-to-average power ratio}
\newacronym{usrp}{USRP}{Universal Software Radio Peripheral}
\newacronym{irr}{IRR}{image rejection ratio}
\newacronym{lo}{LO}{local oscillator}
\newacronym{vm}{VM}{vector modulator}
\newacronym{mmwave}{mmWave}{millimeter wave}
\newacronym{eirp}{EIRP}{effective isotropic radiated power}
\newacronym{rsrp}{RSRP}{reference signal received power}

\newacronym{csma}{CSMA}{carrier-sense multiple access}
\newacronym{csmaca}{CSMA/CA}{carrier-sense multiple access with collision avoidance}
\newacronym{csmacd}{CSMA/CD}{carrier-sense multiple access with collision detection}
\newacronym{mac}{MAC}{medium access control}
\newacronym{phy}{PHY}{physical layer}
\newacronym{4g}{4G}{fourth generation}
\newacronym{lte}{LTE}{Long-Term Evolution}
\newacronym{4glte}{4G LTE}{\gls{4g} \gls{lte}}
\newacronym{5g}{5G}{fifth generation}
\newacronym{nr}{NR}{New Radio}
\newacronym{5gnr}{5G NR}{5G New Radio}
\newacronym{ieee}{IEEE}{Institute of Electrical and Electronics Engineers}
\newacronym{wifi}{Wi-Fi}{IEEE 802.11}
\newacronym{lan}{LAN}{local area network}
\newacronym{wlan}{WLAN}{wireless local area network}
\newacronym[plural=BSs]{bs}{BS}{base station}
\newacronym[plural=SBSs]{sbs}{SBS}{small-cell base station}
\newacronym[plural=FD-SBSs]{fdsbs}{FD-SBS}{\gls{fd}-enabled \gls{sbs}}
\newacronym[plural=MBSs]{mbs}{MBS}{macrocell base station}
\newacronym[plural=UEs]{ue}{UE}{user equipment}
\newacronym{ul}{UL}{uplink}
\newacronym{dl}{DL}{downlink}
\newacronym{qos}{QoS}{Quality of Service}
\newacronym{fcc}{FCC}{Federal Communications Commission}
\newacronym{iab}{IAB}{integrated access and backhaul}
\newacronym{fab}{FAB}{fixed access and backhaul}
\newacronym{hetnet}{HetNet}{heterogeneous network}

\newacronym{siso}{SISO}{single-input single-output}
\newacronym{mimo}{MIMO}{multiple-input multiple-output}
\newacronym{sumimo}{SU-MIMO}{single-user \gls{mimo}}
\newacronym{mumimo}{MU-MIMO}{multi-user \gls{mimo}}
\newacronym{bf}{BF}{beamforming}
\newacronym{ca}{CA}{constant amplitude}
\newacronym{ula}{ULA}{uniform linear array}
\newacronym{upa}{UPA}{uniform planar array}
\newacronym[\glslongpluralkey={angles of arrival}]{aoa}{AoA}{angle of arrival}
\newacronym[\glslongpluralkey={angles of departure}]{aod}{AoD}{angle of departure}
\newacronym{dof}{DoF}{degrees of freedom}
\newacronym{csi}{CSI}{channel state information}
\newacronym{csit}{CSIT}{\gls{csi} at the transmitter}
\newacronym{csir}{CSIR}{\gls{csi} at the receiver}
\newacronym{cs}{CS}{compressed sensing}

\newacronym{fd}{FD}{in-band full-duplex}
\newacronym{hd}{HD}{half-duplex}
\newacronym{si}{SI}{self-interference}
\newacronym{sic}{SIC}{self-interference cancellation}
\newacronym{soi}{SoI}{signal of interest}
\newacronym{asic}{A-SIC}{analog \acrlong{sic}}
\newacronym{dsic}{D-SIC}{digital \gls{sic}}
\newacronym{star}{STAR}{simultaneous transmit and receive}
\newacronym{warp}{WARP}{Wireless Open-Access Research Platform}
\newacronym{bfc}{BFC}{beamforming cancellation}
\newacronym{ipi}{IPI}{inter-panel-interference}
\newacronym{ipic}{IPIC}{inter-panel-interference cancellation}

\newacronym{qcqp}{QCQP}{quadratically-constrained quadratic programming}
\newacronym{pdf}{PDF}{probability density function}
\newacronym{cdf}{CDF}{cumulative density function}
\newacronym{iid}{i.i.d.}{independently and identically distributed}

\newacronym{elf}{ELF}{extremely low frequency}
\newacronym{slf}{SLF}{super low frequency}
\newacronym{ulf}{ULF}{ultra low frequency}
\newacronym{vlf}{VLF}{very low frequency}
\newacronym{lf}{LF}{low frequency}
\newacronym{mf}{MF}{medium frequency}
\newacronym{hf}{HF}{high frequency}
\newacronym{vhf}{VHF}{very high frequency}
\newacronym{uhf}{UHF}{ultra high frequency}
\newacronym{shf}{SHF}{super high frequency}
\newacronym{ehf}{EHF}{extremely high frequency}
\newacronym{thf}{THF}{tremendously high frequency}

\newacronym{wncg}{WNCG}{Wireless Networking and Communications Group}
\newacronym{linc}{LINC}{Laboratory of Informatics, Networks, and Communications}
\newacronym{ut}{UT Austin}{The University of Texas at Austin}
\newacronym{uiuc}{UIUC}{University of Illinois at Urbana-Champaign}
\newacronym{usc}{USC}{University of Southern California}
\newacronym{mit}{MIT}{Massachusetts Institute of Technology}
\newacronym{berkeley}{UC Berkeley}{University of California, Berkeley}
\newacronym{osu}{OSU}{Ohio State University}

\newcommand{\upa}{\gls{upa}\xspace}
\newcommand{\upas}{\glspl{upa}\xspace}

\newcommand{\mmwave}{\gls{mmwave}\xspace}
\newcommand{\mimo}{\gls{mimo}\xspace}

\newcommand{\ue}{\gls{ue}\xspace}

\newcommand{\dl}{\acrlong{dl}\xspace}
\newcommand{\ul}{\acrlong{ul}\xspace}

\newcommand{\iab}{\gls{iab}\xspace}

\newcommand{\gcdf}{\gls{cdf}\xspace}

\newcommand{\gsnr}{\gls{snr}\xspace}
\newcommand{\ginr}{\gls{inr}\xspace}
\newcommand{\gsinr}{\gls{sinr}\xspace}

\newcommand{\gpsnr}{\glspl{snr}\xspace}

\newcommand{\gpsinr}{\glspl{sinr}\xspace}

\newcommand{\tdd}{\gls{tdd}\xspace}

\newcommand{\naive}{naive\xspace}
\newcommand{\Naive}{Naive\xspace}

\newcommand{\Naively}{Naively\xspace}
\newcommand{\dldl}{DL-DL\xspace}
\newcommand{\ulul}{UL-UL\xspace}

\newcommand{\secref}[1]{Section~\ref{#1}}

\newcommand{\tabref}[1]{Table~\ref{#1}}
\newcommand{\figref}[1]{\figurename~\ref{#1}}
\newcommand{\algref}[1]{Algorithm~\ref{#1}}

\usepackage{mathtools}

\newcommand{\steer}{\textsc{Steer}\xspace}

\begin{document}
	
\title{\textsc{Steer}: Beam Selection for Full-Duplex\\Millimeter Wave Communication Systems}


\author{%
	Ian~P.~Roberts,~%
	Aditya Chopra,~%
	Thomas Novlan,~\\%
    Sriram Vishwanath,~%
	and Jeffrey~G.~Andrews%
	% and Sriram Vishwanath%
	\thanks{I.~P.~Roberts, S.~Vishwanath, and J.~G.~Andrews are with the 6G@UT Research Center and the Wireless Networking and Communications Group at the University of Texas at Austin. T.~Novlan is with the Advanced Wireless Technologies Group at AT\&T Labs. A.~Chopra was with AT\&T Labs during this work; he is now with Project Kuiper, Amazon.} % Last updated: \today.}%
	% \thanks{I.~P.~Roberts is supported by the National Science Foundation Graduate Research Fellowship Program under Grant No.~DGE-1610403. Any opinions, findings, and conclusions or recommendations expressed in this material are those of the author(s) and do not necessarily reflect the views of the National Science Foundation.}%
}

\maketitle

% \pagebreak

% \setcounter{tocdepth}{1} % 0,1,2,...

% \tableofcontents

% \pagebreak

\begin{abstract}

% Like half-duplex \mmwave communication systems, ones with full-duplex capability will rely on beam alignment to close the links with devices being served.
% Like half-duplex \mmwave communication systems, ones with 
% A full-duplex \mmwave communication system will presumably conduct beam alignment on its transmit link and receive link to close the link with the devices it serves. % without the need for channel knowledge.
% To deliver sufficient link margin, 
% A full-duplex \mmwave communication system will presumably conduct beam alignment with the device it transmits to and with the device it receives from.

%To reliably deliver high beamforming gain, a full-duplex \mmwave communication system will presumably conduct beam alignment with the device it transmits to and with the device it receives from.

% Modern \mmwave communication systems rely on beam alignment to achieve sufficient beamforming gain.
% to close the links with the devices it transmits to and receives from. % without the need for channel knowledge.
Modern \mmwave communication systems rely on beam alignment to deliver sufficient beamforming gain to close the link between devices.
We present a novel beam selection methodology for multi-panel, full-duplex \mmwave systems, which we call \steer, that delivers high beamforming gain while significantly reducing the full-duplex self-interference coupled between the transmit and receive beams.
\steer does not necessitate changes to conventional beam alignment methodologies nor additional over-the-air feedback, making it compatible with existing cellular standards.
% Instead, \steer uses conventional beam alignment to identify the general directions beams should be steered to deliver sufficient link margin. 
% Rather than relying on self-interference channel knowledge that is difficult to obtain, 
% Rather than estimate the self-interference channel, which is difficult, \steer makes use of explicit measurements of self-interference to jointly select transmit and receive beams that deliver high gain in these directions while coupling low self-interference.
Instead, \steer uses conventional beam alignment to identify the general directions beams should be steered, and then it makes use of a minimal number of self-interference measurements to jointly select transmit and receive beams that deliver high gain in these directions while coupling low self-interference.
% Then, based on some design parameters, transmit and receive beams are jointly selected to deliver service and simultaneously minimize the degree of self-interference they couple. 
We implement \steer on an industry-grade 28 GHz phased array platform and use further simulation to show that full-duplex operation with beams selected by \steer can notably outperform both half-duplex and full-duplex operation with beams chosen via conventional beam selection. 
% To validate \steer, we use measurements with an industry-grade phased array platform along with further simulation to show that full-duplex operation with beams selected by \steer can significantly outperform both half-duplex and full-duplex operation with beams chosen via conventional beam selection. 
% For example, the SNR gain can easily exceed 10 dB and the sum spectral efficiency gain can be nearly $2\times$.
For instance, \steer can reliably reduce self-interference by more than $20$ dB and improve SINR by more than $10$ dB, compared to conventional beam selection.
% For instance, \steer typically introduces SINR gains beyond $10$ dB that can translate to nearly doubling the sum spectral efficiency.
% We show that full-duplex operation with beams selected by \steer can outperform both half-duplex and also full-duplex operation where the transmit and receive beams are chosen independently via conventional beam selection. % , even in the presence of cross-link interference.
% are configured the transmit and receive links independently via conventional beam selection, even in the presence of cross-link interference.
Our experimental results highlight that beam alignment can be used not only to deliver high beamforming gain in full-duplex \mmwave systems but also to mitigate self-interference to levels near or below the noise floor, rendering additional self-interference cancellation unnecessary with \steer.
% These results validate the practical efficacy of full-duplex \mmwave systems that rely solely on beamforming to mitigate self-interference to levels near or below the noise floor, meaning no additional cancellation may be necessary with \steer.
\end{abstract}

% \pagebreak

% \input{sec-peer-review-title.tex}

% \input{sec-keywords.tex}

\glsresetall

% \pagebreak

\section{Introduction} \label{sec:introduction}

\comment{
Enhancing \iab with full-duplex capability is an ongoing effort in 3GPP to overcome \mmwave deployment hurdles while offering low latency \cite{iab,3GPP_IAB,cudak21integrated}.
A multi-panel \iab node with full-duplex capability could receive backhaul from a fiber-connected donor node with one of its panels and simultaneously deliver access to a downlink user using another---making full use of the available \mmwave spectrum. % taking full advantage of the same \mmwave spectrum.
Along with improved spectral efficiency, full-duplexing access and backhaul eliminates the delays associated with time-division duplexing, allowing the network to better serve low-latency applications.
% , especially in several-hop networks.
}

To equip \mmwave transceivers with full-duplex capability, recent work has proposed leveraging dense antenna arrays to mitigate self-interference spatially via strategic beamforming \cite{riihonen_loopback,xia_2017,roberts_wcm,everett_softnull_2016,roberts_2021_robustcb,liu_beamforming_2016,lopez-valcarce_beamformer_2019,lopez_prelcic_2019_analog,prelcic_2019_hybrid,satyanarayana_hybrid_2019,zhu_uav_joint_2020,da_silva_2020,lopez_analog_2022,koc_ojcoms_2021,cai_robust_2019,roberts_bflrdr,roberts_equipping_2020}.
For instance, in \cite{prelcic_2019_hybrid,satyanarayana_hybrid_2019,zhu_uav_joint_2020,roberts_bflrdr,cai_robust_2019,da_silva_2020,lopez_analog_2022,koc_ojcoms_2021}, designs are presented that tailor hybrid beamformers at a \mmwave transceiver to mitigate self-interference while maintaining transmission and reception.
% In addition to relying solely on beamforming, proposed solutions have also suggested using analog and digital self-interference cancellation \cite{roberts_equipping_2020,smida_asic,Alexandropoulos_Duarte_2017,singh_2020_acm,Zhang_Luo_2019,dinc_60_2016} to enable \mmwave full-duplex; solutions like these require additional hardware, demand complex digital signal processing, and/or do not scale well to systems with many antennas. %---making them neither practically viable nor desirable.
Proposed solutions \cite{roberts_equipping_2020,smida_asic,Alexandropoulos_Duarte_2017,singh_2020_acm,Zhang_Luo_2019,dinc_60_2016,suk_iab_2022} have also suggested using analog and digital self-interference cancellation to enable \mmwave full-duplex; solutions like these require additional hardware, demand complex digital signal processing, and/or do not scale well to systems with many antennas. %---making them neither practically viable nor desirable.
For these reasons, the scope of the present paper focuses on using beamforming alone to mitigate self-interference spatially, which does not require dedicated hardware nor complex signal processing. 
% , unlike analog and digital self-interference cancellation.
% ---and makes use of dense antenna arrays to mitigate self-interference spatially.
% If successfully equipped with full-duplex capability, \mmwave communication systems could see impressive rate and latency enhancements at the physical layer, which can further magnify at the network level \cite{gupta_fdiab_arxiv} and facilitate \iab network deployments.
If successfully equipped with full-duplex capability, \mmwave communication systems could see impressive throughput and latency enhancements, which magnify at the network level and facilitate \iab deployments \cite{gupta_fdiab_arxiv}.

% impressive rate and latency enhancements at the physical layer, which can further magnify at the network level \cite{gupta_fdiab_arxiv} and facilitate \iab network deployments.

% deliver higher rates with lower latency

Most proposed beamforming-based solutions to mitigate self-interference are not well-suited for practical systems for a few reasons.
First, many practical \mmwave systems are equipped with analog beamforming but lack digital beamforming, rendering proposed hybrid beamforming designs (e.g., \cite{prelcic_2019_hybrid,satyanarayana_hybrid_2019,zhu_uav_joint_2020,roberts_bflrdr,cai_robust_2019,da_silva_2020,lopez_analog_2022,koc_ojcoms_2021}) unfit for such systems.
This is especially problematic since digital beamforming mitigates the large majority of self-interference in most of these designs. 
% , meaning full-duplex operation would not be worthwhile without such.
Proposed designs \cite{liu_beamforming_2016,roberts_2021_robustcb,lopez-valcarce_beamformer_2019,lopez_prelcic_2019_analog} that rely solely on analog beamforming (rather than hybrid beamforming) are also impractical for a few reasons.
These rely on instantaneous knowledge of the self-interference channel---a high-dimensional \mimo channel---whose real-time estimation is currently impractical due to complications posed by analog beamforming and its sheer size.
Furthermore, \cite{liu_beamforming_2016,satyanarayana_hybrid_2019,lopez-valcarce_beamformer_2019,lopez_prelcic_2019_analog,prelcic_2019_hybrid,cai_robust_2019,zhu_uav_joint_2020,da_silva_2020,lopez_analog_2022,koc_ojcoms_2021} require instantaneous knowledge of the \dl and \ul \mimo channels between a full-duplex transceiver and the devices it serves; even this is currently impractical in \mmwave networks, which circumvent \mimo channel estimation via beam alignment.
Some designs 
\cite{prelcic_2019_hybrid,liu_beamforming_2016,satyanarayana_hybrid_2019,lopez_prelcic_2019_analog,lopez-valcarce_beamformer_2019,zhu_uav_joint_2020,cai_robust_2019} do not account for phase shifters and/or attenuators with limited resolution that practical analog beamforming networks are subjected to.

Many proposed beamforming designs for full-duplex \mmwave systems do not accommodate beam alignment and subsequent analog beam selection \cite{liu_beamforming_2016,satyanarayana_hybrid_2019,prelcic_2019_hybrid,lopez_prelcic_2019_analog,lopez-valcarce_beamformer_2019,zhu_uav_joint_2020,da_silva_2020,lopez_analog_2022,koc_ojcoms_2021}.
Beam alignment is a critical component of practical \mmwave systems to provide sufficient link margin to sustain communication without the need for \ul/\dl \mimo channel knowledge \cite{heath_overview_2016,ethan_beam}. %, which is difficult to obtain \cite{heath_overview_2016,ethan_beam}.
Using measurements from beam alignment, a \mmwave system can configure its analog beamformers through \textit{beam selection}.
% Like half-duplex \mmwave systems, ones with full-duplex capability will need to execute beam selection based on beam alignment measurements.
% Since a full-duplex transceiver juggles a transmit beam and receive beam concurrently, beam selection will need to be executed for both beams.
Conventional half-duplex systems typically aim to overcome severe path loss by selecting beams that maximize received \gsnr.
Like half-duplex \mmwave systems, a full-duplex one will presumably execute beam selection and will do so on its transmit link and receive link.
\Naively applying conventional beam selection on the two links independently, however, does not account for self-interference that couples between the transmit and receive beams when operating in a full-duplex fashion.
This has motivated us to create the first beam selection methodology for full-duplex systems.   
The two principal contributions of this paper are summarized as follows.

% since it provides a robust means to configure analog beams without the need for uplink/downlink channel knowledge, which is difficult to obtain.

% Summarize beam selection for conventional half-duplex systems, how it is done.
% Why conventional beam selection is problematic for full-duplex systems.

% how to do fd iab

% we will mainly compare against beamforming-based solutions to show
% issues with existing solutions
% - don't support beam alignment
% - require self-interference channel knowledge
% - require digital beamforming
% - have been proven using highly idealized self-interference channel models, we use measurements to validate our work

%\subsection*{Contributions}

% our proposed solution
% In this work, we present \steer, a beam selection methodology for full-duplex \mmwave systems.
% This motivates the work herein, where 

\textbf{A beam selection methodology for full-duplex systems.}
We present \steer, the first beam selection methodology for jointly selecting the transmit and receive beams of a full-duplex \mmwave transceiver. % that reduces self-interference while delivering high gain on the \ul and \dl.
To do so, we leverage our observations from a recent measurement campaign of \mmwave self-interference \cite{roberts_2021_att_measure,roberts_att_angular}, which showed that small shifts in the steering directions of the transmit and receive beams (on the order of one degree) can lead to noteworthy reductions in self-interference.
% \steer is compatible with conventional beam alignment schemes and introduces no additional over-the-air feedback.
\steer makes use of self-interference measurements across small spatial neighborhoods to jointly select transmit and receive beams at the full-duplex device that offer reduced self-interference while delivering high beamforming gain on the \ul and \dl.
% In this work, we are particularly interested in the application of \steer in full-duplex \iab networks, where juggling access and backhaul simultaneously can increase spectral efficiency and significantly reduce latency, compared to half-duplex operation \cite{gupta_fdiab_arxiv}.
\edit{Following its formulation, we present an algorithm for executing \steer with a minimal number of self-interference measurements.}
% that minimizes the overhead associated with measuring self-interference.
The execution of \steer takes place only at the full-duplex device, introducing no changes to the devices being served nor additional over-the-air feedback, making it compatible with existing beam alignment schemes.

\textbf{Validation of \steer through measurement and simulation.}
We validate \steer by combining simulation with self-interference measurements from an industry-grade 28 GHz phased array platform.
These experimental results illustrate that full-duplex operation with \steer can offer a sum spectral efficiency notably higher than both half-duplex and full-duplex operation with beams from conventional beam alignment.
%These numerical results illustrate that full-duplex operation with \steer can offer a sum spectral efficiency notably higher than:
%\begin{itemize}
%    \item half-duplex operation with beams from conventional beam alignment
%    \item full-duplex operation with beams from conventional beam alignment.
%\end{itemize}
In fact, in most cases, \steer can reduce self-interference to levels such that no additional cancellation is warranted (i.e., near or below the noise floor). % and full-duplex operation can be achieved without introducing dedicated hardware or complex signal processing.
With \steer, beamforming alone can deliver self-interference mitigation sufficient for full-duplex while importantly accommodating beam alignment, even in the presence of cross-link interference arising when serving devices simultaneously and in-band.
% \steer demonstrates the practical efficacy of using beamforming alone to bring full-duplex operation to \mmwave systems.

% No additional self-interference cancellation may be needed.

% Emphasize latency improvement of \iab

\section{System Model} \label{sec:system-model}

\begin{figure}
    \centering
    \includegraphics[width=\linewidth,height=0.2\textheight,keepaspectratio]{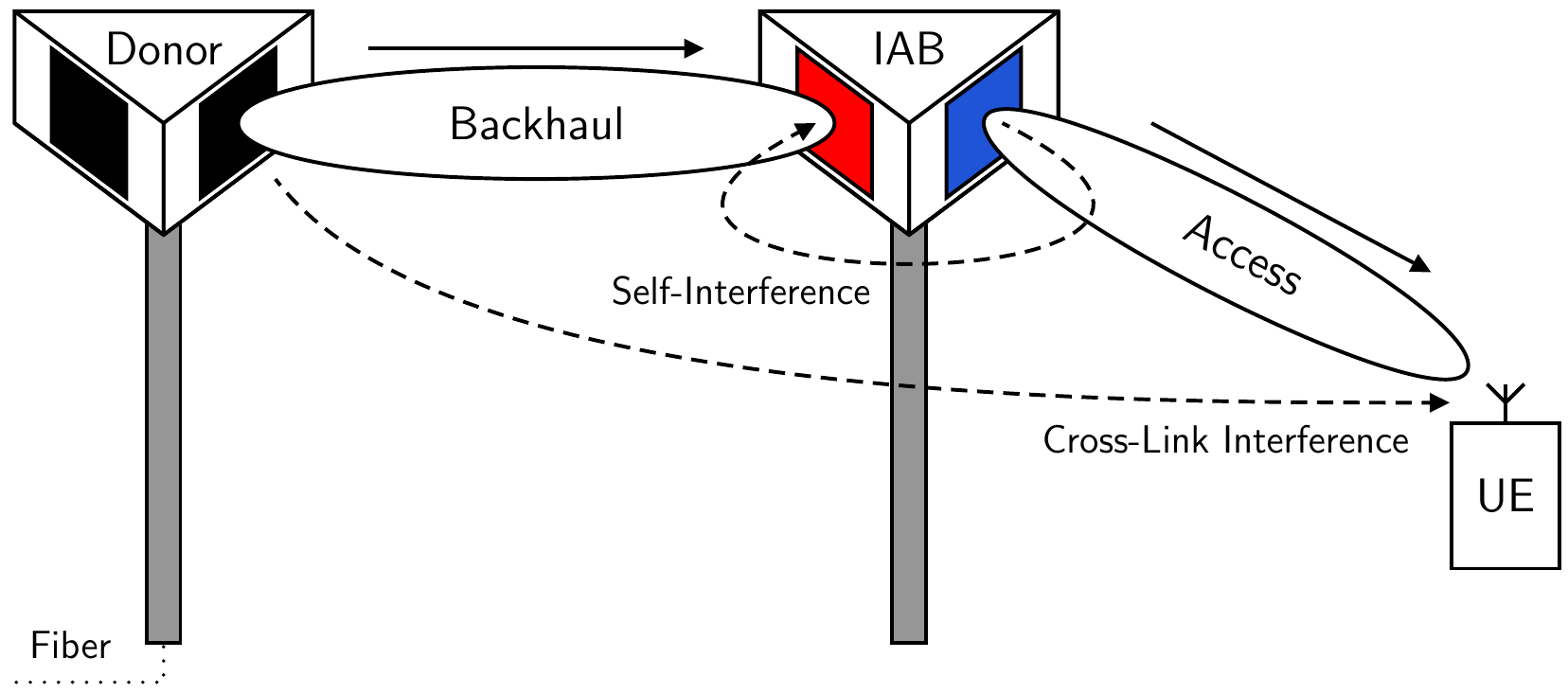}
    \caption{A full-duplex \iab node receives backhaul from a fiber-connected donor node while simultaneously transmitting access to a \ue, giving rise to self-interference at the \iab node and cross-link interference at the \ue. We refer to this as the \dldl operating mode. We also consider the \ulul mode where the \ue transmits uplink access and the \iab node transmits backhaul.}
    \label{fig:network}
\end{figure}

As a relevant application of this work, we consider the \mmwave communication system shown in \figref{fig:network}, where a sectorized, multi-panel \iab node maintains backhaul and serves access in a full-duplex fashion (i.e., on the same time-frequency resources) \cite{cudak21integrated}.
This is particularly attractive application of full-duplex in \mmwave cellular systems \cite{iab,3GPP_IAB}, but the contributions of this work are not limited to \iab.
% In this work, we are particularly interested in the application of \steer in full-duplex \iab networks, where juggling access and backhaul simultaneously can increase spectral efficiency and significantly reduce latency, compared to half-duplex operation \cite{gupta_fdiab_arxiv}.
Illustrated in \figref{fig:network} is the downlink-downlink (\dldl) operating mode, where an \iab node receives backhaul from a fiber-connected donor and transmits access to a \ue.
In this work, we present a formulation and design that generalizes to both the \dldl operating mode and the analogous uplink-uplink (\ulul) mode where the \iab node receives access and transmits backhaul.
We are particularly interested in these full-duplex operating modes since they unlock scheduling opportunities that can reduce latency in \iab networks while also increasing throughput \cite{gupta_fdiab_arxiv}.
It is important to consider and evaluate both full-duplexing modes since there may exist significant disparities between the donor and \ue---most notably transmit power, noise power, and number of antennas.
% \edit{Unlocking the \dldl and \ulul operating modes with full-duplex can translate to impressive latency and throughput gains at the network level by expanding scheduling opportunities.}

Separate \upas are present at the \iab node, each of which can either transmit or receive and can be independently configured via a network of analog beamforming weights.
In this multi-panel full-duplex setting, one array will transmit while the other receives.
To simplify notation between the \dldl and \ulul modes, we assume each array at the \iab node is equipped with $\Na$ antennas.
We denote the vector of transmit beamforming weights at the \iab node as $\vf \in \setvectorcomplex{\Na}$.
Likewise, the receive beamforming vector at the \iab node is denoted $\vw \in \setvectorcomplex{\Na}$.
For transmit power and noise power normalizations, we assume that the beamforming weights have unit power as $\normtwo{\vf}^2 = \normtwo{\vw}^2 = 1$.
Extending this work to systems with multiple beams at the transmitter and receiver would be great future work.

For simplicity, we assume the \ue is a single-antenna device, though the work herein could extend naturally to those with multiple antennas.
Let the row vector $\vhac\ctrans \in \setmatrixcomplex{1}{\Na}$ be the access channel between the transmit array of the \iab node and the \ue.
Practically, the donor will have an antenna array through which it serves backhaul.
With an array at the donor and an array at the \iab node, we use $\mHbh$ to denote the \mimo backhaul channel matrix between the donor and the receive panel of the \iab node.
We assume the donor transmits with some beamforming weights $\vv$ and instead consider henceforth the column vector
\begin{align}
\vhbh \triangleq \mHbh \vv \in \setmatrixcomplex{\Na}{1}
\end{align}
which is the effective backhaul channel from the beamformed donor to the \iab node---abstracting out beamforming at the donor.
We normalize the access and backhaul channel vectors as $\normtwo{\vhac}^2 = \normtwo{\vhbh}^2 = \Na$
% \begin{align}
% \normtwo{\vhac}^2 &= \Na, \quad \normtwo{\vhbh}^2 = \Na
% \end{align}
and abstract out their large-scale path gains (inverse path loss) as $\Gac^2$ and $\Gbh^2$, respectively.
% As such, we abstract out beamforming at the donor.
% Let the column vector $\vhbh \in \setmatrixcomplex{\Nr}{1}$ be the channel channel between the receive array of the \iab and the donor it intends to receive from.
We invite readers to assume access and backhaul channels that are \gls{los} for simplicity, but this work does not depend on such.
% In this work, we assume the donor and \ue are in \gls{los} of the \iab node---this offers simplification that could be relaxed with slight modifications in future work.

In the \dldl operating mode---when transmitting access and receiving backhaul from the \iab node in a full-duplex fashion---a \mimo self-interference channel $\mHsi \in \setmatrixcomplex{\Na}{\Na}$ manifests between the transmit and receive arrays of the \iab node.
% During \ulul mode, the direction of self-interference is reversed, meaning its channel becomes $\mHsi\ctrans$.
We similarly abstract out its large-scale path gain as $\Gsi^2$ and enforce $\normfro{\mHsi}^2 = \Na^2$.
In addition, during \dldl, since the donor transmits backhaul while the \ue receives access, a cross-link interference channel $\vhiui\ctrans$ (a row vector) manifests between the donor's antenna array and the \ue.
Having conditioned on some beamforming weights $\vv$ at the donor, the effective cross-link interference channel is the scalar
\begin{align}
\hiui \triangleq \vhiui\ctrans \vv \in \setvectorcomplex{1}.
\end{align}
% In the \ulul mode, the cross-link interference channel reverses to become $\hiui\ctrans$.
We similarly abstract out its large-scale gain as $\Giui^2$ by letting $\bars{\hiui}^2 = 1$.
Symbols are transmitted by the donor, \iab node, and \ue with powers (in watts) $\powertxdonor$, $\powertxiab$, and $\powertxue$, respectively.
Additive noise incurred at the donor, \iab node, and \ue have respective powers (in watts) $\powernoisedonor$, $\powernoiseiab$, and $\powernoiseue$.
With these definitions in hand, we can formulate the quality of each link in the \dldl and \ulul modes.

\comment{
---

The \glspl{snr} of the backhaul and access links during \dldl are respectively
\begin{align}
\snrbhdl &= \frac{\powertxdonor \cdot \Gbh^2 \cdot \bars{\vw\ctrans \vhbh}^2}{\powernoiseiab}, \quad
\snracdl = \frac{\powertxiab \cdot \Gac^2 \cdot \bars{\vhac\ctrans \vf}^2}{\powernoiseue}.
\end{align}
Likewise, during \ulul mode, we have
\begin{align}
\snrbhul &= \frac{\powertxiab \cdot \Gbh^2 \cdot \bars{\vhbh\ctrans \vf}^2}{\powernoisedonor}, \quad
\snracul = \frac{\powertxue \cdot \Gac^2 \cdot \bars{\vw\ctrans \vhac}^2}{\powernoiseiab}.
\end{align}
By virtue of the fact that
\begin{align}
% \Na &= \max_{\vf} \ \bars{\vh\ctrans \vf}^2 \ \st \normtwo{\vf}^2 = 1, \normtwo{\vh}^2 = \Na \\
% \Na &= \max_{\vw} \ \bars{\vw\ctrans \vh}^2 \ \st \normtwo{\vw}^2 = 1, \normtwo{\vh}^2 = \Na
\Na &= \max_{\vx} \ \bars{\vh\ctrans \vx}^2 \ \st \normtwo{\vx}^2 = 1, \normtwo{\vh}^2 = \Na
\end{align}
the maximum \gpsnr, which we denote with an overline, are 
%\begin{align}
%\snrbhdlbar &= \frac{\powertxdonor \cdot \Gbh^2 \cdot \Na}{\powernoiseiab} \geq \snrbhdl \\
%\snracdlbar &= \frac{\powertxiab \cdot \Gac^2 \cdot \Na}{\powernoiseue} \geq \snracdl \\
%\snrbhulbar &= \frac{\powertxiab \cdot \Gbh^2 \cdot \Na}{\powernoisedonor} \geq \snrbhul \\
%\snraculbar &= \frac{\powertxue \cdot \Gac^2 \cdot \Na}{\powernoiseiab} \geq \snracul.
%\end{align}
\begin{gather}
\snrbhdlbar = \frac{\powertxdonor \cdot \Gbh^2 \cdot \Na}{\powernoiseiab} \geq \snrbhdl, \quad
\snracdlbar = \frac{\powertxiab \cdot \Gac^2 \cdot \Na}{\powernoiseue} \geq \snracdl \\
\snrbhulbar = \frac{\powertxiab \cdot \Gbh^2 \cdot \Na}{\powernoisedonor} \geq \snrbhul, \quad
\snraculbar = \frac{\powertxue \cdot \Gac^2 \cdot \Na}{\powernoiseiab} \geq \snracul.
\end{gather}
These are achieved by beamforming \textit{directly} toward the donor and \ue to deliver maximum gain.

During \dldl mode, backhaul is corrupted by self-interference and access is corrupted by cross-link interference, leading to the \ginr of each being
% The \ginr of the backhaul and access links during downlink-downlink mode are
\begin{align}
\inrbhdl &= \frac{\powertxiab \cdot \Gsi^2 \cdot \bars{\vw\ctrans \mHsi \vf}^2}{\powernoiseiab}, \quad
\inracdl = \frac{\powertxdonor \cdot \Giui^2 \cdot \bars{\hiui}^2}{\powernoiseue}.
\end{align}
In \ulul mode, backhaul is corrupted by cross-link interference and access is corrupted by self-interference, yielding
\begin{align}
\inrbhul &= \frac{\powertxue \cdot \Giui^2 \cdot \bars{\hiui\ctrans}^2}{\powernoisedonor}, \quad
\inracul = \frac{\powertxiab \cdot \Gsi^2 \cdot \bars{\vw\ctrans \mHsi\ctrans \vf}^2}{\powernoiseiab}.
\end{align}
Notice that the degree of self-interference depends on its channel $\mHsi$ and the beamformers $\vf$ and $\vw$ at the \iab node.
Cross-link interference, on the other hand, does not depend on $\vf$ nor $\vw$ and is fixed for a given setting, having conditioned on the donor's beamformer $\vv$.

\begin{align}
\snrtx = 
\frac{\powertxiab \cdot \Gtx^2 \cdot \bars{\vhtx\ctrans \vf}^2}{\powernoiseue} \\
\quad
\snrrx = 
\begin{dcases}
\frac{\powertxdonor \cdot \Gbh^2 \cdot \bars{\vw\ctrans \vhbh}^2}{\powernoiseiab}, & \casedl \\
\frac{\powertxue \cdot \Gac^2 \cdot \bars{\vw\ctrans \vhac}^2}{\powernoiseiab}, & \caseul
\end{dcases}
\end{align}

\begin{align}
\vhtx = 
\begin{cases}
\vhac, & \casedl \\
\vhbh, & \caseul
\end{cases}
\end{align}

\begin{align}
\snrtxbar &= 
\begin{dcases}
\frac{\powertxiab \cdot \Gac^2 \cdot \bars{\vhac\ctrans \vf}^2}{\powernoiseue}, & \casedl \\
\frac{\powertxiab \cdot \Gbh^2 \cdot \bars{\vhbh\ctrans \vf}^2}{\powernoisedonor}, & \caseul
\end{dcases} \\
\snrrxbar &= 
\begin{dcases}
\frac{\powertxdonor \cdot \Gbh^2 \cdot \bars{\vw\ctrans \vhbh}^2}{\powernoiseiab}, & \casedl \\
\frac{\powertxue \cdot \Gac^2 \cdot \bars{\vw\ctrans \vhac}^2}{\powernoiseiab}, & \caseul
\end{dcases}
\end{align}

\begin{align}
\snrtx = 
\begin{dcases}
\frac{\powertxiab \cdot \Gac^2 \cdot \bars{\vhac\ctrans \vf}^2}{\powernoiseue}, & \casedl \\
\frac{\powertxiab \cdot \Gbh^2 \cdot \bars{\vhbh\ctrans \vf}^2}{\powernoisedonor}, & \caseul
\end{dcases},
\quad
\snrrx = 
\begin{dcases}
\frac{\powertxdonor \cdot \Gbh^2 \cdot \bars{\vw\ctrans \vhbh}^2}{\powernoiseiab}, & \casedl \\
\frac{\powertxue \cdot \Gac^2 \cdot \bars{\vw\ctrans \vhac}^2}{\powernoiseiab}, & \caseul
\end{dcases}
\end{align}
}

\section{Problem Formulation and Motivation} \label{sec:problem-formulation}

\begin{table}[!t]
    \small
    \centering
    \caption{The transmit and receive links during different full-duplexing modes.}
    \label{tab:tl-rl}
    \begin{tabular}{|ccc|}
        \hline
        \textbf{Mode} & \textbf{Transmit Link} & \textbf{Receive Link} \\
        \hline
        Downlink-Downlink & Access & Backhaul \\
        \hline
        Uplink-Uplink & Backhaul & Access \\
        \hline
    \end{tabular}
\end{table}

We leverage our formulations of the \dldl and \ulul modes to present a general formulation of the system.
Taking the perspective of the full-duplex \iab node, we introduce the terms \textit{transmit link} and \textit{receive link}, which correspond to either access or backhaul depending on if the system operates in \dldl or \ulul mode, as summarized in \tabref{tab:tl-rl}.
% To generalize between the two modes, we take the perspective of the \iab node and refer to the transmit and receive links.
In the \dldl mode, for instance, the \gsnr of the transmit link and receive link can be expressed respectively as
\begin{align}
\snrtx &= \frac{\powertxiab \cdot \Gac^2 \cdot \bars{\vhac\ctrans \vf}^2}{\powernoiseue}, \qquad
\snrrx = \frac{\powertxdonor \cdot \Gbh^2 \cdot \bars{\vw\ctrans \vhbh}^2}{\powernoiseiab}.
\end{align}
By virtue of the fact that $\Na = \max_{\vx} \ \bars{\vh\ctrans \vx}^2 \ \st \normtwo{\vx}^2 = 1, \normtwo{\vh}^2 = \Na$,
%\begin{align}
%% \Na &= \max_{\vf} \ \bars{\vh\ctrans \vf}^2 \ \st \normtwo{\vf}^2 = 1, \normtwo{\vh}^2 = \Na \\
%% \Na &= \max_{\vw} \ \bars{\vw\ctrans \vh}^2 \ \st \normtwo{\vw}^2 = 1, \normtwo{\vh}^2 = \Na
%\Na &= \max_{\vx} \ \bars{\vh\ctrans \vx}^2 \ \st \normtwo{\vx}^2 = 1, \normtwo{\vh}^2 = \Na
%\end{align}
the maximum \gpsnr during the \dldl mode, which we denote with an overline, are 
\begin{gather}
\snrtxbar = \frac{\powertxiab \cdot \Gac^2 \cdot \Na}{\powernoiseue} \geq \snrtx, \qquad
\snrrxbar = \frac{\powertxdonor \cdot \Gbh^2 \cdot \Na}{\powernoiseiab} \geq \snrrx.
\end{gather}
These are achieved by beamforming {directly} toward the \ue and donor to deliver maximum gain.
In the \dldl mode, access is corrupted by cross-link interference and backhaul is corrupted by self-interference, leading to the \ginr of the transmit and receive links being respectively
\begin{align}
\inrtx &= \frac{\powertxdonor \cdot \Giui^2 \cdot \bars{\hiui}^2}{\powernoiseue}, \qquad
\inrrx = \frac{\powertxiab \cdot \Gsi^2 \cdot \bars{\vw\ctrans \mHsi \vf}^2}{\powernoiseiab}. \label{eq:inr-tx-rx}
\end{align}
Notice that the degree of self-interference depends on its channel $\mHsi$ and the beamformers $\vf$ and $\vw$ at the \iab node.
Cross-link interference, on the other hand, does not depend on $\vf$ nor $\vw$ and is fixed for a given setting, having conditioned on the donor's beamformer $\vv$.
All terms presented here for the \dldl mode can be defined analogously for the \ulul mode, with backhaul as the transmit link and access as the receive link.
Note that, regardless of operating mode, the transmit link is plagued by cross-link interference and the receive link is plagued by self-interference.

With these formulations of the transmit and receive links, their \gpsinr can be written as
\begin{align}
\sinrtx = \frac{\snrtx}{1+\inrtx}, \qquad 
\sinrrx = \frac{\snrrx}{1+\inrrx}. \label{eq:sinr-tx-rx}
\end{align}
Treating interference as noise, the maximum achievable spectral efficiencies on each link are
\begin{align}
\setx &= \logtwo{1 + \sinrtx}, \qquad
\serx = \logtwo{1 + \sinrrx} \label{eq:se-tx-rx}
\end{align}
whereas the individual Shannon capacities of each of these links are
\begin{align}
\captx &= \logtwo{1 + \snrtxbar}, \qquad
\caprx = \logtwo{1 + \snrrxbar}. % \\ \capsum &= \capul + \capdl
\end{align}
% In this paper, our goal is to present a methodology for beam selection that yields high $\setx$ and $\serx$.
% In other words, we seek a means to strategically choose $\vf$ and $\vw$ such that $\setx$ and $\serx$ can be improved over conventional/\naive approaches.
In this paper, we seek a means to strategically select beams $\vf$ and $\vw$ such that self-interference can be significantly reduced and spectral efficiency can be improved over conventional/\naive approaches.
Taking a full-duplex perspective, we desire a sum spectral efficiency $\setx + \serx$ that approaches the full-duplex capacity $\captx + \caprx$.
In this pursuit, we account for a number of practical considerations in this work, most notably limited channel knowledge and practical codebook-based beam alignment.

\begin{figure*}
    \centering
    \includegraphics[width=\linewidth,height=0.3\textheight,keepaspectratio]{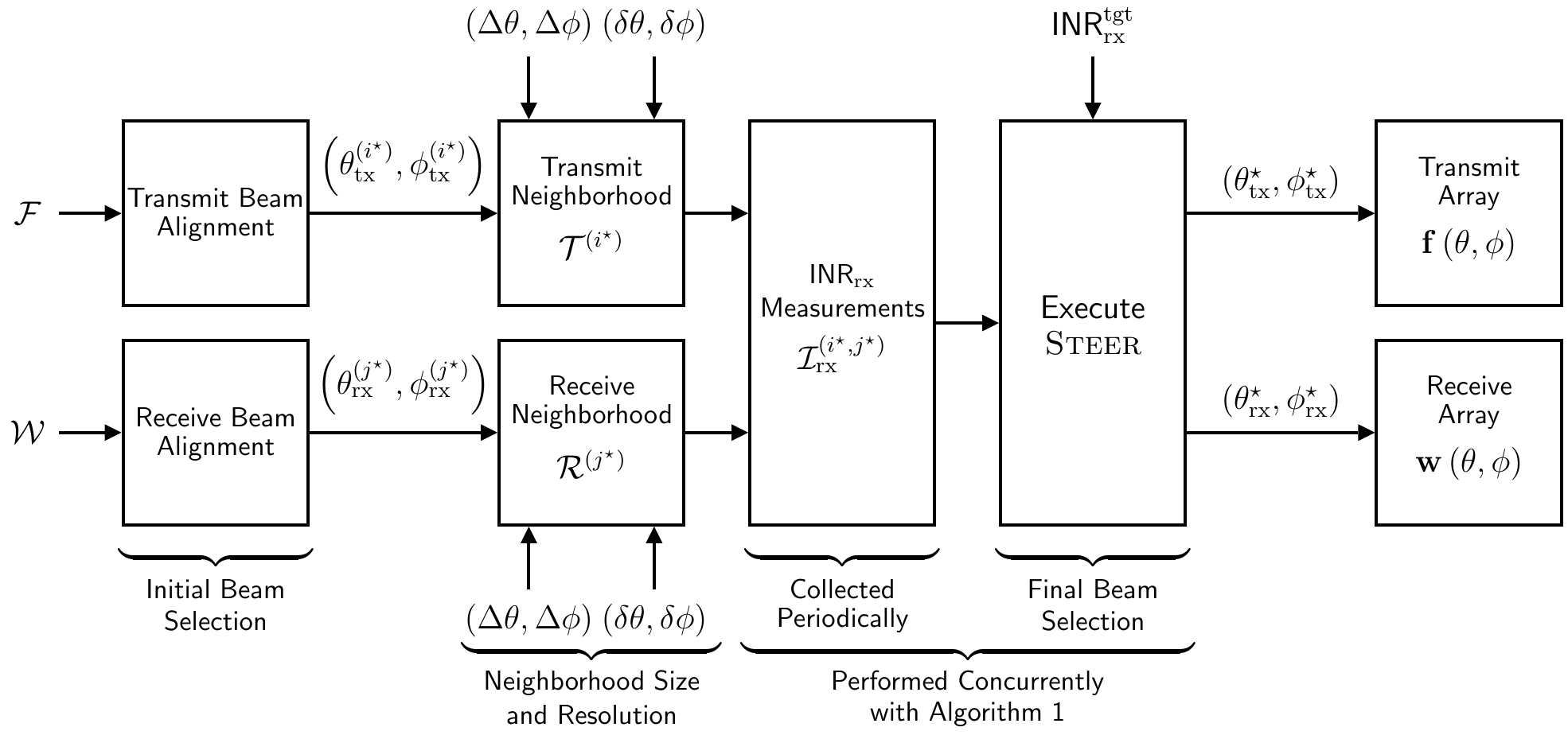}
    \caption{A block diagram summarizing beam selection via \steer, which jointly selects transmit and receive beams to reduce self-interference. Conventional beam selection chooses transmit and receive beams independently, ignoring self-interference.}
    \label{fig:block-diagram}
\end{figure*}

% In our recent measurement campaign of \mmwave self-interference \cite{roberts_2021_att_measure,roberts_att_angular}, we observed that small shifts in the steering directions of the transmit and receive beams (on the order of one degree) can lead to noteworthy reductions in self-interference (i.e., $\inrrx$).
% In this work, we aim to leverage this phenomena by slightly shifting transmit and receive beams so that self-interference can be reduced while still delivering high beamforming gain (i.e., high $\snrtx$ and $\snrrx$) in the desired directions.
% In our recent measurement campaign of \mmwave self-interference \cite{roberts_2021_att_measure,roberts_att_angular}, we observed that small shifts in the steering directions of the transmit and receive beams (on the order of one degree) can lead to noteworthy reductions in self-interference (i.e., $\inrrx$).
In this work, we aim to leverage small-scale phenomena observed in our recent measurement campaign \cite{roberts_2021_att_measure,roberts_att_angular} by slightly shifting transmit and receive beams so that self-interference (i.e., $\inrrx$) can be reduced while still delivering high beamforming gain (i.e., high $\snrtx$ and $\snrrx$) in the desired directions.
To do this in a systematic manner, we introduce the first beam selection methodology specifically for full-duplex \mmwave systems, called \steer, that makes use of self-interference measurements to choose beams that reduce self-interference and facilitate full-duplex operation.
\edit{In fact, we show that \steer can reduce self-interference to levels sufficiently low for full-duplex operation without the need for any additional analog nor digital cancellation. This is particularly desirable because it removes the need for additional hardware and signal processing that conventional self-interference cancellation strategies demand.}
In the sections that follow, we present the three components of \steer:
\begin{enumerate}
    \item initial beam selection via conventional beam alignment (\secref{sec:initial-beam-selection});
    \item measurements of self-interference across small spatial neighborhoods (\secref{sec:measure}); these are not necessarily taken in real-time but rather periodically as needed;
    \item final beam selection to minimize self-interference (\secref{sec:steer}).
\end{enumerate}
A block diagram summarizing \steer is shown in \figref{fig:block-diagram}, whose details will become clear as we present our design in the next three sections.

\comment{
---

We leverage our formulations of the \dldl and \ulul modes to present a general formulation of the system.
Taking the perspective of the full-duplex \iab node, we introduce the terms \textit{transmit link} and \textit{receive link}, which correspond to either access or backhaul depending on if the system operates in \dldl or \ulul mode, as summarized in \tabref{tab:tl-rl}.
We use this to define the \gsnr of the transmit and receive links as follows.
\begin{align}
\snrtx = 
\begin{cases}
\snracdl, & \casedl \\
\snrbhul, & \caseul
\end{cases},
\quad
\snrrx = 
\begin{cases}
\snrbhdl, & \casedl \\
\snracul, & \caseul
\end{cases}
\end{align}
% In a similar fashion, 
The \ginr on the transmit and receive links during the two modes can be expressed similarly, where
%\begin{align}
%\inrtx = 
%\begin{cases}
%\inracdl, & \casedl \\
%\inrbhul, & \caseul
%\end{cases},
%\quad
%\inrrx = 
%\begin{cases}
%\inrbhdl, & \casedl \\
%\inracul, & \caseul
%\end{cases}.
%\end{align}
$\inrrx$ corresponds to self-interference and $\inrtx$ corresponds to cross-link interference.
}

\section{Initial Beam Selection: Conventional Beam Alignment} \label{sec:initial-beam-selection}

% Practical \mmwave communication systems rely on beam alignment. % to support initial access and to configure their beams in a manner that delivers sufficient link margin without the need for explicit channel knowledge, which is difficult to obtain.
% Beam alignment schemes typically involve sweeping candidate beams, measuring the \gls{rsrp} for each candidate, and delivering feedback before determining the beam(s) for the \mmwave link \cite{heath_overview_2016,ethan_beam}.
Practical \mmwave communication systems rely on beam alignment schemes to deliver high beamforming gain.
These schemes typically involve sweeping candidate beams, measuring the \gls{rsrp} for each candidate, and delivering feedback before determining the beam(s) for the \mmwave link \cite{heath_overview_2016,ethan_beam}.
Candidate beams often come from a codebook, which is constructed by first defining a service region (some portion of space based on an assumed user distribution) and then discretizing it based on the desired number of beams in the codebook or their beamwidth.
% Naturally, to maximize coverage, one would desire many highly directional beams so that users can be served with near maximum gain regardless of their location.
% wherever users are located, they can be served with high gain.
% Practically, however, codebook-based beam alignment must sacrifice some degree of coverage for a desirably low number of beams in favor of reduced beam alignment overhead.
% keeping the number of beams low to avoid the costly overhead associated with searching through the codebook during beam alignment.

In a traditional half-duplex fashion, we suppose the \iab node conducts beam alignment on its transmit link with $\Ntx$ beams and on its receive link with $\Nrx$ beams.
The $\Ntx$ transmit beams and $\Nrx$ receive beams are spatially distributed over their desired coverage regions, where each beam is responsible for serving some portion of its respective region.
Describing the steering direction of each beam in an azimuth-elevation fashion, the collection of transmit directions $\txdirsetcb$ and receive directions $\rxdirsetcb$ we write as
\begin{align}
\txdirsetcb &= \braces{\parens{\thetatx\idx{i},\phitx\idx{i}} : i = 1,\dots,\Ntx}, \qquad 
\rxdirsetcb = \braces{\parens{\thetarx\idx{j},\phirx\idx{j}} : j = 1,\dots,\Nrx}.
\end{align}
% For instance, as illustrated in \figref{fig:codebook}, 
For instance, suppose $\txdirsetcb$ and $\rxdirsetcb$ are each comprised of $\Ntx = \Nrx$ directions distributed uniformly from $-60^\circ$ to $60^\circ$ in azimuth and from $-30^\circ$ to $30^\circ$ in elevation.
% For instance, suppose $\txdirsetcb$ and $\rxdirsetcb$ are each comprised of $\Ntx = \Nrx = 105$ directions distributed uniformly in azimuth from $-56^\circ$ to $56^\circ$ in $8^\circ$ steps and in elevation from $-24^\circ$ to $24^\circ$ also in $8^\circ$ steps.

%\begin{figure}
%    \centering
%    \includegraphics[width=\linewidth,height=0.18\textheight,keepaspectratio]{plots/main_frisbee_v01_04}
%    \caption{Projected onto the Cartesian plane are the steering directions (i.e., $\txdirsetcb$,  $\rxdirsetcb$) of a codebook of $105$ beams used for beam alignment (i.e., $\precb$,  $\comcb$), ranging from $-56^\circ$ to $56^\circ$ in azimuth and $-24^\circ$ to $24^\circ$ in elevation, each with $8^\circ$ spacing.}
%    % \caption{A conventional codebook of $105$ beams used for beam alignment (i.e., $\precb$ or $\comcb$), ranging from $-56^\circ$ to $56^\circ$ in azimuth and $-24^\circ$ to $24^\circ$ in elevation, each with $8^\circ$ spacing.}
%    \label{fig:codebook}
%\end{figure}

Practical phased array systems are often equipped with a mapping from desired steering direction to beamforming weights based on some beam design methodology.\footnote{It is not uncommon for this mapping to be proprietary and to account for nonidealities in the array pattern.} % , which can be thought of as effectively constructing transmit and receive beam codebooks.
% In other words, beam weights are configured solely based on the desired steering direction and this mapping. % and cannot be mapped
%As such, let $\precb$ and $\comcb$ be the transmit and receive codebooks used for conventional beam alignment
%\begin{align}
%\precb &= \braces{\vf\parens{\theta,\phi} : \thph \in \txdirsetcb} \\
%\comcb &= \braces{\vw\parens{\theta,\phi} : \thph \in \rxdirsetcb}
%\end{align}
As such, let $\precb = \braces{\vf\parens{\theta,\phi} : \thph \in \txdirsetcb}$ and $\comcb = \braces{\vw\parens{\theta,\phi} : \thph \in \rxdirsetcb}$ be the transmit and receive codebooks used for conventional beam alignment, where $\vf\thph$ and $\vw\thph$ are transmit and receive weights designed to steer toward some $\thph$.
Let $\vhtx\ctrans$ and $\vhrx$ be the transmit and receive channels corresponding to the particular operating mode.
Conventional beam alignment on the transmit link aims to solve (or approximately solve)
\begin{align}
% \thphtxbar = \arg \max_{\thph \in \txdirsetcb} \ \bars{\vhtx\ctrans \vf\thph}^2 \\
i\opt = \argmax_{i \in \braces{1,\dots,\Ntx}} \ \bars{\vhtx\ctrans \vf\thphtxi}^2
\end{align}
to identify the transmit beam $\vf\thphtxiopt \in \precb$ that maximizes beamforming gain delivered on the transmit link.
% the transmit direction $\thphtxiopt \in \txdirsetcb$ and the corresponding transmit beam $\vf\thphtxiopt \in \precb$ that maximizes beamforming gain delivered on the transmit link.
%\begin{align}
%\thphtxbar \triangleq \thphtxiopt
%\end{align}
In other words, the transmit link user---either the donor or the \ue depending on the mode---is approximately located in the direction $\thphtxiopt$ from the transmit panel of the \iab node.
Likewise, beam alignment on the receive link aims to solve
\begin{align}
% \thphrxbar = \arg \max_{\thph \in \rxdirsetcb} \ \bars{\vw\thph\ctrans \vhrx}^2 \\
j\opt = \argmax_{j \in \braces{1,\dots,\Nrx}} \ \bars{\vw\thphrxj\ctrans \vhrx}^2 
\end{align}
which identifies the approximate direction of the receive link user.
%\begin{align}
%\thphrxbar \triangleq \thphrxjopt
%\end{align}
In practice, solving these optimization problems is typically done through a series of of \gls{rsrp} measurements and feedback between the \iab node and the user it aims to serve; recall, we do not have knowledge of $\vhtx$ nor $\vhrx$.
As the first stage of our design, we propose that beam alignment be executed in a half-duplex fashion to yield some $\thphtxiopt$ and $\thphrxjopt$, though we do not suggest any particular scheme for doing so.
As such, our design can accommodate existing beam alignment schemes without changes (including hierarchical schemes) and does not introduce any additional over-the-air feedback. % along the transmit link or receive link beyond what is inherent to the scheme.
If using the beams from conventional beam alignment, the \textit{nominal} \gpsnr of the transmit and receive links are 
%\begin{align}
%\snrtxnom &\triangleq \snrtxbar \cdot \frac{\bars{\vhtx\ctrans \vf\thphtxiopt}^2}{\Na} \\
%\snrrxnom &\triangleq \snrrxbar \cdot \frac{\bars{\vw\thphrxjopt\ctrans \vhrx}^2}{\Na} .
%\end{align}
\begin{align}
\snrtxnom &\triangleq \snrtxbar \cdot \frac{\bars{\vhtx\ctrans \vf\thphtxiopt}^2}{\Na}, \quad
\snrrxnom \triangleq \snrrxbar \cdot \frac{\bars{\vw\thphrxjopt\ctrans \vhrx}^2}{\Na}. \label{eq:snr-tx-rx-nom}
\end{align}
%and the capacities achieved by these beams are
%\begin{align}
%\captxcb &= \logtwo{1 + \snrtxnom} \leq \captx \\
%\caprxcb &= \logtwo{1 + \snrrxnom} \leq \caprx \\
%\capsumcb &= \captxcb + \caprxcb \leq \captx + \caprx
%\end{align}
These \gpsnr are some fraction of the maximum link \gpsnr based on how effectively the selected beams from conventional beam alignment steer toward the transmit and receive users, which naturally depends on their locations, the environment, and the beam codebooks.
The beams output by conventional beam selection will initialize \steer's pursuit to find beams that offer high \gsnr and reduced self-interference.
In doing so, \steer relies on measurements outlined in the following section.

\section{Measuring Self-Interference Across Small Spatial Neighborhoods} \label{sec:measure}

% In our recent measurement campaign of \mmwave self-interference \cite{roberts_2021_att_measure,roberts_att_angular}, we observed that small shifts in the steering directions of the transmit and receive beams (on the order of one degree) can lead to noteworthy reductions in self-interference.
% In this work, we aim to leverage this phenomena by shifting transmit and beams slightly from those output by conventional beam selection in an effort to reduce self-interference while still delivering high beamforming gain.
% Recall that our design of \steer is motivated by the small-scale phenomena observed in our recent measurement campaign of \mmwave self-interference \cite{roberts_2021_att_measure,roberts_att_angular}, where we saw that small shifts in the steering directions of the transmit and receive beams (on the order of one degree) can lead to noteworthy reductions in self-interference.
% \steer is motivated by our recent measurement campaign of \mmwave self-interference \cite{roberts_2021_att_measure,roberts_att_angular}, in which we discovered that slight shifts in the steering directions of the transmit and receive beams (on the order of one degree) can greatly reduce self-interference.
% \steer is motivated by our recent measurement campaign \cite{roberts_2021_att_measure,roberts_att_angular}, in which we discovered that slightly shifting the steering directions of the transmit and receive beams (on the order of one degree) can greatly reduce self-interference.
Our recent measurement campaign \cite{roberts_2021_att_measure,roberts_att_angular} illustrated that slightly shifting the steering directions of the transmit and receive beams (on the order of one degree) can greatly reduce self-interference.
% This motivates us to consider 
% The transmit and receive beams output by our design will steer in \textit{approximately} the same directions as those identified by conventional beam alignment in the previous section.
% In other words, the final transmit and receive beam selections made by \steer will be approximately toward $\thphtxiopt$ and $\thphrxjopt$.
To find promising transmit and receive beams (ones that offer reduced self-interference) that steer in {approximately} the same directions as those identified by conventional beam alignment, we explicitly measure self-interference incurred by a number of candidate beams over a small spatial neighborhood. % , which we outline in this section.
As we will cover shortly, these measurements will not necessarily be taken in real-time (upon each beam selection) but rather may be collected periodically then referenced in real-time, according to the dynamics of self-interference.

If transmitting toward $\thphtx$ and receiving toward $\thphrx$, the \iab node incurs some degree of self-interference, which can be theoretically computed based on \eqref{eq:inr-tx-rx},
% as
% \begin{align} \label{eq:estimate-si}
% \powertxiab \cdot \Gsi^2 \cdot \bars{\vw\thphrx\ctrans \mHsi \vf\thphtx}^2
% \end{align}
assuming knowledge of $\vf\thphtx$, $\vw\thphrx$, $\Gsi$, and $\mHsi$.
Practically, however, it is difficult to efficiently and accurately estimate the self-interference channel matrix $\mHsi$ (which is large). 
% Moreover, knowledge of practical \mmwave self-interference is extremely limited and there do not yet exist measurement-backed \mimo self-interference channel models for \mmwave, meaning its estimation and the efficiency and reliability of such are currently uncertain.
Moreover, characterization and modeling of \mmwave self-interference is extremely limited, and it is currently impractical for a system to predict what level of self-interference it would incur with a particular transmit beam $\vf\thphtx$ and receive beam $\vw\thphrx$. % \footnote{Assuming one were to know the actual beamforming weights set by a phased array's built-in mapping.}.
% As a result, it is currently impractical for a system to predict what levels of self-interference it would incur with a particular transmit beam $\vf\thphtx$ and receive beam $\vw\thphrx$\footnote{Assuming one were to know the actual beamforming weights set by a phased array's built-in mapping.}.
% This is especially true since nonidealities in the arrays (e.g., phase or amplitude inconsistencies across elements) and/or slight imperfections in the beamforming weights (e.g., due to limited resolution phase shifters and attenuators) can lead to unpredictable levels of self-interference.
% This is especially true since nonidealities in the arrays and/or slight imperfections in the beamforming weights can lead to unpredictable levels of self-interference.
All of this---combined with the fact that minor errors in self-interference power can make a significant difference in full-duplex system performance---motivates us to explicitly measure self-interference incurred at the \iab node for particular transmit and receive beams, rather than attempt to estimate it. % via \eqref{eq:estimate-si}.

% Specifically, we will measure self-interference across a number of transmit and receive beam combinations

% $\Gsi \cdot \mHsi$ and the beams $\vf\thphtxi$ and $\vw\thphrxj$.

\begin{figure}
    \centering
    \includegraphics[width=\linewidth,height=0.15\textheight,keepaspectratio]{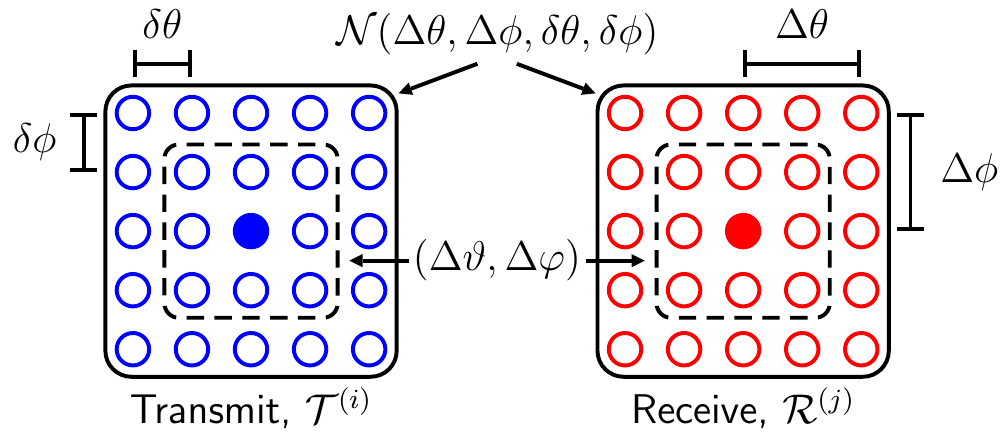}
    \caption{The spatial neighborhoods surrounding a given transmit direction and receive direction (shown as filled circles). The size of the neighborhoods is dictated by \nbr and their resolution by \nbrd. \nbrv will be relevant in \secref{sec:steer}.}
    \label{fig:neighborhood}
\end{figure}

To identify attractive steering directions for full-duplex operation, we are interested in measuring the self-interference incurred when transmitting and receiving around the \textit{spatial neighborhoods} surrounding a given transmit direction and receive direction, as illustrated in \figref{fig:neighborhood}.
Quantifying the size of these spatial neighborhoods, let $\Deltatheta$ and $\Deltaphi$ be maximum absolute azimuthal and elevational deviations from the given transmit direction and receive direction. %  allowed by our design.
% The spatial neighborhood can be thought of as living within the codebook beam spacing; for instance, $\nbr = \nbrtwotwo$ when the codebook beam spacing is $(8^\circ,8^\circ)$.
The spatial neighborhood can be thought of as living within the codebook beam spacing; for instance, $\nbr = \nbrtwotwo$ when the codebook beams are separated by $(8^\circ,8^\circ)$.
Discretizing these neighborhoods, let $\deltatheta$ and $\deltaphi$ be the measurement resolution in azimuth and elevation, respectively, which should not be larger than $\nbr$---e.g., $\nbrd = \nbroneone$.
% This concept of discretized transmit and receive spatial neighborhoods is mathematically defined as follows.
The spatial neighborhood $\setnbr$ surrounding a transmit/receive direction can be expressed using
the azimuthal neighborhood $\setnbrth$ and elevational neighborhood $\setnbrph$ defined as
\begin{align}
\setnbrth\nbrnbrdth &= \braces{m \cdot \deltatheta : m \in \brackets{-\floor{\frac{\Deltatheta}{\deltatheta}},\floor{\frac{\Deltatheta}{\deltatheta}}}} \\
\setnbrph\nbrnbrdph &= \braces{n \cdot \deltaphi : n \in \brackets{-\floor{\frac{\Deltaphi}{\deltaphi}},\floor{\frac{\Deltaphi}{\deltaphi}}}}
\end{align}
where $\floor{\cdot}$ is the floor operation and $\brackets{a,b} = \braces{a,a+1,\dots,b-1,b}$.
The complete neighborhood is the Cartesian product of the azimuthal and elevational neighborhoods as
\begin{align}
\setnbr\nbrnbrd 
&= \setnbrth\nbrnbrdth \times \setnbrph\nbrnbrdph \\
&= \braces{\thph : \theta \in \setnbrth\nbrnbrdth, \phi \in \setnbrph\nbrnbrdph}.
\end{align}

The spatial neighborhoods $\txdirsetmeas\idx{i\opt}$ and $\rxdirsetmeas\idx{j\opt}$ surrounding the transmit and receive directions output by conventional beam alignment from the previous section are respectively written as
%  $i\opt$-th transmit direction and $j\opt$-th receive direction, respectively, are written as
\begin{align}
\txdirsetmeas\idx{i\opt} &= \thphtxiopt + \setnbr\nbrnbrd \\
% \rxdirsetmeas\idx{j\opt} &= \thphrxjopt + \setnbr\nbrnbrd.
\rxdirsetmeas\idx{j\opt} &= \underbrace{\thphrxjopt}_{\mathsf{initial~selection}} + \underbrace{\setnbr\nbrnbrd}_{\mathsf{neighborhood}}.
\end{align}
The size of these sets is
\begin{align}
\card{\txdirsetmeas\idx{i\opt}} = \card{\rxdirsetmeas\idx{j\opt}} = \card{\setnbr\nbrnbrd} =
\parens{2 \cdot \Kth + 1} \cdot \parens{2 \cdot \Kph + 1}
\end{align}
where $\Kth = \floor{\frac{\Deltatheta}{\deltatheta}}$ and $\Kph = \floor{\frac{\Deltaphi}{\deltaphi}}$, 
% \begin{align}
% \Kth &= \floor{\frac{\Deltatheta}{\deltatheta}}, \quad 
% \Kph = \floor{\frac{\Deltaphi}{\deltaphi}}
% \end{align}
indicating that neighborhoods naturally grow with widened \nbr or finer resolution \nbrd.

%When steering its transmit beam toward $\thphtx$ and receive beam toward $\thphrx$, the \iab incurs an \ginr of
%\begin{align}
%\inrrx\thphtxrx = 
%\begin{dcases}
%% \inrbhdl = 
%% \frac{\powertxiab \cdot \Gsi^2 \cdot \bars{\vw\thphrx\ctrans \mHsi \vf\thphtx}^2}{\powernoiseiab}, & \casedl \\
%\underbrace{\frac{\powertxiab \cdot \Gsi^2 \cdot \bars{\vw\thphrx\ctrans \mHsi \vf\thphtx}^2}{\powernoiseiab}}_{\inrbhdl}, & \casedl \\
%% \inracul = 
%% \frac{\powertxiab \cdot \Gsi^2 \cdot \bars{\vw\thphrx\ctrans \mHsi\ctrans \vf\thphtx}^2}{\powernoiseiab}, & \caseul
%\underbrace{\frac{\powertxiab \cdot \Gsi^2 \cdot \bars{\vw\thphrx\ctrans \mHsi\ctrans \vf\thphtx}^2}{\powernoiseiab}}_{\inracul}, & \caseul
%\end{dcases}
%\end{align}
%which depends on if the network is operating in \dldl or \ulul mode.

When steering its transmit beam toward $\thphtx$ and receive beam toward $\thphrx$, the \iab node incurs an \ginr of $\inrrx\thphtxrx$. % , which depends on if the network is operating in \dldl or \ulul mode.
In \dldl mode, $\inrrx$ can be expressed as
\begin{align}
\inrrx\thphtxrx = \frac{\powertxiab \cdot \Gsi^2 \cdot \bars{\vw\thphrx\ctrans \mHsi \vf\thphtx}^2}{\powernoiseiab}
\end{align}
and that during \ulul mode can be stated analogously.
%\begin{align}
%\inrrx\thphtxrx = 
%\begin{dcases}
%% \inrbhdl = 
%% \frac{\powertxiab \cdot \Gsi^2 \cdot \bars{\vw\thphrx\ctrans \mHsi \vf\thphtx}^2}{\powernoiseiab}, & \casedl \\
%\underbrace{\frac{\powertxiab \cdot \Gsi^2 \cdot \bars{\vw\thphrx\ctrans \mHsi \vf\thphtx}^2}{\powernoiseiab}}_{\inrbhdl}, & \casedl \\
%% \inracul = 
%% \frac{\powertxiab \cdot \Gsi^2 \cdot \bars{\vw\thphrx\ctrans \mHsi\ctrans \vf\thphtx}^2}{\powernoiseiab}, & \caseul
%\underbrace{\frac{\powertxiab \cdot \Gsi^2 \cdot \bars{\vw\thphrx\ctrans \mHsi\ctrans \vf\thphtx}^2}{\powernoiseiab}}_{\inracul}, & \caseul
%\end{dcases}
%\end{align}
%which depends on if the network is operating in \dldl or \ulul mode.
% Following conventional beam alignment, 
For each potential initial beam selection $\parens{i\opt,j\opt}$, we propose that the \iab node measure and record $\inrrx\thphtxrx$ for all transmit-receive combinations across the neighborhoods $\txdirsetmeas\idx{i\opt}$ and $\rxdirsetmeas\idx{j\opt}$ to populate 
% We suggest that the \iab collect the following set of measurements by until all transmit-receive combinations have been recorded.
\begin{align}
\setinrmeas\idx{i\opt,j\opt} &= \braces{\inrrx\thphtxrx : \thphtx \in \txdirsetmeas\idx{i\opt}, \thphrx \in \rxdirsetmeas\idx{j\opt}}.
\end{align}
The total number of \ginr measurements collected in $\setinrmeas\idx{i\opt,j\opt}$ is
\begin{align}
\card{\setinrmeas\idx{i\opt,j\opt}} 
&= \card{\txdirsetmeas\idx{i\opt}} \cdot \card{\rxdirsetmeas\idx{j\opt}} 
= \parens{2 \cdot K_\theta + 1}^2 \cdot \parens{2 \cdot K_\phi + 1}^2
\end{align}
which is equal for all $(i\opt,j\opt)$ since we have assumed a fixed neighborhood size.
% This collection of \ginr measurements 
% and will enable the next stage of our design.
This set of receive link \ginr measurements $\setinrmeas\idx{i\opt,j\opt}$ will enable the next stage of our proposed design.\footnote{In \secref{sec:implement}, we present an algorithm that can dramatically reduce the number of measurements needed by \steer, requiring only a fraction of $\setinrmeas\idx{i\opt,j\opt}$ to be measured.}

\begin{remark}[Measurement overhead and frequency]
Naturally, conducting these \ginr measurements at the full-duplex device may become practically prohibitive if the number of measurements grows too large.
This depends on the neighborhood size $\nbr$ and spatial resolution $\nbrd$, along with how frequently these measurements need to be collected.
System engineers can throttle the neighborhood size and/or spatial resolution to reduce the measurement overhead, though this may reduce the effectiveness of \steer, as we will see.
In addition to neighborhood size and spatial resolution, the time-variability of self-interference will heavily dictate the overhead of these measurements.
In the extreme case, a nearly static self-interference demands infrequent self-interference measurements. % , suggesting that the choice of neighborhood size and spatial resolution would be of less concern.
Highly dynamic self-interference, on the other hand, will demand more frequent measurements for reliability.
Notice that these measurements need not be taken strictly following beam alignment; instead, they can be collected for all $(i,j)$ and referenced for particular $(i\opt, j\opt)$, assuming a sufficiently static self-interference channel.
In such a case, the set of all \ginr measurements can be written as $\setinrmeas 
= \bigcup_{i=1}^{\Ntx} \ \bigcup_{j=1}^{\Nrx} \ \setinrmeas\idx{i,j}$, which has cardinality $\card{\setinrmeas} = \Ntx \cdot \Nrx \cdot \parens{2 \cdot \Kth + 1}^2 \cdot \parens{2 \cdot \Kph + 1}^2$
assuming no overlapping transmit neighborhoods or receive neighborhoods.
%\begin{align}
%\setinrmeas 
%&= \bigcup_{i=1}^{\Ntx} \ \bigcup_{j=1}^{\Nrx} \ \setinrmeas\idx{i,j} .
%% = \braces{\inrrx\thphtxrx : \thphtx \in \set{M}_{\mathrm{tx}}, \thphrx \in \set{M}_{\mathrm{rx}}}
%\end{align}
% where combining all transmit neighborhoods and receive neighborhoods yields
% \begin{align}
% \set{M}_{\mathrm{tx}} &= \bigcup_{i=1}^{\Ntx} \ \txdirsetmeas\idx{i}, \quad \set{M}_{\mathrm{rx}} = \bigcup_{j=1}^{\Nrx} \ \rxdirsetmeas\idx{j}.
% \end{align}
% Thus, the total number of \ginr measurements taken would be at most
% \begin{align}
% \card{\setinrmeas} 
% &= 
% \card{\set{M}_{\mathrm{tx}}} \cdot \card{\set{M}_{\mathrm{rx}}} 
% = 
% \Ntx \cdot \Nrx \cdot \parens{2 \cdot \Kth + 1}^2 \cdot \parens{2 \cdot \Kph + 1}^2
% \end{align}
% assuming no overlapping transmit neighborhoods nor receive neighborhoods.
%\Ntx \cdot \Nrx \cdot
It is important to keep in mind that there is no over-the-air feedback associated with these measurements since they are taken between the transmit and receive panels of the full-duplex \iab node.
% Additionally, there may exist the potential to collect some of these measurements during beam alignment by opportunistically sweeping the transmit beams when
% Note that there are no over-the-air feedback costs associated with 
\edit{Reliably measuring $\inrrx$ is key to the methodology that follows, though small measurement errors would be inherently tolerated; exploring in detail how reliable \ginr measurements must be would be interesting future work.}
\edit{Note that measuring $\inrrx$ for some beam pairs may lead to levels of self-interference that saturate the receive chain of the full-duplex transceiver, complicating measurement. It would be valuable future work to develop a means to estimate $\inrrx$ in such cases (e.g., via transmit power control to avoid saturation), though accuracy would not be especially important, as these beam pairs coupling high self-interference would presumably not be selected by \steer, as we will see.}
\end{remark}

%\begin{remark}
%Note that there are no over-the-air feedback costs associated with making these measurements since the transmitter and receiver coexist at the full-duplex \bs.
%These measurements, however, may introduce significant overhead to a practical system.
%The amount of overhead depends heavily on how frequently the \ginr measurements need to be taken (i.e., how static the self-interference channel is, which is still an open question) and how dense the neighborhoods are.
%As such, limiting the neighborhood size $\nbr$ and the resolution $\nbrd$ will be important system design choices to ensure these measurements do not consume resources prohibitively.
%There may exist the potential to take these \ginr measurements during \dl and \ul beam alignment.
%\end{remark}

% \section{\textsc{Steer}: Minimizing Self-Interference\\through Joint Transmit-Receive Beam Selection}
\section{\textsc{Steer}: Joint Transmit-Receive Beam Selection} \label{sec:steer}
In this section, we present \steer, our methodology for choosing beams that the \iab node uses to serve the transmit link and receive link.
\steer incorporates self-interference during beam selection rather than blindly aiming to maximize \gsnr, as is typically done in conventional beam alignment.
To do so, \steer relies on conventional beam alignment from \secref{sec:initial-beam-selection} to identify the general directions in which beams should be steered. % to deliver sufficient link margin to the donor and \ue. % $\thphtxiopt$ and $\thphrxjopt$.
Then, based on some design parameters, transmit and receive beams ($\vf$ and $\vw$) are {jointly} selected to serve access and backhaul while simultaneously minimizing the degree of self-interference they couple.
This is done by leveraging self-interference measurements taken at the full-duplex \iab node as described in \secref{sec:measure}.
Using the initial beam selections $\thphtxiopt$ and $\thphrxjopt$ from beam alignment, along with the \ginr measurements $\setinrmeas\idx{i\opt,j\opt}$, the objective of \steer is to fetch the transmit beam and receive beam from the neighborhoods $\txdirsetmeas\idx{i\opt}$ and $\rxdirsetmeas\idx{j\opt}$ that offers full-duplexing gains (in terms of spectral efficiency) over simply using $\vf\thphtxiopt$ and $\vw\thphrxjopt$ to serve the transmit and receive links.
% maximizes full-duplex performance superior to that of simply using $\vf\thphtxiopt$ and $\vw\thphrxjopt$ to serve the transmit and receive links.
Naturally, when the \gpsnr of the transmit and receive links are maximized and the interference on each are simultaneously driven to zero, the achievable spectral efficiencies approach their capacities.
%\begin{align}
%\snrtx \to \snrtxbar, \inrtx \to 0 &\implies \setx \to \captx \\
%\snrrx \to \snrrxbar, \inrrx \to 0 &\implies \serx \to \caprx 
%\end{align}
In pursuit of appreciable $\setx$ and $\serx$, we therefore aim to achieve high \gsnr on each link while reducing interference.
However, note that the \ginr on the transmit link $\inrtx$ (i.e., cross-link interference) is fixed since we have conditioned on the donor's beamforming weights.
Thus, \steer aims to select $\vf$ and $\vw$---the transmit and receive beams at the \iab node---so that high \gsnr can be achieved on each link while simultaneously reducing self-interference.
% \figref{fig:interplay} illustrates the interplay of our transmit and receive beams and their link qualities.
% In our attempt to improve the sum spectral efficiency $\setx + \serx$, our design aims to reduce the receive link \ginr without severely degrading the \glspl{snr} of the transmit and receive links.
% As a result, we have the potential to greatly improve the receive link \gsinr while maintaining the transmit link \gsnr and, consequently, its \gsinr.
% Recall that the transmit link \ginr is fixed since it does not depend on the \iab beamforming vectors $\vf$ and $\vw$.
Note that we do not require the final beam selections output by \steer to be from the codebooks $\precb$ and $\comcb$ but rather will be drawn to steer within $\txdirsetmeas\idx{i\opt}$ and $\rxdirsetmeas\idx{j\opt}$.

\subsection{Target Self-Interference Level}
Suppose there exists some target receive link \ginr threshold $\inrrxthresh$ our system desires.
For instance, one may choose $\inrrxthresh \approx 0$ dB to ensure self-interference does not overwhelmingly exceed noise. %, or one may even choose $\inrrxthresh$ based on the capabilities of additional digital self-interference cancellation.
Our design that follows does not guarantee that $\inrrx \leq \inrrxthresh$ but rather attempts to meet this target and does not incentivize \steer to provide an $\inrrx$ further below it. 
% For example, having $\inrrx = -40$ dB is virtually the same as having an $\inrrx = -10$ dB in the context of $\sinrrx$; this is reinforced further when considering spectral efficiency.
As we will see in our results in \secref{sec:simulation-results}, choosing a modest $\inrrxthresh$ will help \steer yield a more fair and globally optimal solution in terms of sum spectral efficiency since it throttles the sacrifices made in its effort to reduce self-interference.
Nonetheless, to force \steer to minimize $\inrrx$, engineers can use $\inrrxthresh = -\infty$ dB.
% Choosing a suitable 
% The optimal $\inrrxthresh$ (in a sum spectral efficiency sense) cannot be stated analytically since the receive link and transmit link are coupled by self-interference and the goal of our full-duplex system is to adequately maintain both backhaul and access, among other reasons.
% In \secref{sec:simulation-results}, we find an optimal heuristically, which shows that $\inrrxthresh \approx -7$ dB is generally near-optimal.

\comment{
To approximate a satisfactory $\inrrxthresh$ based on meaningful quantities, one can use
\begin{align}
\inrrxthresh 
= \snrrx \cdot \parens{2^{p \cdot \logtwo{1 + \snrrx}} - 1}\inv - 1
% = \frac{\snrul}{2^{p \cdot \logtwo{1 + \snrul}} - 1} - 1
\end{align}
which is the receive link \ginr that achieves a fraction $p$ of the receive link capacity at an \gsnr of $\snrrx$ defined as
\begin{align}
p 
= \frac{\logtwo{1 + \sinrrx}}{\logtwo{1 + \snrrx}}.
\end{align}
Since the receive link and transmit link are coupled by self-interference and the goal of our full-duplex system is to adequately maintain both backhaul and access, the optimal \inrulthresh (in a sum spectral efficiency sense) cannot be stated analytically.
Nonetheless, in \secref{sec:simulation-results}, we find an optimal heuristically, which suggests that $\inrrxthresh \approx -7$ dB is generally suitable.

\begin{figure*}
    \centering
    \subfloat[Caption a.]{\includegraphics[width=0.475\linewidth,height=0.26\textheight,keepaspectratio]{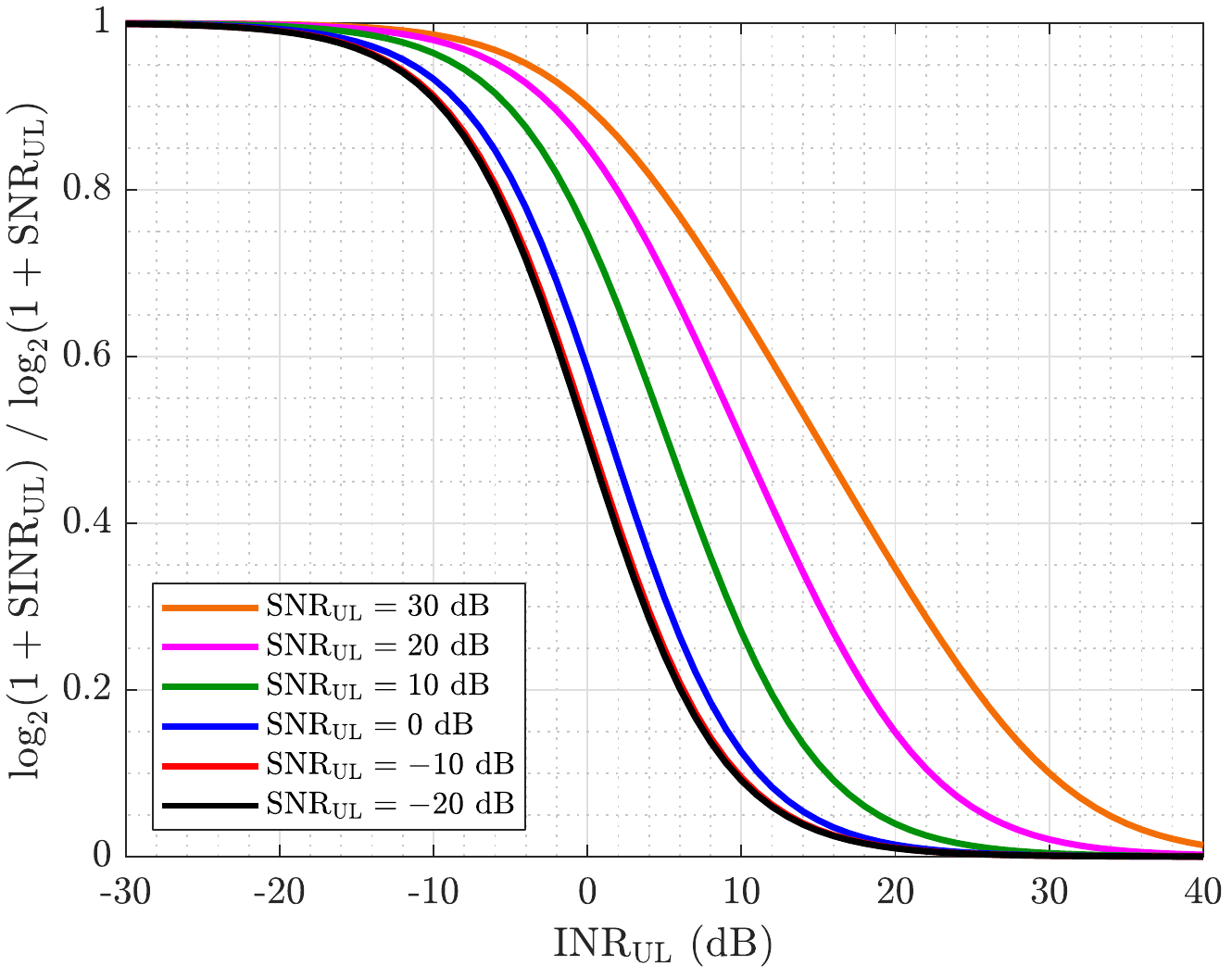}
        \label{fig:inr-thresh-a}}
    \quad
    \subfloat[Caption b.]{\includegraphics[width=0.475\linewidth,height=0.26\textheight,keepaspectratio]{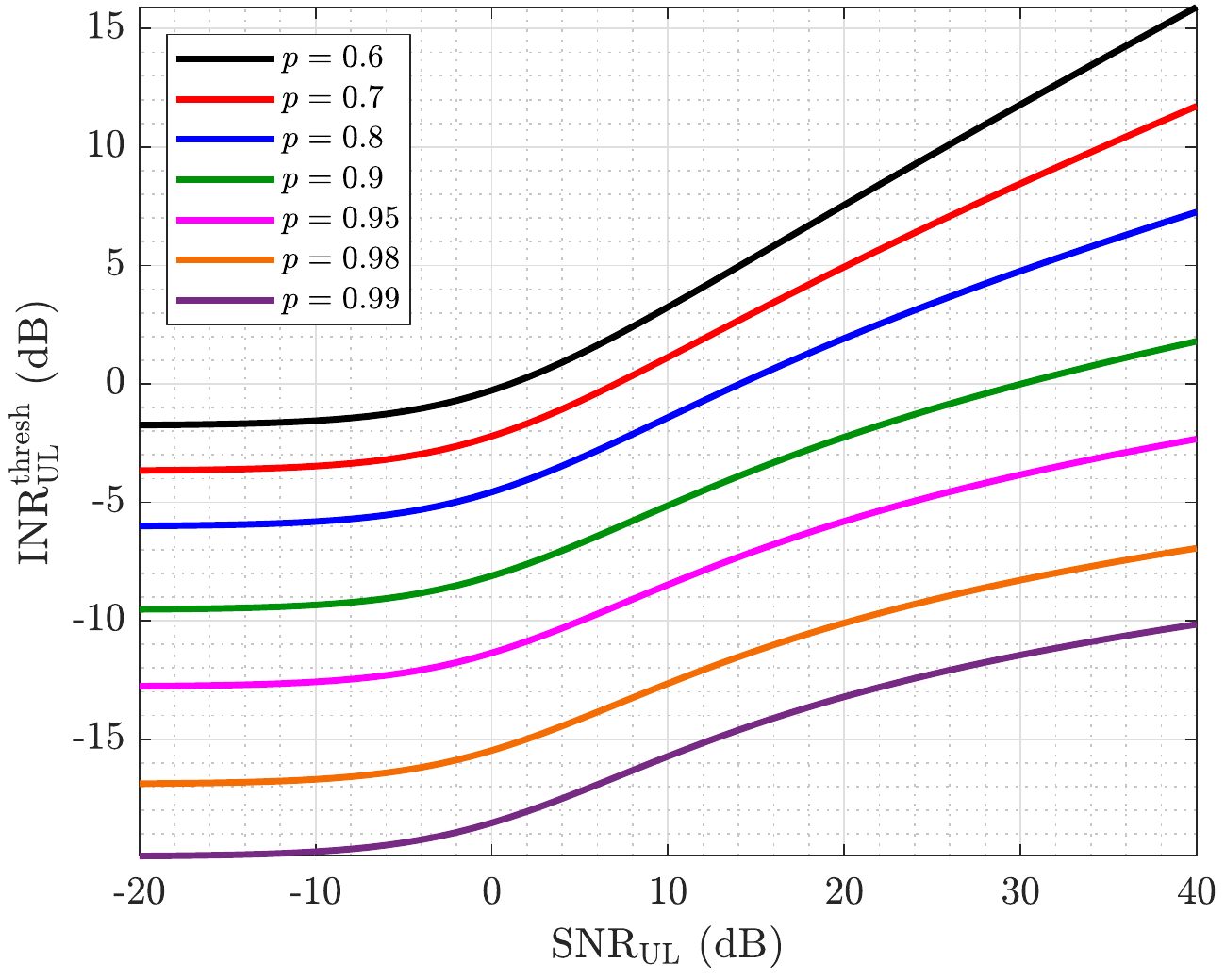}
        \label{fig:inr-thresh-b}}
    \caption{Caption here.}
    \label{fig:inr-thresh}
\end{figure*}

\figref{fig:inr-thresh} provides some insight on choosing $\inrrxthresh$.
In \figref{fig:inr-thresh-a}, we plot $p$ versus $\inrrx$ for various $\snrrx$.
Naturally, at high $\inrrx$, $p \to 0$ since $\sinrrx \to 0$.
At low $\inrrx$, $p \to 0$ since $\sinrrx \to \snrrx$.
In between, small changes in $\inrrx$ (e.g., $5$--$10$ dB) can drastically impact $p$.
This sensitivity is lessened with higher $\snrrx$, where simultaneously a lower $\inrrx$ is required to meet a particular $p$.
In \figref{fig:inr-thresh-b}, we plot $\inrrxthresh$ as a function of $\snrrx$ for various $p$ to provide further intuition on selecting an $\inrrxthresh$.
Notice that an $\inrrxthresh \geq 0$ dB is only suitable for low $p$ or at very high $\snrrx$.
Rather, $\inrrxthresh$ around $-5$ dB to $-10$ dB offers high $p$ even at low $\snrrx$.
Desiring $p$ beyond $0.95$ imposes extremely strict requirements on $\inrrx$ at low $\snrrx$.
From this it is clear that choosing a modest $p$ and hence a modest $\inrrxthresh$ can relieve the requirements of our full-duplex system, which we will see in \secref{sec:simulation-results}.
}

\subsection{Joint Transmit-Receive Beam Selection}
Before beginning our design process, we record the receive link \ginr when steering along the initial beam selections $\thphtxiopt$ and $\thphrxjopt$, which we call the \textit{nominal} receive link \ginr and express as
\begin{align}
\inrrxnom \triangleq \inrrx\thphtxrxijopt.
\end{align}
This is the receive link \ginr incurred if our full-duplex system were to use conventional beam selection and will thus be a useful benchmark to compare against.
Desirably, our final beam selections will yield $\inrrx < \inrrxnom$ to make our design worthwhile.
Note that if $\inrrxnom \leq \inrrxthresh$, we need not proceed with our design for this particular $\thphtxrxijopt$ since the target is met inherently by the beams from conventional beam alignment.
In such a case, we can simply set the transmit beam and receive beam as $\vf\thphtxiopt$ and $\vw\thphrxjopt$.
% In such a case, we can simply let the output of our design (transmit beam $\vf$ and receive beam $\vw$) be 
% \begin{align}
% \vf\opt &= \vf\thphtxiopt, \quad \vw\opt = \vw\thphrxjopt.
% \end{align}
When this is not the case, we proceed with our design as follows.

We begin by denoting the minimum \ginr over the measured spatial neighborhood as 
\begin{align}
\inrrxmin \triangleq \minop{\setinrmeas\idx{i\opt,j\opt}}.
\end{align}
Then, we form the following beam selection problem to retrieve the transmit direction $\thphtxopt$ and receive direction $\thphrxopt$ that the full-duplex \iab node will steer toward.
\begin{subequations} \label{eq:problem-ii}
    \begin{align}
    \thphtxopt, \thphrxopt = 
    \argmin_{\substack{\thphtx\\\thphrx}} & \min_{\parens{\Deltavartheta,\Deltavarphi}} \ \Deltavartheta^2 + \Deltavarphi^2 \\
    \st \ 
    & \inrrx\thphtxrx \leq \maxop{\inrrxthresh,\inrrxmin} \label{eq:target-inr-constraint} \\
    % & \thphtx \in \txdirsetmeas, \ \thphrx \in \rxdirsetmeas \\
    & \thphtx \in \thphtxiopt + \setnbr\nbrvnbrd \\
    & \thphrx \in \thphrxjopt + \setnbr\nbrvnbrd \\
    % & \txdirsetmeas = \quantize{}{\braces{\thph : \anglediff{\theta,\thetatxbar} \leq \delta_\theta, \anglediff{\phi,\phitxbar} \leq \delta_\phi}} \\
    % & \rxdirsetmeas = \quantize{}{\braces{\thph : \anglediff{\theta,\thetarxbar} \leq \delta_\theta, \anglediff{\phi,\phirxbar} \leq \delta_\phi}} \\
    & 0 \leq \Deltavartheta \leq \Deltatheta, \ 0 \leq \Deltavarphi \leq \Delta\phi
    \end{align}
\end{subequations}
The outer maximization aims to find the transmit and receive steering directions $\thphtx$ and $\thphrx$ that abide by three constraints.
First, the steering directions must satisfy \eqref{eq:target-inr-constraint},
%\begin{align}
%\inrrx\thphtxrx \leq \maxop{\inrrxthresh,\inrrxmin} 
%\end{align}
meaning the resulting receive link \ginr should either be below the desired target $\inrrxthresh$ or be the minimum \ginr offered in the surrounding \nbr-neighborhood (i.e., $\inrrxmin$).
Second, the transmit direction $\thphtx$ should be within the $\nbrv$-neighborhood surrounding the initial transmit beam selection $\thphtxiopt$, as illustrated in \figref{fig:neighborhood}.
Third, the receive direction $\thphrx$ should be within the $\nbrv$-neighborhood surrounding the initial receive beam selection $\thphrxjopt$.
To constrain the distance of $\thphtx$ from $\thphtxiopt$ and of $\thphrx$ from $\thphrxjopt$, we minimize the neighborhood size $\nbrv$ by minimizing $\Deltavartheta^2  + \Deltavarphi^2$; other distance measures could also be used.
Solving this problem will find the transmit and receive steering directions that meet the \ginr threshold while minimally deviating from those output by conventional beam alignment.
\edit{In the next section, we present an algorithm for solving problem \eqref{eq:problem-ii} more efficiently and with fewer \ginr measurements than an exhaustive search.}

Notice that, since $0 \leq \Deltavartheta \leq \Deltatheta$ and $0 \leq \Deltavarphi \leq \Delta\phi$, we have
\begin{align}
\thphtxiopt + \setnbr\nbrvnbrd &\subseteq \txdirsetmeas\idx{i\opt} \\
\thphrxjopt + \setnbr\nbrvnbrd &\subseteq \rxdirsetmeas\idx{j\opt}
\end{align}
meaning $\thphtx \in \txdirsetmeas\idx{i\opt}$ and $\thphrx \in \rxdirsetmeas\idx{j\opt}$ for feasible $\thphtx$ and $\thphrx$, and thus $\inrrx\thphtxrx \in \setinrmeas\idx{i\opt,j\opt}$.
% \begin{align}
% \inrrx\thphtxrx \in \setinrmeas\idx{i\opt,j\opt}.
% \end{align}
Hence, solving problem \eqref{eq:problem-ii} simply requires referencing the receive link \ginr measurements $\setinrmeas\idx{i\opt,j\opt}$ to find the transmit direction $\thphtxopt$ and receive direction $\thphrxopt$ that satisfies the \ginr target while minimizing the deviation from $\thphtxiopt$ and $\thphrxjopt$.
After solving this problem, our design concludes by setting the beamforming weights as $\vf\thphtxopt$ and $\vw\thphrxopt$.
%\begin{align}
%\vf &\leftarrow \vf\thphtxopt, \quad
%\vw \leftarrow \vw\thphrxopt.
%\end{align}
When transmitting and receiving with these beams output by \steer, we net $\snrtxours$ and $\snrrxours$ analogous to those in \eqref{eq:snr-tx-rx-nom}
%\begin{align}
%\snrtxours &\triangleq \snrtxbar \cdot \frac{\bars{\vhtx\ctrans \vf\thphtxopt}^2}{\Na} \\
%\snrrxours &\triangleq \snrrxbar \cdot \frac{\bars{\vw\thphrxopt\ctrans \vhrx}^2}{\Na} 
%\end{align}
% \begin{align}
% \snrtxours &\triangleq \snrtxbar \cdot \frac{\bars{\vhtx\ctrans \vf\thphtxopt}^2}{\Na}, \quad
% \snrrxours \triangleq \snrrxbar \cdot \frac{\bars{\vw\thphrxopt\ctrans \vhrx}^2}{\Na}
% \end{align}
and a receive link \ginr of
\begin{align}
% \inrrxnom &\triangleq \inrrx\thphtxrxijopt \\
\inrrxours &\triangleq \inrrx\thphtxrxopt \leq \inrrx\thphtxrxijopt = \inrrxnom. \label{eq:inr-rx-ours-def}
\end{align}
The receive link \ginr achieved by \steer is guaranteed to be no more than that with conventional beam selection.
% \begin{align}
% \inrrx\thphtxrxopt \leq \inrrx\thphtxrxijopt
% \end{align}
Equality in \eqref{eq:inr-rx-ours-def} holds when the target $\inrrxthresh$ is inherently met by the beams output by beam selection. % , meaning we have $\Deltavartheta = \Deltavarphi = 0^\circ$ as the solution to the inner minimization.
When this target is not inherently met by initial beam selection, \steer will deviate from the initial steering directions $\thphtxiopt$ and $\thphrxjopt$ only if it leads to lower self-interference. % receive link \ginr.
%The \gpsinr on the transmit and receive links with the beams from conventional beam alignment are
%\begin{align}
%\sinrtxnom &\triangleq \frac{\snrtxnom}{1 + \inrtx}, \quad 
%\sinrrxnom \triangleq \frac{\snrrxnom}{1 + \inrrxnom}
%\end{align}
%whereas the \gpsinr with beams output by \steer are
%\begin{align}
%\sinrtxours &\triangleq \frac{\snrtxours}{1 + \inrtx}, \quad
%\sinrrxours \triangleq \frac{\snrrxours}{1 + \inrrxours}.
%\end{align}

The steering directions output by \steer relative to those from conventional beam selection are bounded as 
\begin{align}
\bars{\thetatx\opt - \thetatx\idx{i\opt}}, \bars{\thetarx\opt - \thetarx\idx{j\opt}} &\leq \Deltatheta, \qquad
\bars{\phitx\opt - \phitx\idx{i\opt}}, \bars{\phirx\opt - \phirx\idx{j\opt}} \leq \Deltaphi .
\end{align}
% By throttling $\nbr$, the potential \gsnr loss when steering with beams output by \steer rather than those output by conventional beam selection can be limited.
% Letting $\nbr = \nbrzerozero$ will naturally force \steer to transmit and receive using the beams output by conventional beam selection, resulting in no \gsnr loss.
As such, increasing $\nbr$ may lead to beams that offer lower \gpsnr since \steer may use beams that are shifted slightly further away from the transmit and receive devices.
% The hope is, however, that 
% By choosing a low $\nbr$, this potential \gsnr loss can be constrained, though \nbr needs to be large enough to offer \steer sufficient diversity to reduce \ginr.
However, by throttling $\nbr$ and courtesy of the \ginr variability observed over small neighborhoods (on the order of one degree), this potential \gsnr loss can be constrained and may be greatly outweighed by the reduction in \ginr, netting it an improved \gsinr over conventional beam selection.
\edit{Experimental evaluation of \steer in \secref{sec:simulation-results} confirms this.}
Note that \gsnr actually may improve with \steer since it may output beams that are shifted slightly more {toward} the \dl and \ul devices, compared to those from conventional beam alignment.
% \edit{As we will see, experimental evaluation of \steer in \secref{sec:simulation-results} confirms this is the shows that a neighborhood of size $\nbr = \nbrtwotwo$ can in fact offer significant reductions in \ginr while preserving \gpsnr.}
% To summarize concisely, $\inrrxours \leq \inrrxnom$, $\snrtxours \nleq \snrtxnom$, and $\snrrxours \nleq \snrrxnom$.
%\begin{gather}
%\bars{\vhtx\ctrans \vf\thphtxopt}^2 \nleq \bars{\vhtx\ctrans \vf\thphtxiopt}^2 \implies \snrtxours \nleq \snrtxnom \\
%\bars{\vw\thphrxopt\ctrans \vhrx}^2 \nleq \bars{\vw\thphrxjopt\ctrans \vhrx}^2 \implies \snrrxours \nleq \snrrxnom % \\
%% \inrrx\thphtxrxopt &\leq \inrrx\thphtxrxijopt \\
%% \inrrxours &\leq \inrrxnom
%\end{gather}
% Conventional beam alignment will likely use beams that are evenly distributed in space, meaning \steer may introduce coverage holes relative to a conventional codebook when slightly shifting its beams.
% Nonetheless, by throttling \nbr and courtesy of the \ginr variability observed over small neighborhoods (on the order of one degree), the hope is that the coverage loss introduced by \steer will be greatly outweighed by its reduction in \ginr, netting it an improved \gsinr over conventional beam selection.
% We examine this through measurement and simulation in \secref{sec:simulation-results}.

\begin{remark}[Choosing a target self-interference level]
% The optimal choice of $\inrrxthresh$ (in a sum spectral efficiency sense) cannot be stated analytically since the receive link and transmit link are coupled by self-interference and the goal of our full-duplex system is to adequately maintain both backhaul and access, among other reasons.
The optimal choice of $\inrrxthresh$ (in a sum spectral efficiency sense) cannot be stated analytically since it depends on how much \steer must deviate to meet this target \ginr, which itself depends on the self-interference channel.
% As mentioned, it is sensible to use $\inrrxthresh \leq 0$ dB to ensure self-interference is no stronger than noise.
In \secref{sec:simulation-results}, we find an optimal target heuristically, which shows that $\inrrxthresh \approx -7$ dB is generally near-optimal in maximizing the sum spectral efficiency achieved by \steer.
\edit{It is difficult to offer commentary on choosing a suitable $\inrrxthresh$ that generalizes to systems beyond the platform we used to evaluate \steer, since each will have a unique self-interference profile.}
\end{remark}

\begin{remark}[Design decisions and motivations]
    We would now like to comment on some of the design decisions and motivations behind \steer.
    First, we point out that a more attractive beam selection solution would perhaps be one that maximizes the sum spectral efficiency of the transmit and receive links, rather than minimize the receive link \ginr as is done by \steer.
    It is practically implausible to reliably maximize sum spectral efficiency since such a problem requires knowing the \gsnr achieved by a given beam (which would require prohibitive feedback) or downlink and uplink channel knowledge. % , which would somewhat defeat the purpose of beam selection.
    Minimizing receive link \ginr by \steer was chosen deliberately, as it relieves the beam selection problem from requiring \gsnr knowledge.
    Instead, \steer solely minimizes self-interference through measurements taken at the \iab node and can preserve \gsnr by reducing its deviation from those output by conventional beam alignment.
    We also would like to point out that neighborhood size $\nbr$ and spatial resolution $\nbrd$ could be uniquely defined for transmit and receive beam selection, rather than having a common one as we have assumed herein.
    \edit{In cases where transmit and receive beam selection are not executed at the same time (e.g., due to a fixed backhaul link), one could simply condition on the beam not being selected (fixing its steering direction) when running \steer.}
    In cases where \steer cannot offer sufficiently low \ginr to justify full-duplex operation when serving particular transmit and receive users, there is the potential to serve them instead in a half-duplex fashion.
    % This would prevent full-duplex operation with \steer from ever being outperformed by conventional half-duplex operation.
    % Note that systems employing additional self-interference cancellation measures can choose \inrrxthresh based on its cancellation capability, for instance (e.g., let $\inrrxthresh = 20$ dB if the system can reliably cancel $20$ dB of self-interference).
\end{remark}

\section{Efficiently Implementing \steer in Real Systems} \label{sec:implement}

\edit{
Having presented its core components, we now present an algorithm for efficiently executing \steer (by which we mean solving problem \eqref{eq:problem-ii}), along with commentary on key practical considerations.
We begin with presentation of \algref{alg:algorithm-solve}.
As it was presented in the previous sections, \steer may be executed by first collecting a set of self-interference measurements and then solving problem \eqref{eq:problem-ii}, presumably by exhaustive search.
While this exhaustive search has fairly low computational complexity, the radio resources consumed to conduct self-interference measurements may be a key bottleneck in practical settings.
% This exhaustive search has very low computational complexity, especially in comparison to the radio resources consumed by conducting self-interference measurements in a practical system.
% In other words, collecting self-interference measurements is likely a key bottleneck of \steer in real systems.
Motivated by this, we now present an algorithm that executes \steer with a minimal number of measurements.
}

\edit{
\algref{alg:algorithm-solve} begins by sorting all transmit-receive direction pairs $\set{V} = \txdirsetmeas\idx{i\opt} \times \rxdirsetmeas\idx{j\opt}$ based on their deviation from the nominal directions $\thphtxiopt$ and $\thphrxjopt$ output by beam alignment\footnote{We slightly abuse convention here and assume sets have ordering for the sake of illustration.}.
The indices of this sorting we denote $\set{J}$, and the sorted set of transmit-receive direction pairs we write as $\entry{\set{V}}{\set{J}}$, which can be precomputed and is fixed for some neighborhood.
In other words, the transmit-receive direction pairs in $\entry{\set{V}}{\set{J}}$ increase in distance from the nominal direction pair.
Starting with the nominal direction pair $\thphtxiopt$ and $\thphrxjopt$, the \ginr of each transmit-receive direction pair in $\entry{\set{V}}{\set{J}}$ is measured until a transmit-receive pair yields a measured $\inrrx$ less than the target $\inrrxthresh$.
Once this target is met, the transmit-receive pair is guaranteed to satisfy all constraints of problem \eqref{eq:problem-ii} and minimizes the distance $\Delta\vartheta^2 + \Delta\varphi^2$, courtesy of our sorting of $\set{V}$.
No further measurements are required, having only measured a fraction of the full spatial neighborhood.
If no beam pairs meet the threshold, the entire $\nbr$-neighborhood is measured, with the beam pair offering the lowest \ginr being selected.
Rather than collecting all measurements in $\setinrmeas\idx{i\opt,j\opt}$ and then exhaustively solving problem \eqref{eq:problem-ii}, \algref{alg:algorithm-solve} provides a means to solve problem \eqref{eq:problem-ii} \textit{while} collecting measurements. 
This reduces its computational overhead since its extremely simple logic can be executed while taking measurements, and more importantly, the overhead consumed to collect \ginr measurements can be dramatically reduced.
We illustrate this reduction in measurement overhead in \secref{sec:simulation-results} using an actual 28 GHz phased array platform.
}

\begin{algorithm}[!t]
    \begin{algorithmic}[0]
        \REQUIRE $\thphtxiopt$, $\thphrxjopt$, $\txdirsetmeas\idx{i\opt}$, $\rxdirsetmeas\idx{j\opt}$, $\inrrxthresh$ % $\parens{\deltatheta,\deltaphi}$, $\parens{\Delta\theta,\Delta\phi}$
        \STATE $\inrrxmin = \infty$
        % \STATE $\inrrxnom = \inrrx\thphtxrxijopt$
        \STATE $\set{V} = \txdirsetmeas\idx{i\opt} \times \rxdirsetmeas\idx{j\opt}$
        % \STATE $\set{D}_{\vartheta} = \braces{\Delta\vartheta = \maxop{\bars{\thetatx-\thetatx\idx{i\opt}},\bars{\thetarx-\thetarx\idx{j\opt}}} : \thphtx \in \txdirsetmeas\idx{i\opt}, \thphrx \in \rxdirsetmeas\idx{j\opt}}$
        \STATE $\set{D}_{\vartheta} = \braces{\Delta\vartheta = \maxop{\bars{\thetatx-\thetatx\idx{i\opt}},\bars{\thetarx-\thetarx\idx{j\opt}}} : \parens{\thphtx,\thphrx} \in \set{V}}$
        \STATE $\set{D}_{\varphi} = \braces{\Delta\varphi = \maxop{\bars{\phitx-\phitx\idx{i\opt}},\bars{\phirx-\phirx\idx{j\opt}}} : \parens{\thphtx,\thphrx} \in \set{V}}$
        \STATE $\set{D} = \braces{\Delta\vartheta^2 + \Delta\varphi^2 : \Delta\vartheta \in \set{D}_{\vartheta}, \Delta\varphi \in \set{D}_{\varphi}}$
        \STATE $\brackets{\sim,\set{J}} = \mathrm{sort}\parens{\set{D},\mathsf{ascend}}$
        % \STATE $\set{V} = \braces{\thphtxrx : \thphtx \in \txdirsetmeas\idx{i\opt}, \thphrx \in \rxdirsetmeas\idx{j\opt}}$
        % \STATE $\set{D} = \mathrm{sort}\parens{\txdirsetmeas\idx{i\opt} \times \rxdirsetmeas\idx{j\opt},\thphtxiopt,\thphrxjopt}$
        \FOR{$\parens{\thphtx,\thphrx} \in \entry{\set{V}}{\set{J}}$}
        % \FOR{$\parens{\thphtx,\thphrx} \in \txdirsetmeas\idx{i\opt} \times \rxdirsetmeas\idx{j\opt}$}
        \STATE Measure (or reference) $\inrrx\thphtxrx$.
        \IF{$\inrrx\thphtxrx < \inrrxmin$}
        \STATE $\inrrxmin = \inrrx\thphtxrx$
        \STATE $\thphtxopt = \thphtx$
        \STATE $\thphrxopt = \thphrx$
        \IF{$\inrrx\thphtxrx \leq \inrrxthresh$} % or $\inrulmin > \inrulthresh$}
        % \STATE $\inrulmin = \inrulnom$
        \STATE Break for-loop; target met; no further measurements required.
        \ENDIF
        \ENDIF
        \ENDFOR
        \ENSURE $\thphtxopt$, $\thphrxopt$
    \end{algorithmic}
    \caption{Executing \steer by solving problem \eqref{eq:problem-ii} with a minimal number of measurements.}
    \label{alg:algorithm-solve}
\end{algorithm}

\begin{remark}[Practical considerations]
    \edit{We highlight that the execution of \steer, along with its associated self-interference measurements, take place solely at the full-duplex \iab node.
    In fact, the donor and user presumably need not be informed of the beams selected by \steer since only the beams at the \iab node are slightly shifted from those output by conventional beam alignment.
    This is a practically desirable property of \steer.}
    We would also like to emphasize that a practical implementation of \steer is highly dependent on a number of things.
    First and foremost, it depends heavily on the time-variability of self-interference, which is currently not well investigated.
    % Highly static self-interference will allow the \iab node to conduct \ginr measurements very infrequently, merely referencing its measurements for each execution rather than needing to recollect them.
    If self-interference is highly dynamic, measurements will need to be collected more frequently and, with updated measurements, \steer will need to be rerun.
    \edit{However, it may be preferable to first re-measure existing \steer solutions to locate ones that may have become stale, no longer offering sufficiently low \ginr.}
    % Still, it may be more practical to collect all necessary measurements at once and precompute \steer solutions.
    % This is a particularly good topic for future work to validate the practicality of \steer.
    Note that, when swapping from the \dldl operating mode to the \ulul mode, the self-interference measurements may not be symmetric since the panels presumably swap transmit/receive roles, meaning measurements may need to be collected uniquely for each of the two full-duplexing modes.
    % Efficiently collecting self-interference measurements for \steer, practical implementations of \steer, and investigating self-interference reciprocity are good topics for future work.
    Characterizing the time-variability and reciprocity of self-interference, along with practical implementations of \steer at \mmwave frequencies beyond 28 GHz, are good topics for future work.
    % Note that when transmitting with one panel and receiving with the other, the self-interference measurements may not align with those taken when swapping which panels are used for transmission and reception (i.e., the measurements may not be symmetric across the panels).
    % In such a case, unique measurements must be collected in both directions (i.e. transmitting with panel A while receiving with panel B, transmitting with panel B while receiving with panel A).
\end{remark}

\begin{remark}[Precomputing \steer solutions]
    With collected measurements and known codebooks \precb and \comcb, the \iab node can precompute the solution output by \steer for all possible $(i\opt,j\opt)$, keeping a record of $\thphtxrxopt$ for each.
    Thereafter, the \iab node can directly map initial beam selection indices to the precomputed solution via a lookup table, rather than re-running \steer.
    This is another practically desirable property of \steer.
    \begin{align}
    % \thphtxrxijopt \overset{\mathsf{lookup}}{\longrightarrow} \thphtxrxopt 
    % \thphtxrxijopt \xrightarrow{\mathsf{lookup}} \thphtxrxopt \\
    (i\opt,j\opt) \xrightarrow{\mathsf{lookup}} \thphtxrxopt 
    % \\
    % (i\opt,j\opt) \overset{\mathsf{LUT}}{\longrightarrow} \thphtxrxopt
    \end{align}
    In such a case, the \iab node will need to store $\Ntx \cdot \Nrx \cdot 4$ values for a given $\nbr$ and $\nbrd$.
    Note that, when the phased arrays are equipped with functions that internally map steering direction to beamforming weights, precomputing \steer solutions only requires storing the steering directions  $\thphtxopt$, $\thphrxopt$ and not the explicit beamforming weights $\vf\opt$ and $\vw\opt$, reducing storage requirements.
    % Note that when transmitting with one panel and receiving with the other, the self-interference measurements may not align with those taken when swapping which panels are used for transmission and reception (i.e., the measurements may not be symmetric across the panels).
    % In such a case, unique measurements must be collected in both directions (i.e. transmitting with panel A while receiving with panel B, transmitting with panel B while receiving with panel A).
\end{remark}

% \subsection{Summary}

\begin{remark}[Summary of \steer]
A summary of our entire beam selection methodology is illustrated in \figref{fig:block-diagram} and is outlined in \algref{alg:algorithm-ii}.
\steer begins by executing conventional beam alignment using codebooks $\precb$ and $\comcb$ to yield initial transmit and receive beam selections that steer toward $\thphtxiopt$ and $\thphrxjopt$.
Then, based on some defined neighborhood size \nbr and spatial resolution \nbrd, the spatial neighborhoods surrounding these initial transmit and receive beams are constructed as $\txdirsetmeas\idx{i\opt}$ and $\rxdirsetmeas\idx{j\opt}$.
% The full-duplex \iab node then forms $\setinrmeas\idx{i\opt,j\opt}$ by referencing receive link \ginr measurements for all transmit-receive steering combinations in $\txdirsetmeas\idx{i\opt}$ and $\rxdirsetmeas\idx{j\opt}$.
% The full-duplex \iab node 
% The minimum \ginr over this set $\setinrmeas\idx{i\opt,j\opt}$ is recorded as $\inrrxmin$.
A receive link \ginr target $\inrrxthresh$ is specified by system engineers, likely based on simulation, experimentation, and field trials.
% Solving problem \eqref{eq:problem-ii} will yield the transmit and receive steering directions $\thphtxopt$ and $\thphrxopt$ that minimally deviate from the initial steering directions while attempting to meet the target \ginr.
Solving problem \eqref{eq:problem-ii} via \algref{alg:algorithm-solve} will yield the transmit and receive steering directions $\thphtxopt$ and $\thphrxopt$ that minimally deviate from the initial steering directions while attempting to meet the target \ginr and minimizing the number of \ginr measurements collected.
\end{remark}

\begin{algorithm}[!t]
    \begin{algorithmic}[0]
        \STATE {1. Define transmit and receive coverage regions $\txdirsetcb$ and $\rxdirsetcb$ and corresponding transmit and receive codebooks $\precb$ and $\comcb$ for beam alignment.}
        \STATE {2. Conduct conventional beam alignment to yield $\thphtxiopt$ and $\thphrxjopt$.}
        \STATE {3. Define the measurement spatial resolution $\parens{\deltatheta,\deltaphi}$ and neighborhood size $\parens{\Delta\theta,\Delta\phi}$.}
        \STATE {4. Construct transmit and receive neighborhoods $\txdirsetmeas\idx{i\opt}$ and $\rxdirsetmeas\idx{j\opt}$.}
        % \STATE {5. Reference self-interference measurements:} 
        % \STATE \qquad $\setinrmeas\idx{i\opt,j\opt} = \braces{\inrrx\thphtxrx : \thphtx \in \txdirsetmeas\idx{i\opt}, \thphrx \in \rxdirsetmeas\idx{j\opt}}$
        % \STATE {6. Identify minimum measured receive link \ginr: $\inrrxmin = \minop{\setinrmeas\idx{i\opt,j\opt}}$.}
        \STATE {5. Define a desired receive link \ginr threshold $\inrrxthresh$.}
        \STATE {6. Solve problem \eqref{eq:problem-ii} for $\thphtxopt$ and $\thphrxopt$ using \algref{alg:algorithm-solve}, collecting/referencing a minimal number of measurements of self-interference.}
        % \STATE {6. Solve problem \eqref{eq:problem-ii} for $\thphtxopt$ and $\thphrxopt$ using \algref{alg:algorithm-solve}, collecting/referencing measurements of self-interference only as necessary.}
        \STATE {7. Set transmit weights as $\vf\thphtxopt$ and receive weights as $\vw\thphrxopt$.}
    \end{algorithmic}
    \caption{A summary of our beam selection methodology \steer.}
    \label{alg:algorithm-ii}
\end{algorithm}

\section{Evaluating \textsc{Steer} Through Measurement and Simulation} \label{sec:simulation-results}

% To evaluate \steer, we combine Monte Carlo simulation with measurements taken using a 28 GHz phased array platform; rather than rely on a particular self-interference channel model, we explicitly measure \ginr for more accurate evaluation.
% Then, we combine these measurements with simulation \cite{mfm_arxiv} by randomly dropping a donor and \ue, conducting initial beam alignment, and then executing \steer.

\begin{figure}
    \centering
    \includegraphics[width=\linewidth,height=0.25\textheight,keepaspectratio]{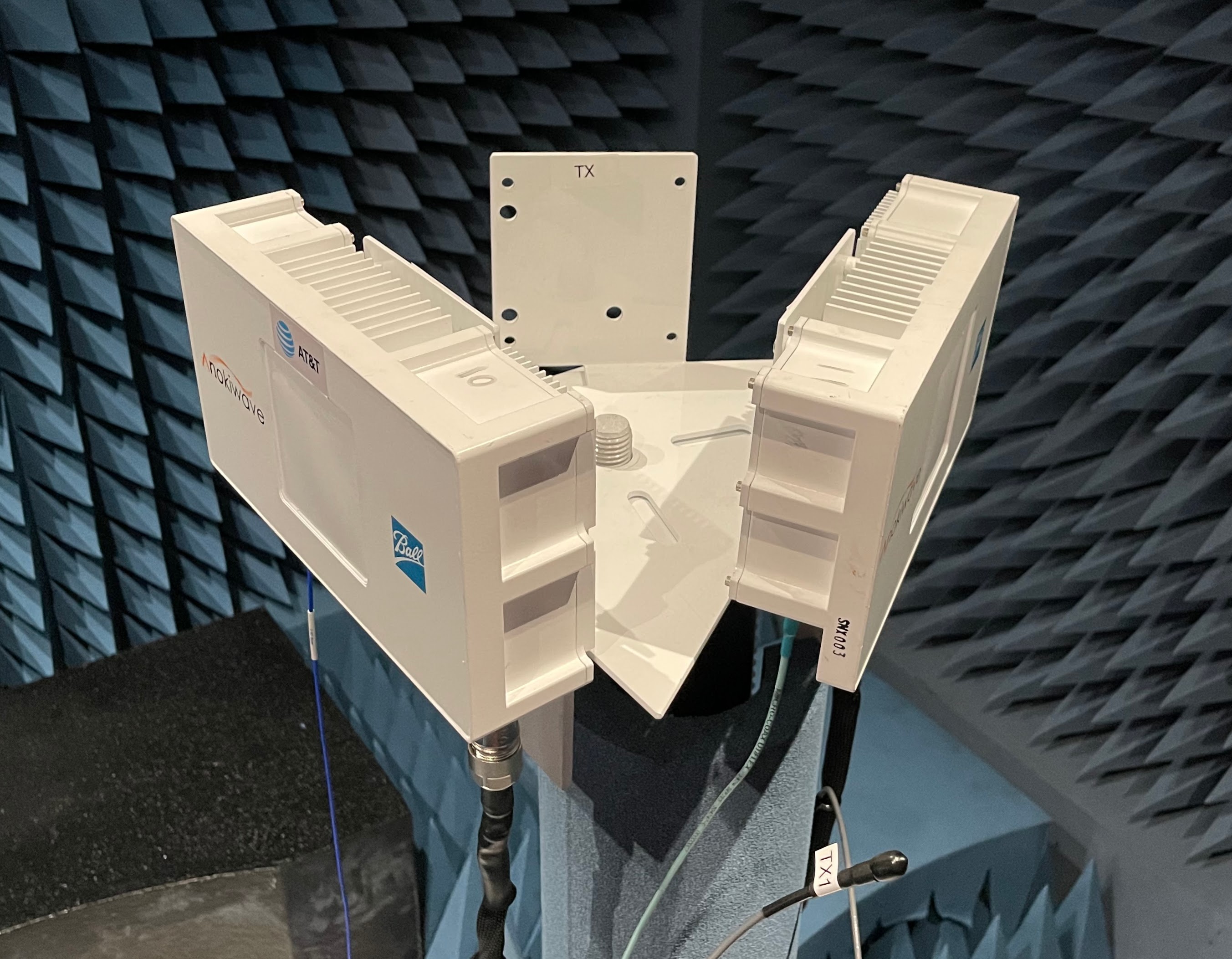}
    \caption{The multi-panel 28 GHz phased array platform used to evaluate \steer through \ginr measurements. Each phased array is a $16 \times 16$ half-wavelength \upa. Measurements were taken in an anechoic chamber, free from significant reflectors.}
    \label{fig:setup}
\end{figure}

\edit{We experimentally evaluate \steer by combining Monte Carlo simulation with \ginr measurements taken with a 28 GHz phased array platform.
A donor and \ue are randomly dropped around an \iab node within simulation \cite{mfm_arxiv}, followed by beam alignment, and then execution of \steer using actual \ginr measurements.
In other words, $\inrrx$ values have been measured, while \gsnr terms are based on simulation.}
We consider the case where a full-duplex \iab node transmits and receives using panels on two sides of a sectorized triangular platform, as illustrated in \figref{fig:setup}.
The transmit and receive arrays are identical $28$ GHz $16 \times 16$ half-wavelength \upas \cite{anokiwave}.
Using the platform in \figref{fig:setup}, we measure $\inrrx$ as described in \secref{sec:measure} using a fixed spatial resolution of $\nbrd = \nbroneone$; it would be valuable future work would explore finer resolutions.
We take measurements in an anechoic chamber to first explore the impacts of the direct coupling between arrays; investigating the effects of reflections off of realistic environments is also a good topic for dedicated future work.
\edit{We transmit upconverted Zadoff-Chu sequences with 100 MHz of bandwidth and apply correlation-based processing to reliably estimate self-interference well below the noise floor.
Our $\inrrx$ measurements are typically accurate to within $1$ dB, based on validation with high-fidelity test equipment \cite{pwrMeter} and stepped attenuators \cite{stepAtten}. 
We refer readers to our prior work \cite{roberts_att_angular} for more details regarding our measurement methodology, which we also employ herein.}
% As illustrated in \figref{fig:codebook}, 
We consider identical transmit and receive codebooks comprised of $\Ntx = \Nrx = 105$ narrow beams that span in azimuth from $-56^\circ$ to $56^\circ$ and in elevation from $-24^\circ$ to $24^\circ$, each with $8^\circ$ spacing.
The transmit and receive beams are steered using conjugate beamforming weights (i.e., equal gain/matched filter beamforming), described as $\vf\thphtxi = \atx{\thetatx\idx{i},\phitx\idx{i}}$ and $\vw\thphrxj = \arx{\thetarx\idx{j},\phirx\idx{j}}$, 
%\begin{align}
%\vf\thphtxi = \atx{\thetatx\idx{i},\phitx\idx{i}}, \quad \vw\thphrxj = \arx{\thetarx\idx{j},\phirx\idx{j}}
%\end{align}
where $\atx{\cdot}$ and $\arx{\cdot}$ are the transmit and receive array response vectors.
The transmit array radiates at an \gls{eirp} of $60$ dBm and the receive array output has a noise floor of $\powernoiseiab = -68$ dBm over $100$ MHz. 
The transmit and receive beams each have a $3$ dB beamwidth of about $7^\circ$.
In a Monte Carlo fashion, we randomly drop a donor and \ue in the coverage region supplied by our codebooks, from $-60^\circ$ to $60^\circ$ in azimuth and from $-28^\circ$ to $28^\circ$ in elevation.
\edit{We assume the donor and \ue are in \gls{los} of the \iab node for simplicity and to more straightforwardly evaluate \steer against conventional beam selection.}
\edit{We make initial beam selections by choosing transmit and receive beams that maximize their \gpsnr (e.g., exhaustive beam search), though \steer could be applied atop any beam alignment scheme.}

% $\setinrmeas\idx{i\opt,j\opt}$ for all $i\opt, j\opt$.

\subsection{Reducing Measurement Overhead via \algref{alg:algorithm-solve}}
\begin{figure}
    \centering
    \subfloat[\edit{Reducing measurement overhead via \algref{alg:algorithm-solve}.}]{\includegraphics[width=\linewidth,height=0.29\textheight,keepaspectratio]{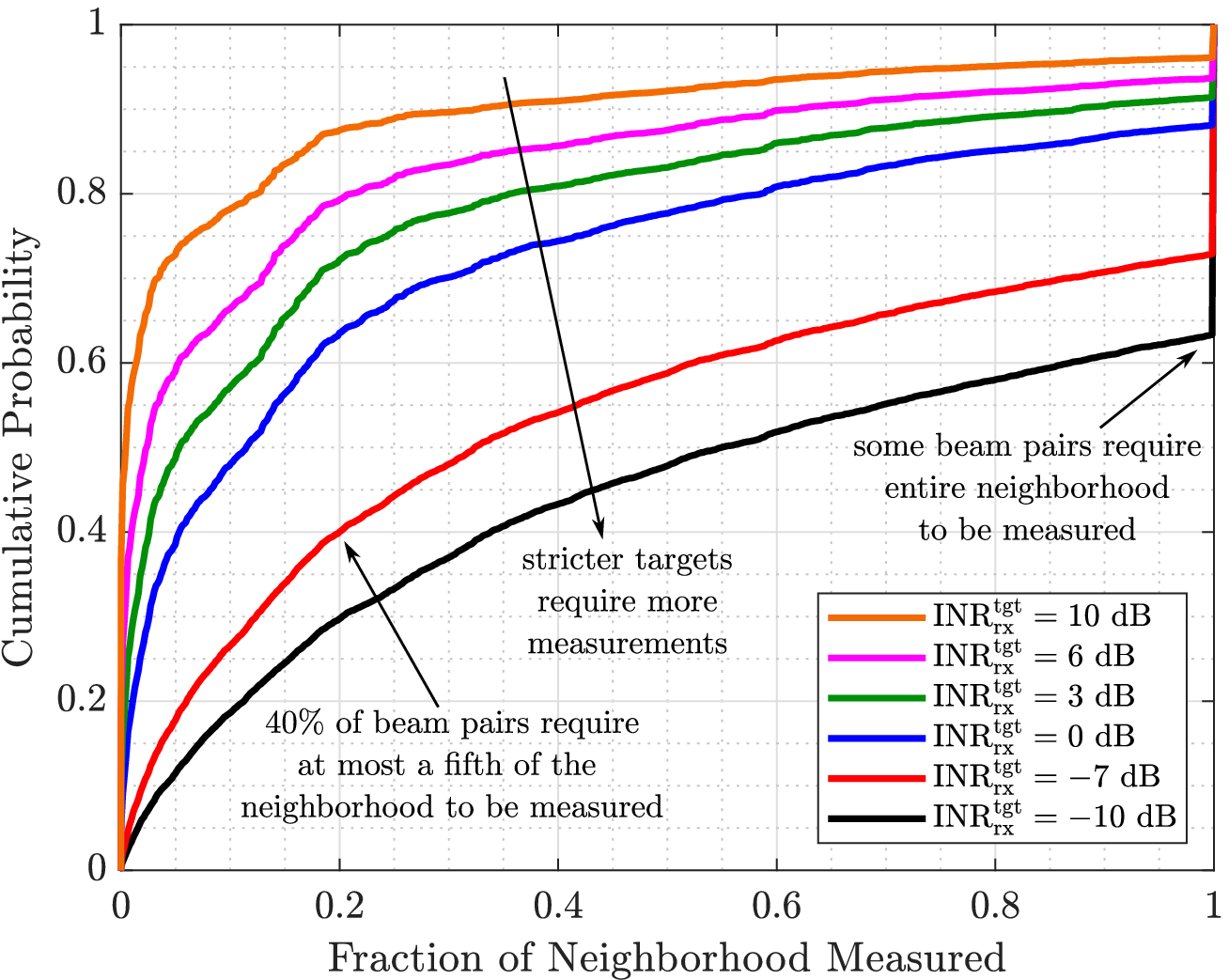}
        \label{fig:alg-a}}
    \qquad
    \subfloat[\edit{Empirical probability of $\nbrv$.}]{\includegraphics[width=\linewidth,height=0.285\textheight,keepaspectratio]{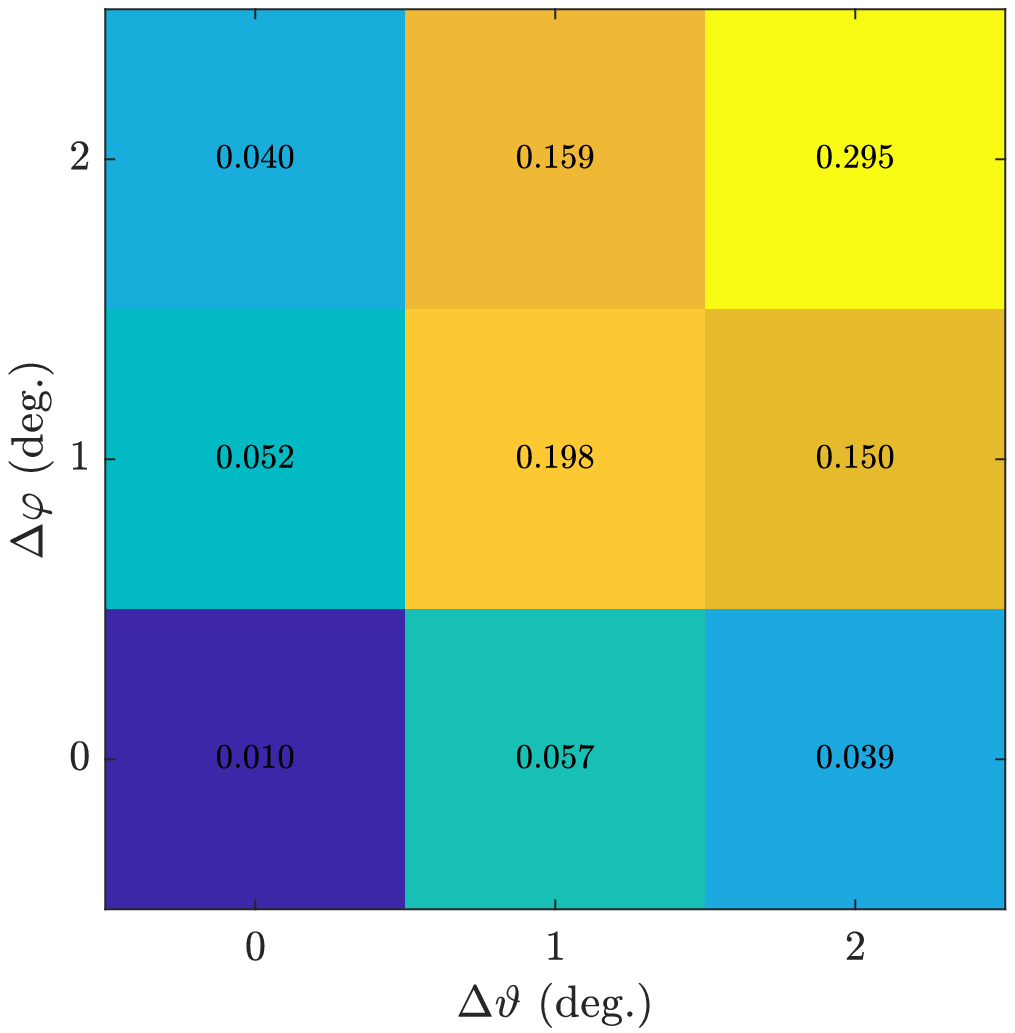}
        \label{fig:alg-b}}
    \caption{(a) Illustrating the reduction in measurement overhead necessitated by \steer when using \algref{alg:algorithm-solve}. $65\%$ of beam pairs only require $20\%$ of measurements to locate a beam pair that meets a target of $\inrrxthresh = 0$ dB. (b) The empirical probability of the resulting $\nbrv$ after executing \steer on the collected \ginr measurements for $\inrrxthresh = -7$ dB.}
    \label{fig:alg}
\end{figure}

\edit{
To begin our evaluation of \steer, we first consider \figref{fig:alg-a}, which highlights the reduction in measurements needed when using \algref{alg:algorithm-solve} to solve problem \eqref{eq:problem-ii}, compared to measuring the entire spatial neighborhood and then applying exhaustive search.
Here, we consider a neighborhood of size $\nbr = \nbrtwotwo$ with resolution $\nbrd = \nbroneone$.
\figref{fig:alg-a} shows the empirical \gls{cdf} of the fraction of $\setinrmeas\idx{i\opt,j\opt}$ measured after running \algref{alg:algorithm-solve} across all possible $(i\opt,j\opt)$.
With $\inrrxthresh = 0$ dB, for instance, nearly $65\%$ of all beam pairs $(i\opt,j\opt)$ require at most $20\%$ of the neighborhood to be measured.
This highlights the impressive savings \algref{alg:algorithm-solve} can offer in terms of measurement overhead, a key practical consideration.
Around $12\%$ of beam pairs $(i\opt,j\opt)$ require the entire neighborhood $\setinrmeas\idx{i\opt,j\opt}$ to be measured for $\inrrxthresh = 0$ dB.
With stricter $\inrrxthresh$, more measurements are required in order to locate a transmit-receive beam pair that can meet the target.
Notice that some fraction of beam pairs require the entire neighborhood to be measured, which is almost exclusively due to the fact that $\inrrxthresh$ cannot be met within the neighborhood.
% For instance, 
In \figref{fig:alg-b}, we show the fraction of beam pairs $(i\opt,j\opt)$ that yield each possible $\nbrv$ when $\nbr = \nbrtwotwo$ and $\inrrxthresh = -7$ dB.
For example, around $29.5\%$ of beam pairs make use of the full $\nbrtwotwo$ tolerance allowed.
Around $20\%$ of beam pairs reach $\inrrx \leq -7$ dB with $\nbroneone$ of shifting.
}

\subsection{Performance Metrics}
We now introduce performance metrics used to evaluate \steer.
Recall, the Shannon capacities of the links we denote $\captx$ and $\caprx$ are based on the inherent link qualities $\snrtxbar$ and $\snrrxbar$, respectively.
Perhaps more practically meaningful, the capacities of the links after conventional beam alignment we refer to as the \textit{codebook capacities}, which are 
\begin{align}
\captxcb &= \logtwo{1 + \snrtxnom} \leq \captx, \quad
\caprxcb = \logtwo{1 + \snrrxnom} \leq \caprx
% \capsumcb &= \captxcb + \caprxcb \leq \captx + \caprx
\end{align}
with these nominal \gpsnr defined in \eqref{eq:snr-tx-rx-nom} and the sum codebook capacity as $\capsumcb = \captxcb + \caprxcb$.
Using conventional beam alignment, the achievable sum spectral efficiency of the transmit and receive links under equal \tdd with fixed power control (i.e., an instantaneous transmit power constraint) are
\begin{align}
% \setxtdd &= 0.5 \cdot \captxcb, \quad 
% \serxtdd = 0.5 \cdot \caprxcb % \\
% \sesumtdd &= 0.5 \cdot \capsumcb
\sesumtdd = 0.5 \cdot \logtwo{1 + \snrtxnom} + 0.5 \cdot \logtwo{1 + \snrrxnom}.
\end{align}
% with sum $\sesumtdd = 0.5 \cdot \capsumcb$.
\edit{
Under an average power constraint, power control can be used during equal \tdd operation to boost transmit power inversely proportional to the transmit duration.
In such a case, the sum spectral efficiency becomes
\begin{align}
\sesumgtdd = 0.5 \cdot \logtwo{1 + 2 \cdot \snrtxnom} + 0.5 \cdot \logtwo{1 + 2 \cdot \snrrxnom}.
\end{align}
Note that this achievable sum spectral efficiency coincides with that of equal \gls{fdd}.
While instantaneous transmit power constraints are more practical, it is still useful for us to compare against an average power constraint to better examine the gains offered by full-duplex.}
The achievable sum spectral efficiency under \steer we denote as $\sesumours = \setxours + \serxours$, which can be computed using \eqref{eq:se-tx-rx} with the \gpsinr achieved by \steer.
The achievable sum spectral efficiency under conventional beam selection is defined analogously.

To normalize these achievable sum spectral efficiencies to the sum codebook capacity, we translate them to quantities denoted by $\capfracsum$ by dividing by $\captxcb + \caprxcb$; for instance, the fraction of the codebook capacity achieved when full-duplexing with \steer is
\begin{align}
\capfracsumours 
= \frac{\sesumours}{\capsumcb}
= \frac{\setxours + \serxours}{\captxcb + \caprxcb}.
\end{align}
Note that $\capfracsum$ is typically less than $1$ but is not truly bounded since the codebook capacity is not a true upper bound on achievable spectral efficiency.
Nonetheless, codebook capacity is a useful metric since it provides insight on best-case full-duplex performance with a conventional beam codebook (i.e., in the presence of no cross-link or self-interference).

\begin{figure}
    \centering
    \includegraphics[width=\linewidth,height=0.27\textheight,keepaspectratio]{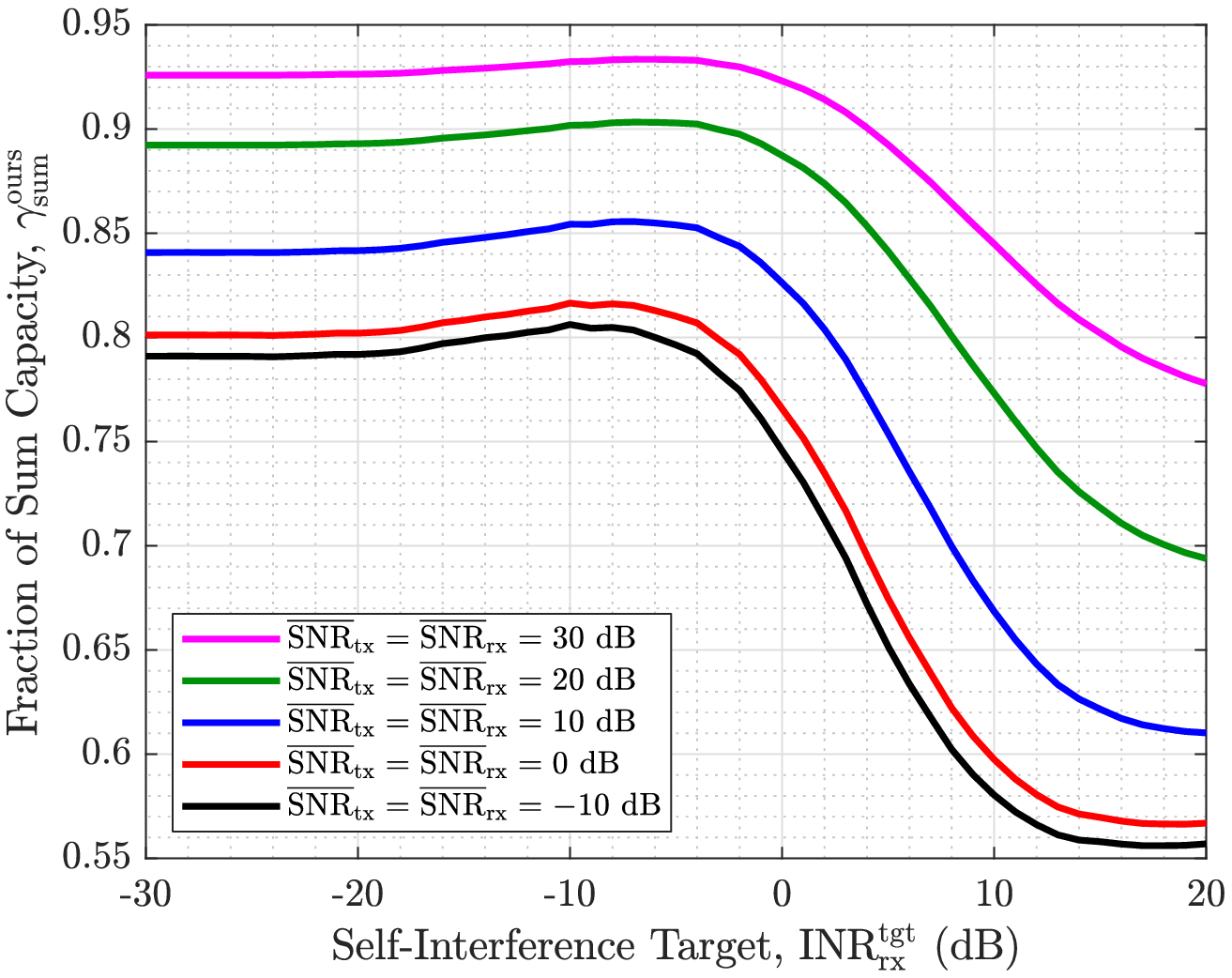}
    \caption{The fraction $\capfracsumours$ of the sum capacity $\capsumcb$ achieved by \steer as a function of the design parameter $\inrrxthresh$ for various $\snrtxbar = \snrrxbar$. Heuristically, $\inrrxthresh = -7$ dB proves to be broadly optimal or near-optimal.}
    \label{fig:thresh}
\end{figure}

\subsection{Choosing the Receive Link \ginr Target, $\inrrxthresh$.}
In \figref{fig:thresh}, we plot the fraction of the sum capacity $\capfracsumours$ achieved by \steer as a function of the design parameter $\inrrxthresh$ for various $\snrtxbar = \snrrxbar$, where we have used $\nbr = \nbrtwotwo$ and $\inrtx = 0$ dB for this illustration.
Heuristically, we observe that the $\inrrxthresh$ which maximizes sum spectral efficiency ranges from $-10$ dB to $0$ dB across this broad range of $\snrtxbar = \snrrxbar$.
While choosing $\inrrxthresh = -\infty$ dB is near-optimal (especially at high \gsnr), the optimal $\inrrxthresh$ is finite.
This is thanks to the fact that choosing a modest $\inrrxthresh$ can throttle the deviation that \steer makes from the nominal beams, reducing the chance the donor and/or \ue sees low beamforming gain.
Henceforth, these numerical results use $\inrrxthresh = -7$ dB since it is observed to be optimal or near-optimal broadly across $\snrtxbar$ and $\snrrxbar$. % (even when $\snrtxbar \neq \snrrxbar$).
% While not explicitly shown, 

% We illustrate that $

\subsection{How Does Full-Duplexing with \steer Compare to Other Multiplexing Strategies?}

%In \figref{fig:compare}, we compare full-duplexing with \steer to the following multiplexing strategies:
%\begin{itemize}
%    \item half-duplexing with beams from conventional beam selection
%    \item full-duplexing with beams from conventional beam selection.
%\end{itemize}

\begin{figure*}
    \centering
    % \subfloat[As a function of link quality, $\snrtxbar = \snrrxbar$.]{\includegraphics[width=\linewidth,height=0.27\textheight,keepaspectratio]{plots/main_frisbee_v04_06_v02}
    \subfloat[As a function of link quality, $\snrtxbar = \snrrxbar$.]{\includegraphics[width=\linewidth,height=0.27\textheight,keepaspectratio]{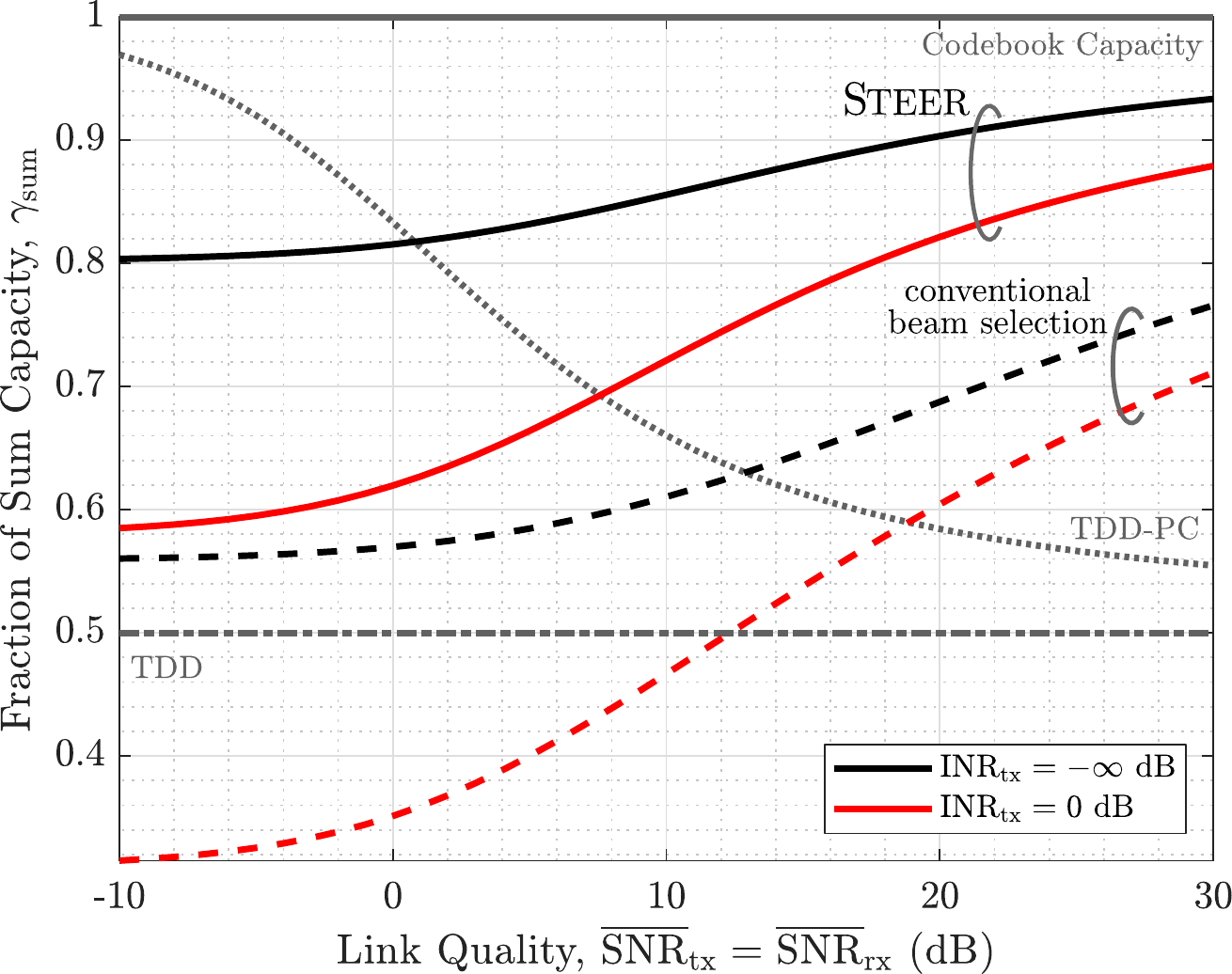}
        \label{fig:compare-a}}
    \quad
    % \subfloat[As a function of cross-link interference, $\inrtx$.]{\includegraphics[width=\linewidth,height=0.27\textheight,keepaspectratio]{plots/main_frisbee_v04_28_v02}
    \subfloat[As a function of cross-link interference, $\inrtx$.]{\includegraphics[width=\linewidth,height=0.27\textheight,keepaspectratio]{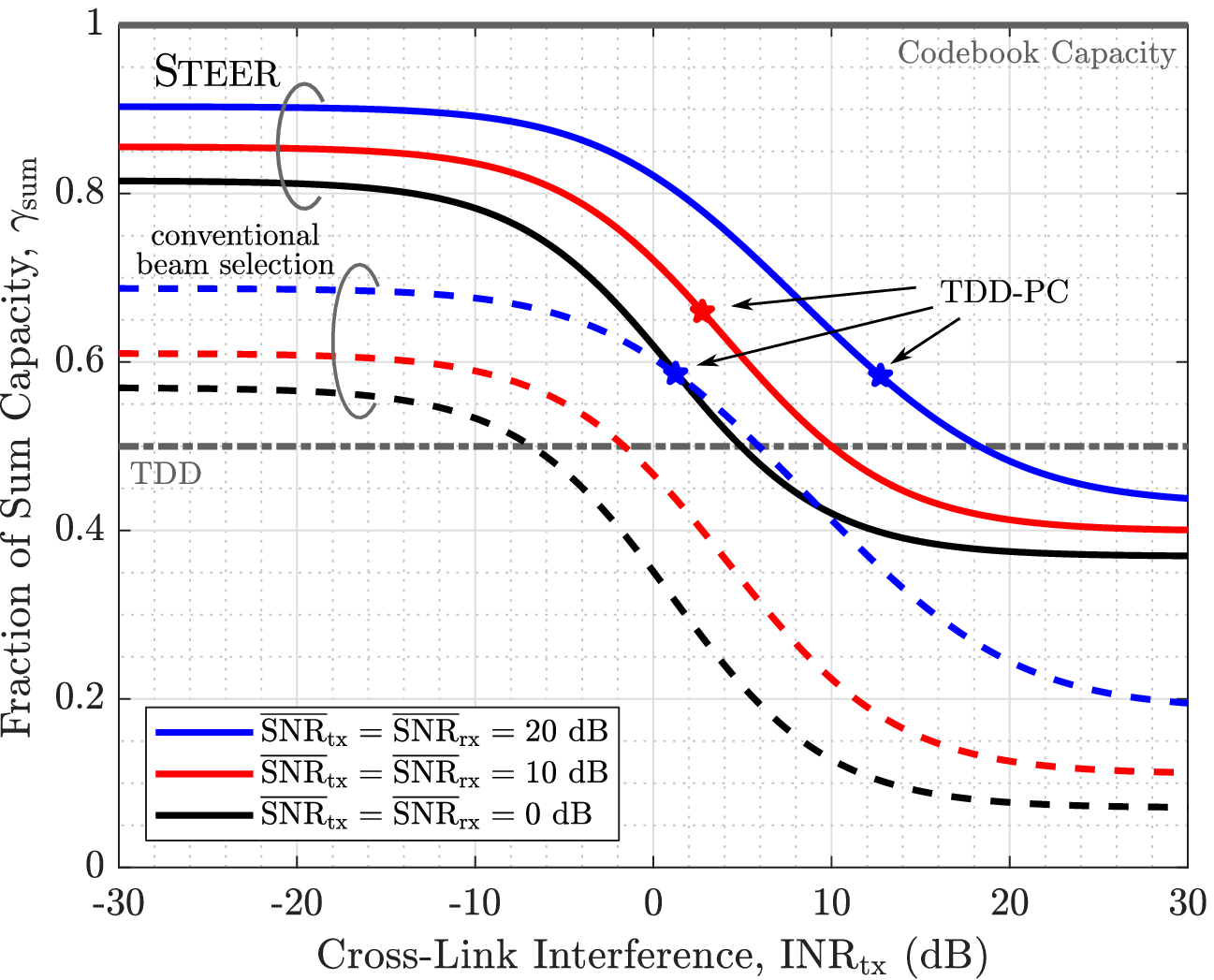}
        \label{fig:compare-b}}
    \caption{(a) The fraction of the sum capacity $\capfracsum$ as a function of $\snrtxbar = \snrrxbar$ for various $\inrtx$. (b) The fraction of the sum capacity $\capfracsum$ as a function of $\inrtx$ for various $\snrtxbar = \snrrxbar$. $\star$ markers indicate intersections with TDD-PC.} % \steer can outperform conventional beam selection and half-duplexing; when doing so, \steer demands lower link quality and can tolerate higher cross-link interference.
    \label{fig:compare}
\end{figure*}

In \figref{fig:compare}, we compare full-duplexing with \steer to both half-duplexing and full-duplexing with beams from conventional beam selection.
We let $\nbr = \nbrtwotwo$ with a spatial resolution of $\nbrd = \nbroneone$ and $\inrrxthresh = -7$ dB when running \steer.
First, let us examine \figref{fig:compare-a}, which shows the fraction of the sum capacity $\capfracsum$ achieved by various multiplexing strategies as a function of $\snrtxbar = \snrrxbar$.
(For now, we let $\snrtxbar = \snrrxbar$ for simplicity and examine $\snrtxbar \neq \snrrxbar$ shortly.)
We aim for the codebook capacity $\capfracsum = 1$ during full-duplex operation, whereas 
% is achieved if self-interference happened to be completely mitigateis completely eliminated.
half of this, $\capfracsum = 0.5$, can be achieved via half-duplexing with equal \tdd.
% With greedy \tdd, small fractions above this can be had, which then converges to equal \tdd at high \gsnr.
% Solid lines indicate the performance of \steer for various cross-link interference levels $\inrtx$ whereas their dashed counterparts indicate that with conventional beam selection.
With TDD-PC, high $\capfracsum$ can be had at low \gsnr thanks to $\log\parens{1+x} \approx x$ at low $x$, though these gains diminish toward $0.5$ as \gsnr increases.
Without cross-link interference (when $\inrtx = -\infty$ dB; shown in black), \steer vastly outperforms \tdd  across \gpsnr, achieving $80$\% of the codebook capacity at low \gsnr and over $90$\% at high \gsnr.
Albeit less practical, TDD-PC can outperform \steer at low \gsnr, where doubling \gsnr approximately doubles spectral efficiency, but falls short for $\snrtxbar = \snrrxbar \geq 1$ dB.
% The dashed black line shows the performance with 
Conventional beam selection also broadly outperforms \tdd but only outperforms TDD-PC at $\snrtxbar = \snrrxbar \geq 13$ dB.
The sizable gap between \steer and conventional beam alignment of around $20$\% of the codebook capacity is attributed to better self-interference mitigation of \steer. %  (since there is no cross-link interference).
With cross-link interference that is equal to noise (when $\inrtx = 0$ dB; shown in red), we naturally see a drop in performance of both \steer and conventional beam selection.
In this case, \steer still broadly outperforms half-duplexing with TDD, while the same cannot be said about conventional beam selection.
Rather, attempting to full-duplex with beams from conventional beam selection falls short of TDD for $\snrtxbar = \snrrxbar \leq 12$ dB.
This is due to higher self-interference with conventional beams, which plagues the receive link, and the presence of cross-link interference, which plagues the transmit link.
\steer can reduce self-interference to levels that make full-duplex operation worthwhile, even in the presence of cross-link interference.
% With cross-link interference that is ten times stronger than noise (when $\inrtx = 10$ dB; shown in blue), this trend continues, illustrating that full-duplexing with \steer is attractive for $\snrtxbar, \snrrxbar \geq 10$ dB.
% Full-duplexing with conventional beam selection, however, is only attractive for $\snrtxbar, \snrrxbar \geq 25$ dB.

%\begin{figure*}
%    \centering
%    \subfloat[Fraction of the sum capacity, $\capfracsum$.]{\includegraphics[width=\linewidth,height=0.27\textheight,keepaspectratio]{plots/main_frisbee_v04_06}
%        \label{fig:compare-a}}
%    \quad
%    \subfloat[Gain in sum spectral efficiency of \steer, $\segainsum$.]{\includegraphics[width=\linewidth,height=0.27\textheight,keepaspectratio]{plots/main_frisbee_v04_09}
%        \label{fig:compare-b}}
%    \caption{(a) The fraction of the sum capacity $\capfracsum$ as a function of $\snrtxbar = \snrrxbar$ for various $\inrtx$. Solid lines show the performance of \steer; dashed lines show that of conventional beam selection. (b) The gain in sum spectral efficiency $\segainsum$ of \steer over half-duplexing strategies as a function of $\snrtxbar = \snrrxbar$ for various $\inrtx$. Exceeding $\segainsum = 1$ indicates full-duplexing gains.}
%    \label{fig:compare}
%\end{figure*}

In \figref{fig:compare-b}, we plot the fraction of the sum capacity $\capfracsum$ achieved by \steer as a function of cross-link interference $\inrtx$ for various $\snrtxbar = \snrrxbar$.
The starred markers indicate the intersection with performance of TDD-PC.
% The corresponding dashed lines represent that of conventional beam selection.
At low cross-link interference, full-duplexing with conventional beam selection outperforms half-duplexing with \tdd across \gpsnr.
With \steer, significantly higher spectral efficiencies are obtained largely thanks to its ability to better mitigate self-interference on the receive link.
As cross-link interference increases, conventional beam selection degrades as its marginal full-duplexing gains are negated, eventually falling below \tdd.
At $\snrtxbar = \snrrxbar = 10$ dB, for instance, conventional beam selection can only tolerate $\inrtx \leq -2$ dB, whereas \steer can tolerate $\inrtx \leq 10$ dB---a gain of about $12$ dB in robustness to cross-link interference.
Here, the rightward and upward shift of \steer compared to conventional beam selection captures its increased robustness to cross-link interference and its improved sum spectral efficiency.
Nonetheless, these results emphasize that justifying full-duplex operation with \steer over half-duplexing strategies depends on cross-link interference levels and the inherent transmit and receive link qualities.
% For example, with $\snrtxbar = \snrrxbar = -10$ dB and $\inrtx = 10$ dB, it is counterproductive to full-duplex using \steer; instead, conventional half-duplexing should be used.
This motivates the need to study and measure practical cross-link interference levels and routes to mitigate it as needed, potentially via user selection, which may also be used to meet \gsnr requirements.
% Keep in mind that devices may be served in a half-duplex fashion as necessary if they lead to cross-link interference levels and \gpsnr that make full-duplex counterproductive.

\comment{
In \figref{fig:compare-b}, we simply plot the gain in sum spectral efficiency with \steer versus equal \tdd and greedy \tdd (i.e., the solid lines in \figref{fig:compare-a} divided by the half-duplex lines).
Note that we desire $\segainsumtdd, \segainsumgtdd > 1$ to make full-duplex operation worthwhile.
As seen in \figref{fig:compare-a}, without cross-link interference or that which is equal to noise, full-duplexing with \steer can nearly match or outperform equal and greedy \tdd even at low \gsnr.
With $\inrtx = 10$ dB, \steer does not outperform half-duplex except at approximately $\snrtxbar, \snrrxbar \geq 10$ dB.
At low \gsnr, the full-duplexing gains do not justify introducing such high cross-link interference.
This ``barrier to entry'' for full-duplex operation with \steer when $\inrrx = 0$ dB is approximately $\snrtxbar, \snrrxbar \geq -7$ dB and vanishes with cross-link interference.
The gains of \steer over half-duplex climb toward $\segainsum = 2$ with increased \gsnr but will fundamentally fall short with the presence of cross-link interference.
These results emphasize that justifying full-duplex operation with \steer over conventional half-duplexing strategies depends heavily on cross-link interference levels and the inherent transmit and receive link qualities.
This motivates the need to study and measure potential cross-link interference levels and routes to mitigate it as needed, potentially via user selection, which may also help meet \gsnr requirements.
Keep in mind that devices may be served in a half-duplex fashion as necessary if they lead to cross-link interference levels and \gpsnr that make full-duplex counterproductive.
}

% \subsubsection{How Does Full-Duplexing with \steer Compare to Half-Duplexing Strategies?}

% \pagebreak

% \subsection{How Does Full-Duplexing with \steer Compare to that with \Naive Beam Selection?}

\begin{figure*}
    \centering
    % \subfloat[Fraction of the sum capacity, $\capfracsum$.]{\includegraphics[width=\linewidth,height=0.27\textheight,keepaspectratio]{plots/main_frisbee_v04_17_v02}
    \subfloat[Fraction of the sum capacity, $\capfracsum$.]{\includegraphics[width=\linewidth,height=0.27\textheight,keepaspectratio]{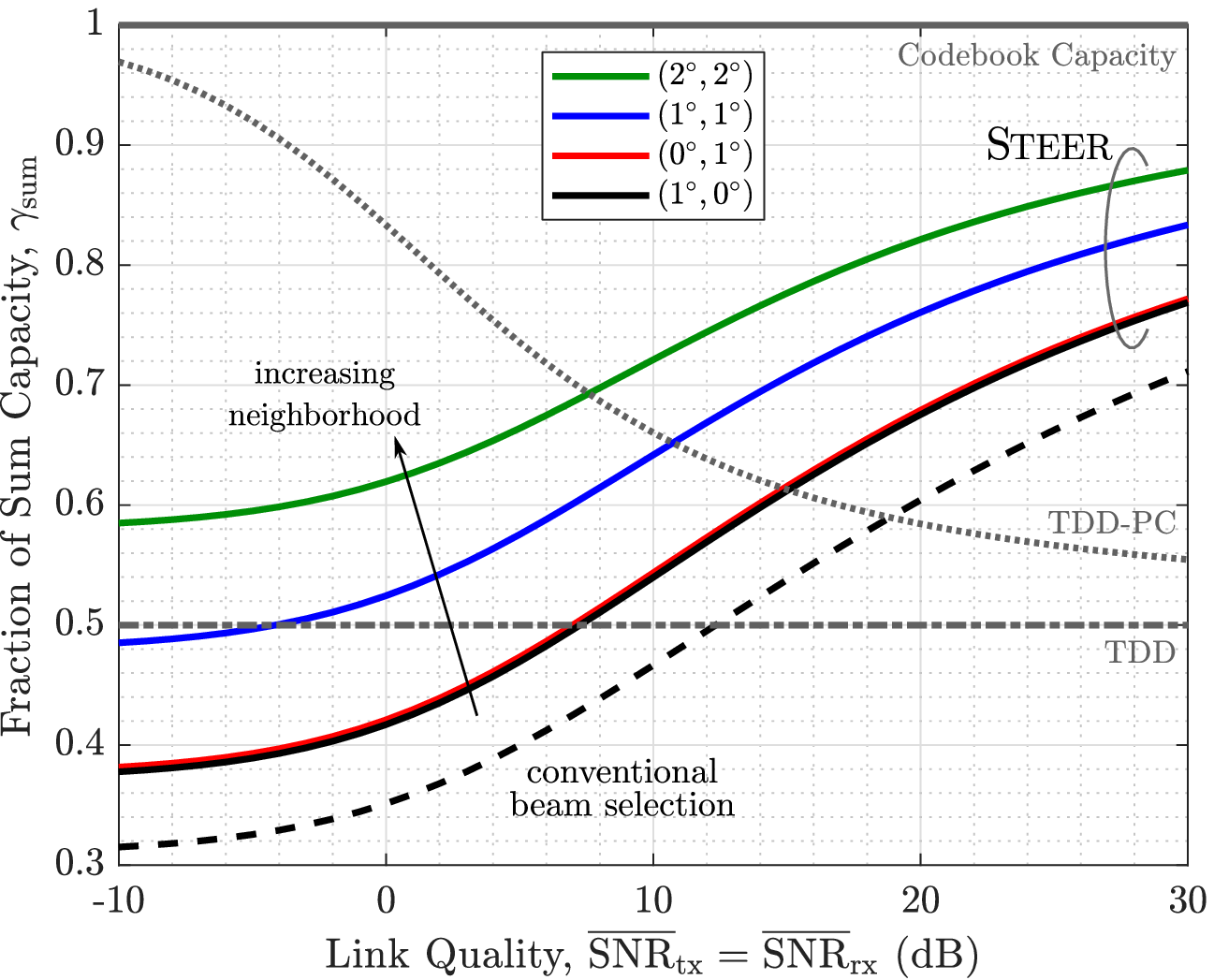}
        \label{fig:nbr-vary-a}}
    \quad
    % \subfloat[Gain in sum spectral efficiency of \steer, $\segainsumnom$.]{\includegraphics[width=\linewidth,height=0.27\textheight,keepaspectratio]{plots/main_frisbee_v04_18}
    \subfloat[\edit{Sum spectral efficiency, $\setx + \serx$.}]{\includegraphics[width=\linewidth,height=0.265\textheight,keepaspectratio]{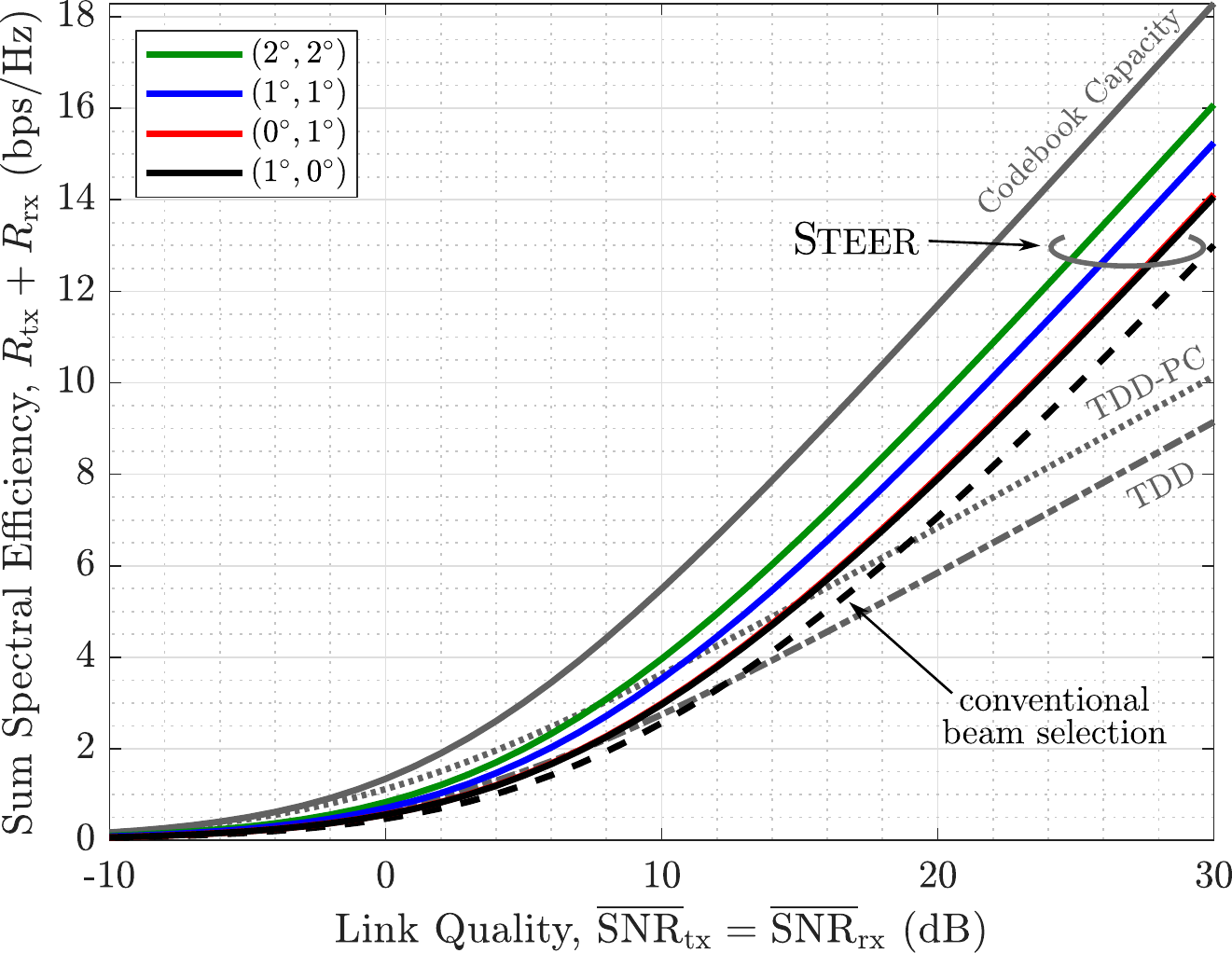}
        \label{fig:nbr-vary-b}}
    \caption{(a) The fraction of the sum capacity $\capfracsum$ as a function of $\snrtxbar = \snrrxbar$ for various $\nbr$, where cross-link interference $\inrtx = 0$ dB. (b) The unnormalized counterpart of (a). Enlarging $\nbr$ offers noteworthy spectral efficiency gains, especially at high \gsnr.} % Enlarging the neighborhood $\nbr$ of \steer can offer noteworthy spectral efficiency gains over conventional beam selection.}
    \label{fig:nbr-vary}
\end{figure*}

\subsection{The Impact of Neighborhood Size $\nbr$ on \steer's Peformance}

Having examined the performance of \steer versus other multiplexing strategies for a fixed neighborhood size $\nbr = \nbrtwotwo$, we now inspect its performance for various neighborhood sizes while fixing the spatial resolution to $\nbrd = \nbroneone$.
A larger neighborhood size $\nbr$ improves \steer's ability to reduce self-interference since it widens the search space, though this comes at the cost of additional measurement overhead and the potential for gaps in coverage (i.e., reduced \gsnr).
% Choosing the neighborhood size is an important design decision that can only be evaluated numerically through measurements and simulation.
In \figref{fig:nbr-vary-a}, we plot the fraction of the sum capacity $\capfracsum$ achieved by \steer for various $\nbr$ as a function of $\snrtxbar = \snrrxbar$, where $\inrtx = 0$ dB.
The dashed line shows the performance with conventional beam selection.
% As before, we plot that with greedy \tdd and equal \tdd, along with the codebook capacity, to compare \steer against these baselines.
In \figref{fig:nbr-vary-b}, we plot the unnormalized counterpart of \figref{fig:nbr-vary-a} (i.e., absolute sum spectral efficiency $\setx + \serx$).

Full-duplexing with conventional beam selection offers gains over TDD only beyond \gpsnr of $12$ dB, which are fairly modest until high \gsnr.
% By allowing \steer to shift beams by a mere $1^\circ$ in either azimuth \textit{or} elevation, we see noteworthy improvement
By allowing \steer to shift beams by at most $1^\circ$ in azimuth and elevation (i.e., $\nbr = \nbroneone$), it can choose beams that greatly reduce self-interference to levels such that full-duplex operation matches or significantly outperforms \tdd. 
Notice, with $\nbroneone$, \steer only requires $\snrtxbar = \snrrxbar \geq -5$ dB to justify full-duplex operation over \tdd.
Compared to full-duplexing with conventional beam alignment, this is an \gsnr gain of over $15$ dB.
It is important to realize that this $\nbroneone$-neighborhood lives well within the $7^\circ$ beamwidth of our beams.
% With a neighborhood of merely $\nbr = \nbrtwotwo$, \steer can choose beams that greatly reduce self-interference to levels that make full-duplex operation that nearly matches or significantly outperforms equal \tdd.
% Notice that \steer requires much lower $\snrtxbar = \snrrxbar$ to justify full-duplex operation, introducing a so-called \textit{\gsnr gain}.
A gain of about $0.17$ in $\capfracsum$ is observed with $\nbroneone$ and this jumps to around $0.27$ with $\nbrtwotwo$.

% The gain in sum spectral efficiency $\segainsumnom$ of \steer over conventional beam selection is shown in \figref{fig:nbr-vary-b}.
% Over $50$\% improvement in sum spectral efficiency can be had with \steer at low \gsnr for $\nbroneone$.
% This jumps to nearly doubling the sum spectral efficiency for $\nbrtwotwo$. 
% As $\snrtxbar = \snrrxbar$ increases, the impacts of self-interference are lessened and, thus, conventional beam alignment approaches the performance of \steer.
% The two converge only at very high $\snrtxbar = \snrrxbar$, suggesting that \steer can offer noteworthy improvement compared to full-duplexing \naively with conventional beam selection for practical \gpsnr.

\begin{figure*}
    \centering
    \subfloat[CDF of $\inrrx$.]{\includegraphics[width=0.475\linewidth,height=\textheight,keepaspectratio]{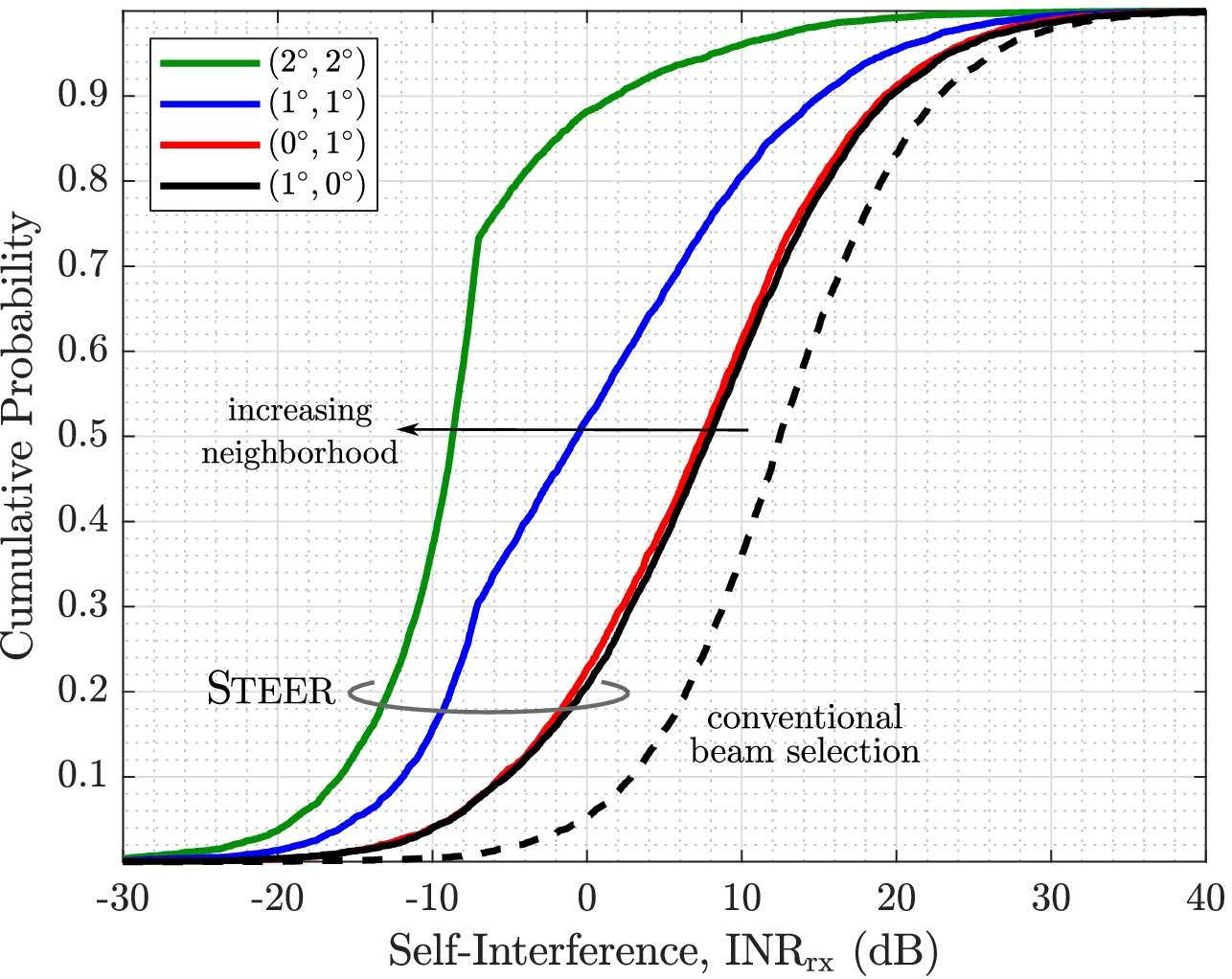}
        \label{fig:cdf-a}}
    \quad
    \subfloat[CDF of $\todB{\sinrrx} - \todB{\snrrxbar}$.]{\includegraphics[width=0.475\linewidth,height=\textheight,keepaspectratio]{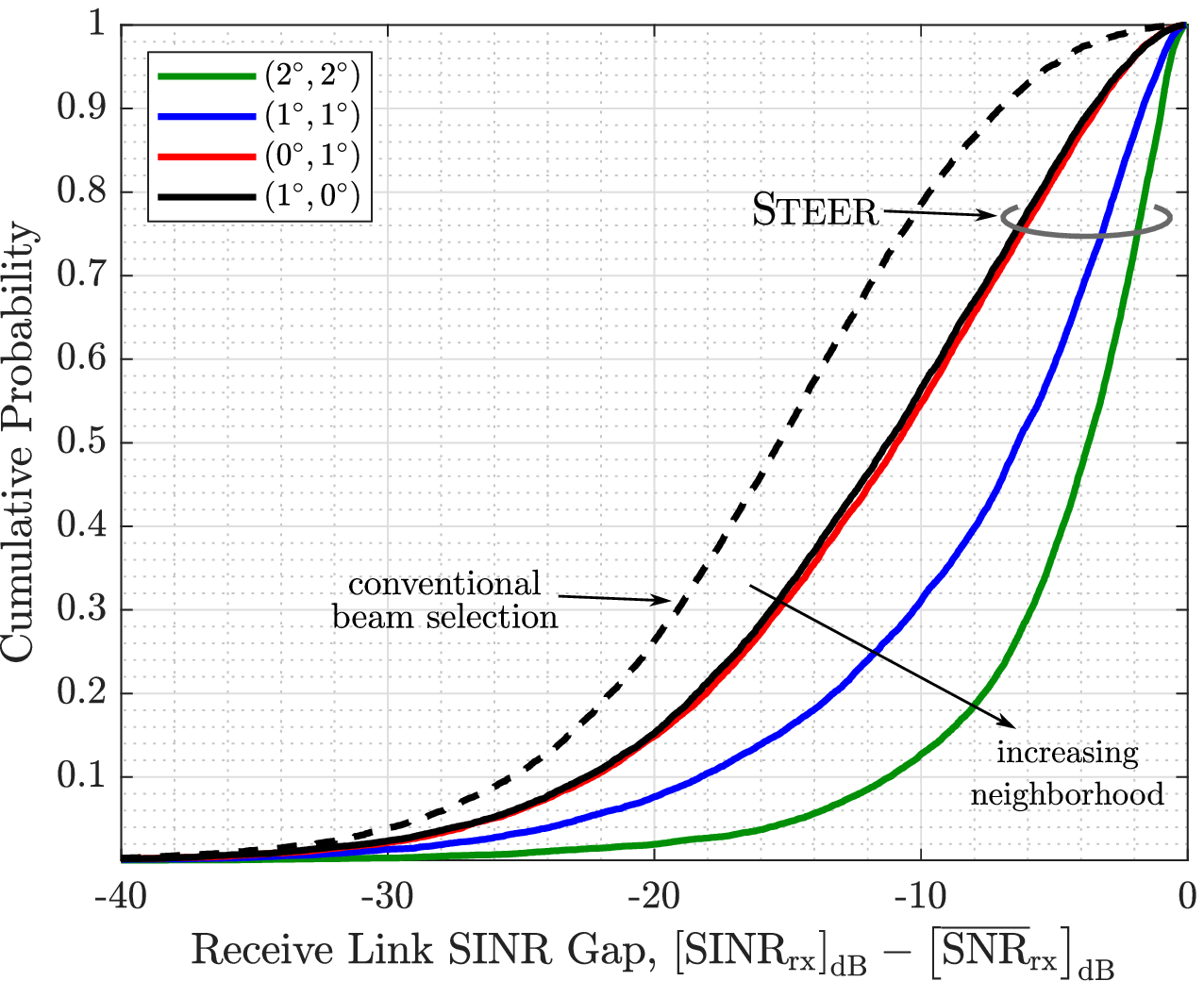}
        \label{fig:cdf-b}}
    \caption{(a) The \gcdf of \inrrx for various neighborhood sizes \nbr. (b) The \gcdf of the gap between $\sinrrx$ and $\snrrxbar$ for various neighborhood sizes $\nbr$. \steer reliably reduces $\inrrx$, as evident in (a), while maintaining high beamforming gain, shifting $\sinrrx$ closer to $\snrrxbar$ (its upper bound) as shown in (b).}
    \label{fig:cdf}
\end{figure*}

In \figref{fig:cdf-a}, we plot the \gcdf of receive link \ginr offered by \steer for various neighborhood sizes $\nbr$. % , with the dashed line being that for conventional beam selection.
Under conventional beam selection, the \ginr distribution undesirably lay largely above $0$ dB, where self-interference is stronger than noise.
% When running \steer, we use an \ginr target of $\inrrxthresh = -7$ dB.
Notice that shifting the transmit and receive beams by at most $1^\circ$ in azimuth and elevation, the \inrrx distribution shifts notably leftward---by about $13$ dB in median. 
% Half of all will see enjoy an $\inrrx \approx 0$ dB (a reduction of around $12$ dB versus conventional beam selection).
Half of all possible transmit-receive user pairs enjoy an $\inrrx \leq 0$ dB with \steer when $\nbr = \nbroneone$. 
With $\nbrtwotwo$, almost $90$\% of user pairs enjoy $\inrrx \leq 0$ dB.
As a result of using $\inrrxthresh = -7$ dB, we see a sharp bend around $\inrrx = -7$ dB since \steer is not incentivized to reduce the \ginr below such.
Note that there exist select user pairs that require further deviation beyond $\nbrtwotwo$ in order to deliver low \ginr.
Around $4$\% of user pairs see $\inrrx \geq 10$ dB even with searching for beams across a $\nbrtwotwo$-neighborhood.
Enlarging the neighborhood $\nbr$ or using a finer spatial resolution $\nbrd$ may facilitate full-duplexing these user pairs or perhaps they are better off served in a half-duplex fashion.

In \figref{fig:cdf-b}, we plot the \gcdf of the difference in $\todB{\sinrrx}$ and $\todB{\snrrxbar}$ (its upper bound) of \steer for various neighborhood sizes $\nbr$.
This difference is useful in capturing two artifacts of \steer: (i) its reduction in \inrrx and (ii) its effectiveness in receive beamforming.
Recall that $\snrrxbar$ is the maximum achievable \gsnr on the receive link and is only achieved by beamforming directly toward the receive device; a conventional codebook would only achieve this \gsnr if the receive device was precisely in the direction of one of its receive beams.
With conventional beam alignment, $\sinrrx$ is typically well over $10$ dB short of $\snrrxbar$ and is not unlikely to fall over $20$ dB short.
When \steer is supplied a $\nbroneone$-neighborhood, the distribution greatly shifts rightward.
Around $40$\% of the time, \steer delivers an $\sinrrx$ that is within $5$ dB of $\snrrxbar$.
This is thanks to \steer's ability to reduce $\inrrx$ and simultaneously deliver high beamforming gain (i.e., high $\snrrx$).
With $\nbrtwotwo$, even better performance is delivered by \steer: around $70$\% of the time $\sinrrx$ is within $6$ dB and around $40$\% of the time, within $3$ dB.
As highlighted before, a tail exists even with a $\nbrtwotwo$-neighborhood since some user pairs simply cannot be offered low $\inrrx$ without further deviation (or potentially finer spatial resolution)---a characteristic of the self-interference channel.
Due to space constraints, we have omitted an examination of the transmit link.
Similar conclusions are drawn: \gsnr loss is throttled by neighborhood size, as was the motivation behind the design of \steer.
It would be valuable future work to investigate how the gains of \steer saturate as the neighborhood $\nbr$ is widened beyond $\nbrtwotwo$ or as the resolution $\nbrd$ is reduced below $\nbroneone$.

\subsection{For Disparate Transmit and Receive Links}

\begin{figure*}
    \centering
    \subfloat[Achieved by conventional beam selection, $\capfracsumnom$.]{\includegraphics[width=\linewidth,height=0.25\textheight,keepaspectratio]{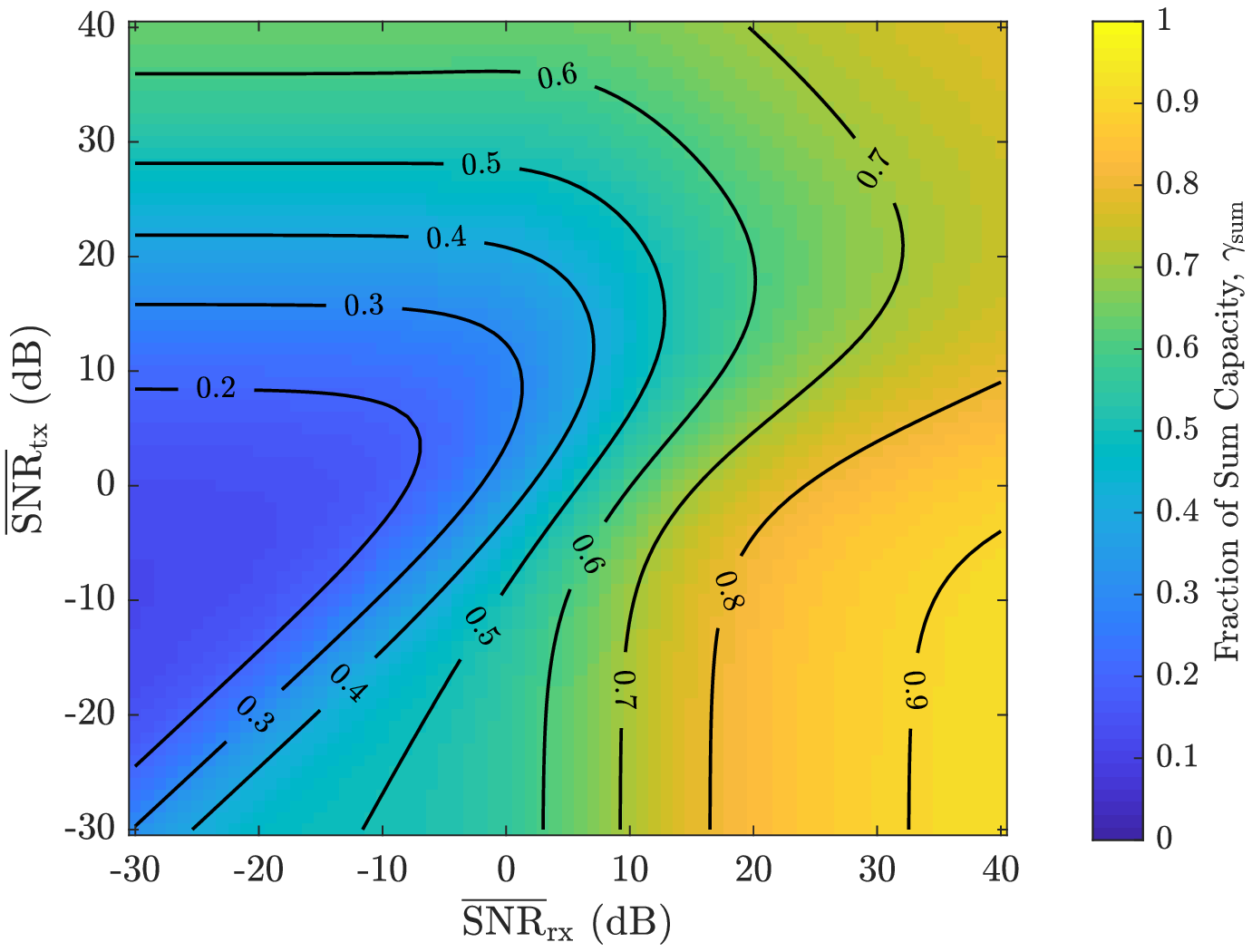}
        \label{fig:heatmap-a}}
    \quad
    \subfloat[Achieved by \steer, $\capfracsumours$.]{\includegraphics[width=\linewidth,height=0.25\textheight,keepaspectratio]{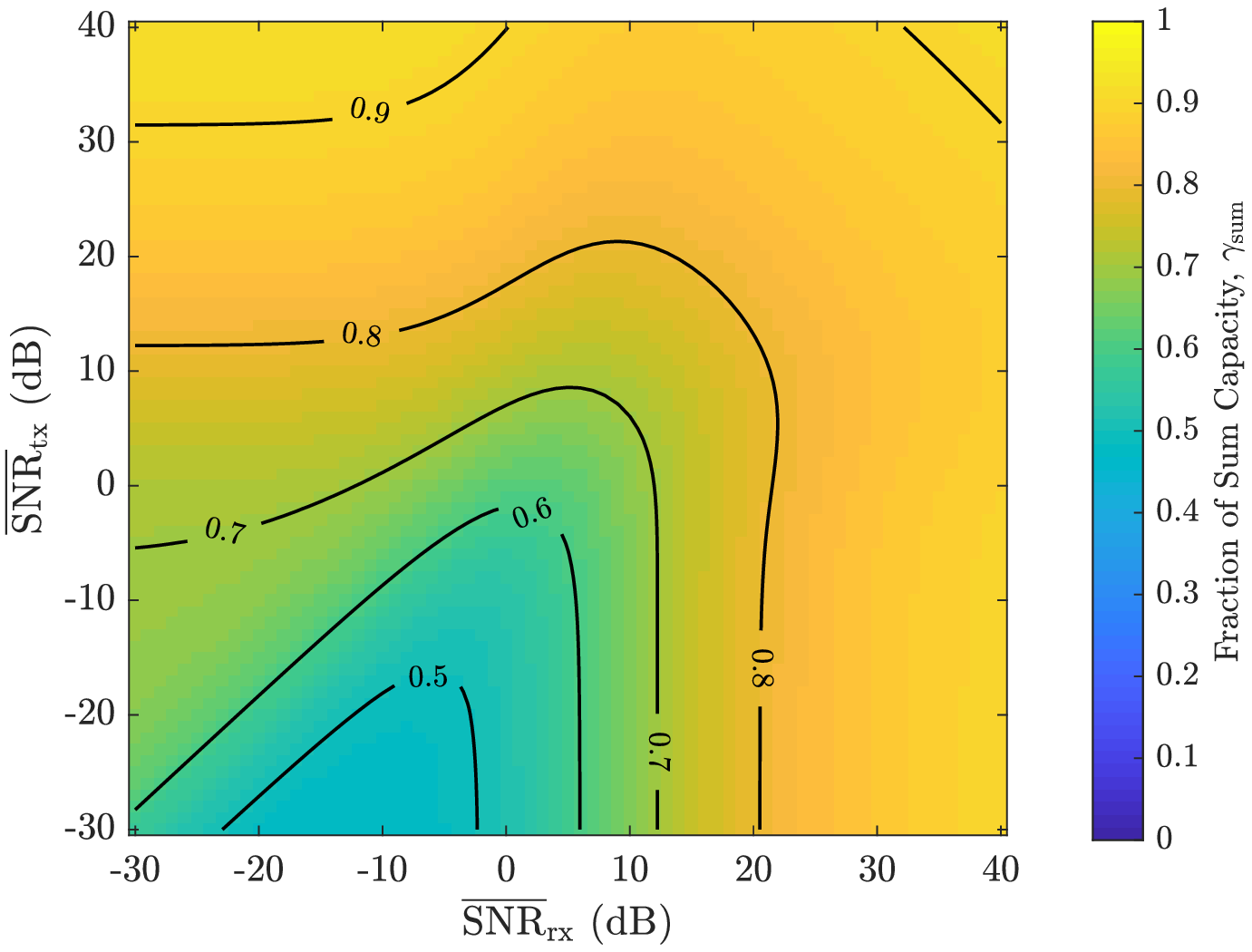}
        \label{fig:heatmap-b}}
    \caption{The fraction of the sum capacity achieved by (a) conventional beam selection and (b) \steer as a function of $\parens{\snrrxbar,\snrtxbar}$, where $\inrtx = 0$ dB. With \steer, greater sum spectral efficiency is achieved broadly across \gpsnr, and the region that nets $\capfrac \leq 0.5$ (where half-duplexing is preferred) is greatly reduced with \steer.}
    \label{fig:heatmap}
\end{figure*}

For our final set of results, we investigate how \steer performs for $\snrtxbar \neq \snrrxbar$.
This is especially important in \iab applications since network-side infrastructure, like fiber-connected donor base stations and \iab nodes, will likely have higher transmit powers, more antennas, and lower noise figures compared to user equipment.
In \figref{fig:heatmap-a}, as a function of $\snrtxbar$ and $\snrrxbar$, we show the fraction of the sum capacity $\capfracsum$ achieved when full-duplexing with beams from conventional beam selection, where $\inrtx = 0$ dB.
In \figref{fig:heatmap-b}, we plot that with beams from \steer.
Recall that $\capfracsum = 0.5$ can always be achieved by half-duplexing with \tdd, meaning we desire full-duplex operation that exceeds this.
% An obvious improvement can be seen with \steer over conventional beam selection, as higher fractions of the sum capacity are achieved broadly across combinations of $\snrtx$ and $\snrrx$. 
With conventional beam selection, high $\capfracsum$ is seen only at high $\snrrxbar$, where self-interference is less impactful due to the diminishing gains of $\logtwo{1+x}$.
Notice that, when $\snrrxbar \leq 0$ dB, conventional beam selection largely yields $\capfracsum$ that is worse than \tdd for practical $\snrtxbar$.
When \steer is employed, on the other hand, an obvious improvement can be seen, as higher fractions of the sum capacity are achieved broadly across combinations of $\snrtxbar$ and $\snrrxbar$. 
Only a small region at low $\snrtxbar$ and low $\snrrxbar$ yields $\capfracsum \leq 0.5$, in which case the system is better off operating using \tdd in terms of sum spectral efficiency.
Recall these results are with $\inrtx = 0$ dB, meaning they would only improve with reduced cross-link interference.

\begin{figure}
    \centering
    \includegraphics[width=\linewidth,height=0.3\textheight,keepaspectratio]{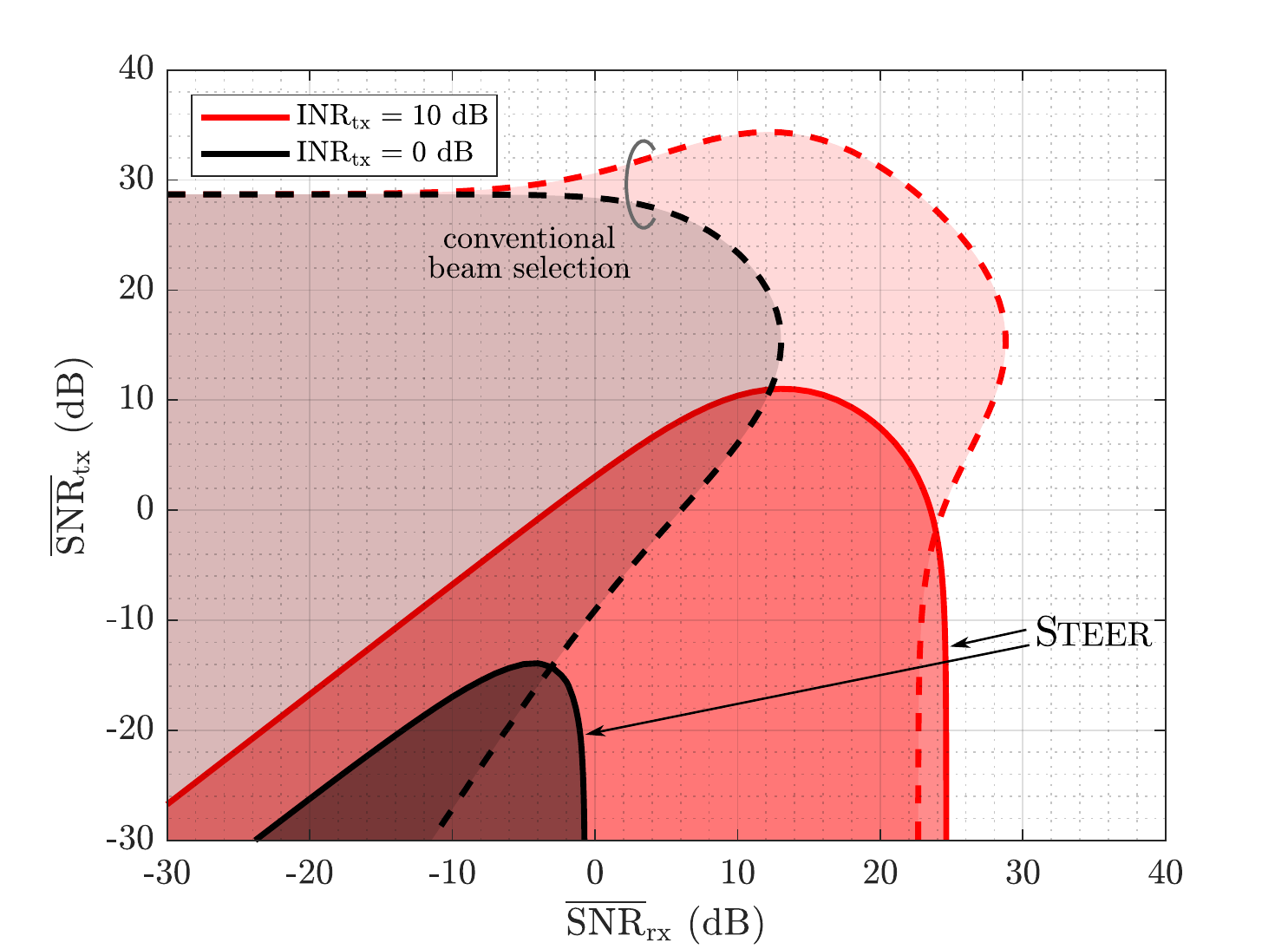}
    % \caption{For various cross-link interference levels $\inrtx$, the boundary of the $\parens{\snrtxbar,\snrrxbar}$ region where the fraction of the sum capacity $\capfracsum \leq 0.5$ (achieved by equal \tdd). The region where $\capfracsum \leq 0.5$ shrinks dramatically with \steer, outside of which full-duplexing is justified. \steer generally demands lower \gpsnr and can tolerate higher cross-link interference to offer gains over half-duplex.}
    \caption{For various cross-link interference levels, the shaded $\parens{\snrtxbar,\snrrxbar}$-region where $\capfracsum \leq 0.5$ (achieved by equal \tdd). This shaded region shrinks dramatically with \steer, outside of which full-duplexing is justified. \steer generally demands lower \gpsnr to offer gains over half-duplex and can tolerate higher cross-link interference .}
    \label{fig:boundary}
\end{figure}

Finally, in \figref{fig:boundary}, we compare the $\parens{\snrtxbar,\snrrxbar}$-regions where $\capfracsum \leq 0.5$ for \steer and for conventional beam selection at various cross-link interference levels. % $\inrtx$.
Within the shaded regions, it is advantageous to operate using \tdd; outside of them, full-duplexing is worthwhile (i.e., $\capfracsum \geq 0.5$).
When $\inrtx = 0$ dB, the dashed and solid black lines correspond to the $\capfracsum = 0.5$ contours in \figref{fig:heatmap-a} and \figref{fig:heatmap-b}, respectively.
The region bounded by the solid black line is notably smaller than that bounded by the dashed black line, highlighting the dramatic \gsnr improvement offered by \steer.
At $\snrrxbar = -10$ dB, for instance, \steer offers an $\snrtxbar$ gain of over $30$ dB.
% This is even more drastic for $\inrtx = -10$ dB, whose boundary is not even visible over this range of $\snrtxbar$ and $\snrrxbar$.
With higher $\inrtx$, the regions naturally grow as cross-link interference erodes some of the full-duplexing gains.
Still, \steer proves to be more robust to cross-link interference as it generally demands lower \gpsnr to outperform half-duplex. % , widening the region where full-duplex is justified.
% These results highlight an important fact: compared to full-duplexing with conventional beam selection, \steer demands lower \gsnr to outperform half-duplex and can tolerate higher cross-link interference, widening the region where full-duplexing is justified.

\comment{
---

\begin{figure}
    \centering
    \includegraphics[width=\linewidth,height=0.3\textheight,keepaspectratio]{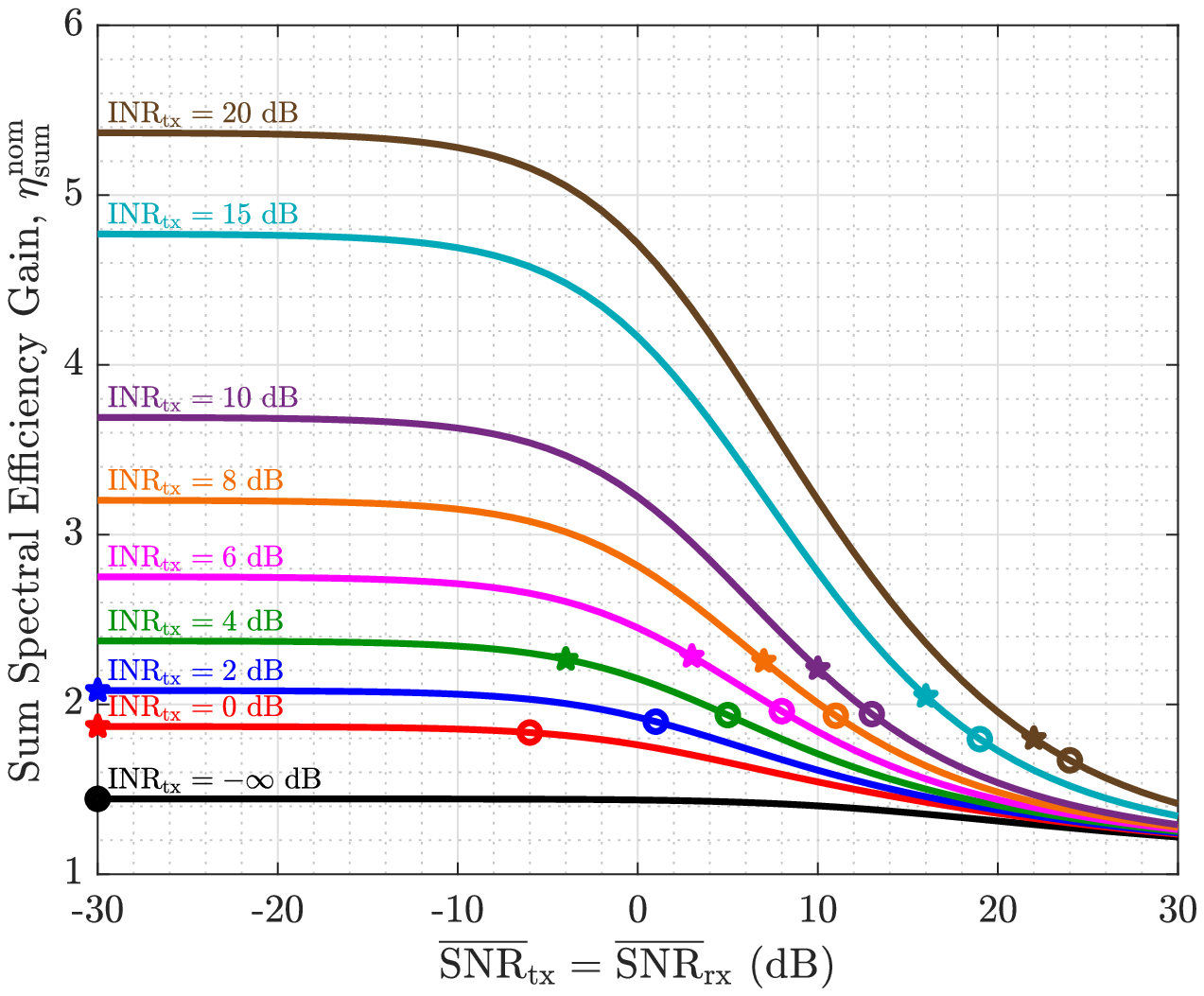}
    \caption{Caption.}
    \label{fig:}
\end{figure}

\begin{figure*}
    \centering
    \subfloat[Caption a.]{\includegraphics[width=0.475\linewidth,height=\textheight,keepaspectratio]{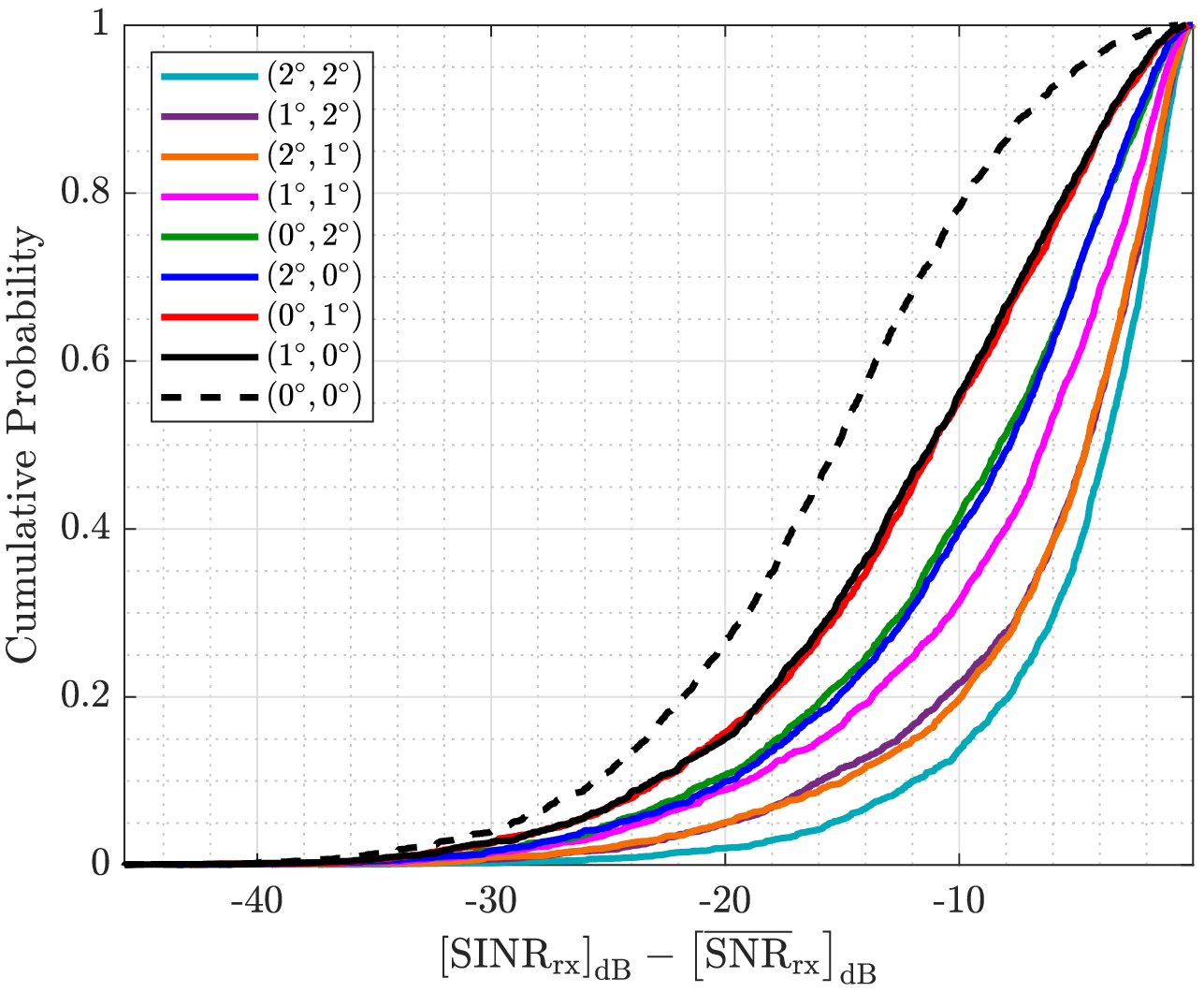}
        \label{fig:subfig-a}}
    \quad
    \subfloat[Caption b.]{\includegraphics[width=0.475\linewidth,height=\textheight,keepaspectratio]{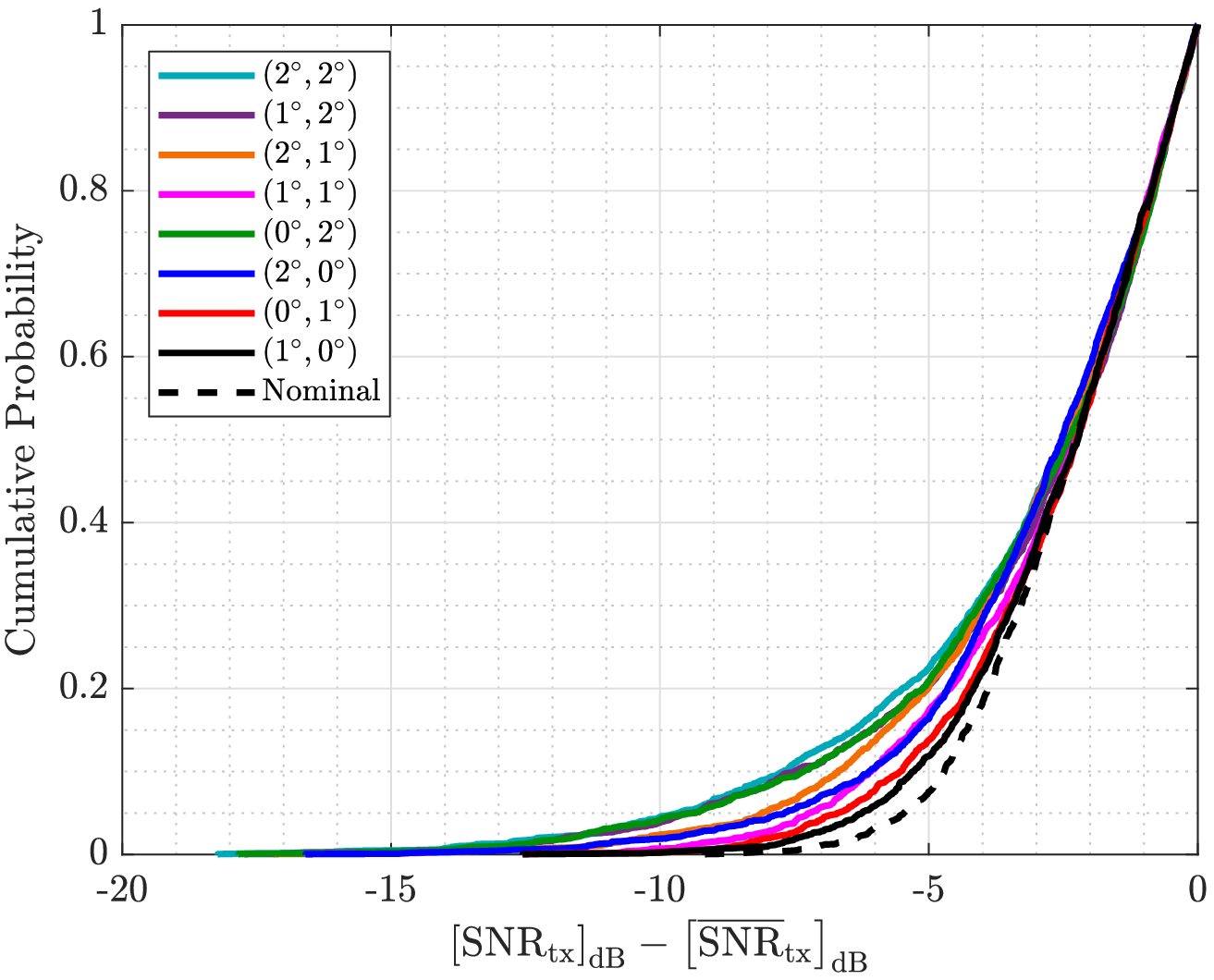}
        \label{fig:subfig-b}}
    \caption{Caption here.}
    \label{fig:subfigs}
\end{figure*}

\begin{figure*}
    \centering
    \subfloat[Caption a.]{\includegraphics[width=0.475\linewidth,height=\textheight,keepaspectratio]{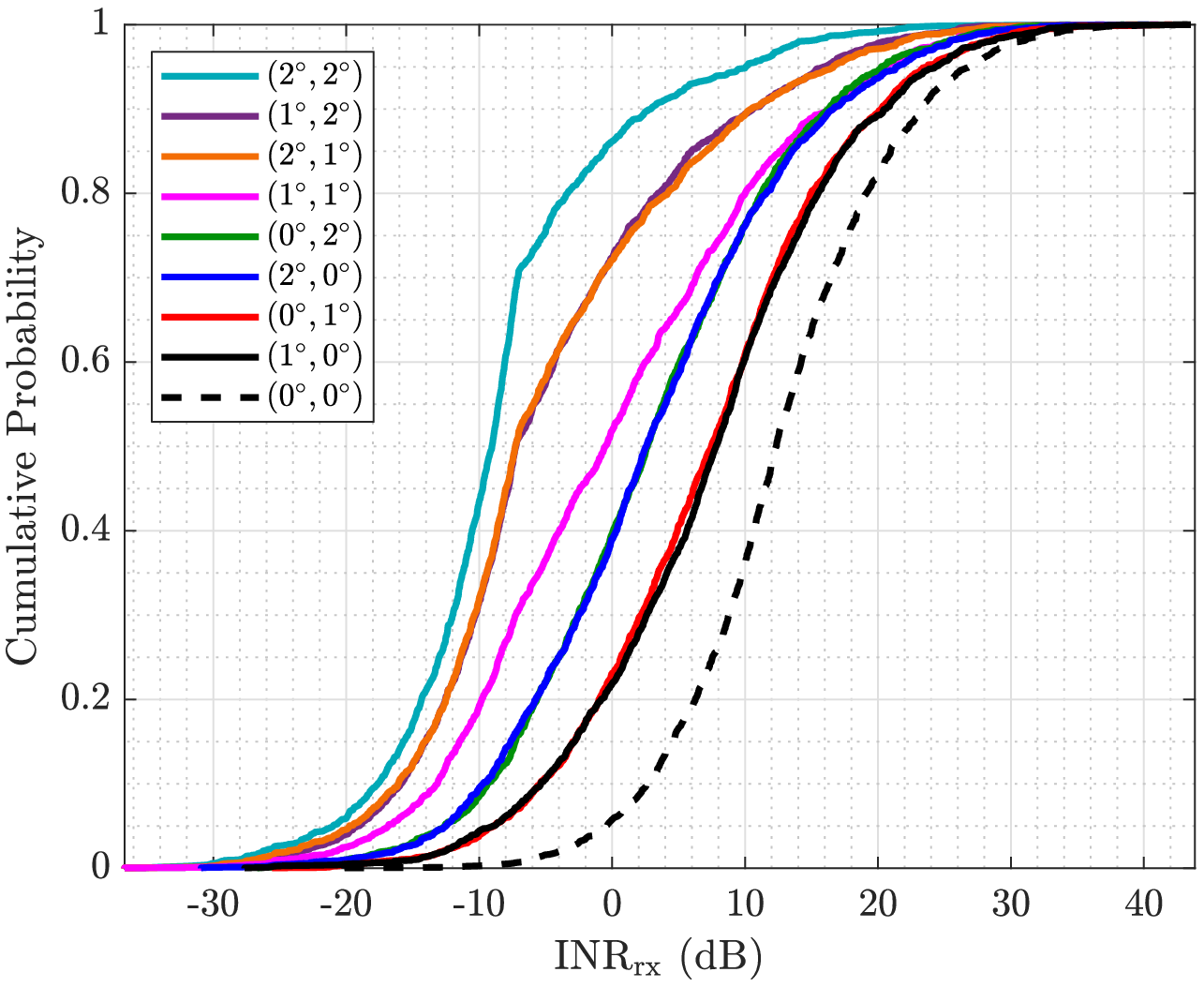}
        \label{fig:subfig-a}}
    \quad
    \subfloat[Caption b.]{\includegraphics[width=0.475\linewidth,height=\textheight,keepaspectratio]{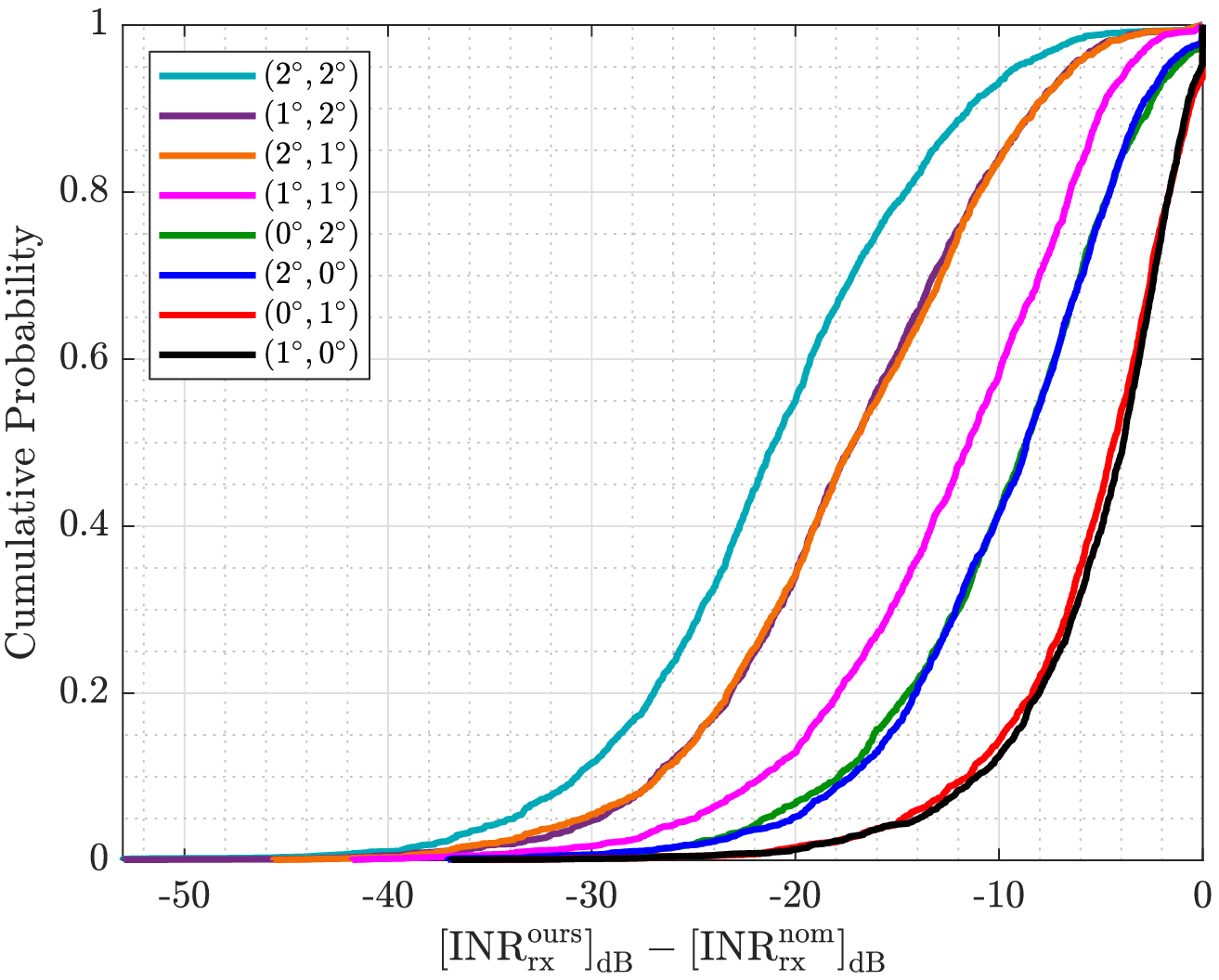}
        \label{fig:subfig-b}}
    \caption{Caption here.}
    \label{fig:subfigs}
\end{figure*}

\begin{figure}
    \centering
    \includegraphics[width=\linewidth,height=0.28\textheight,keepaspectratio]{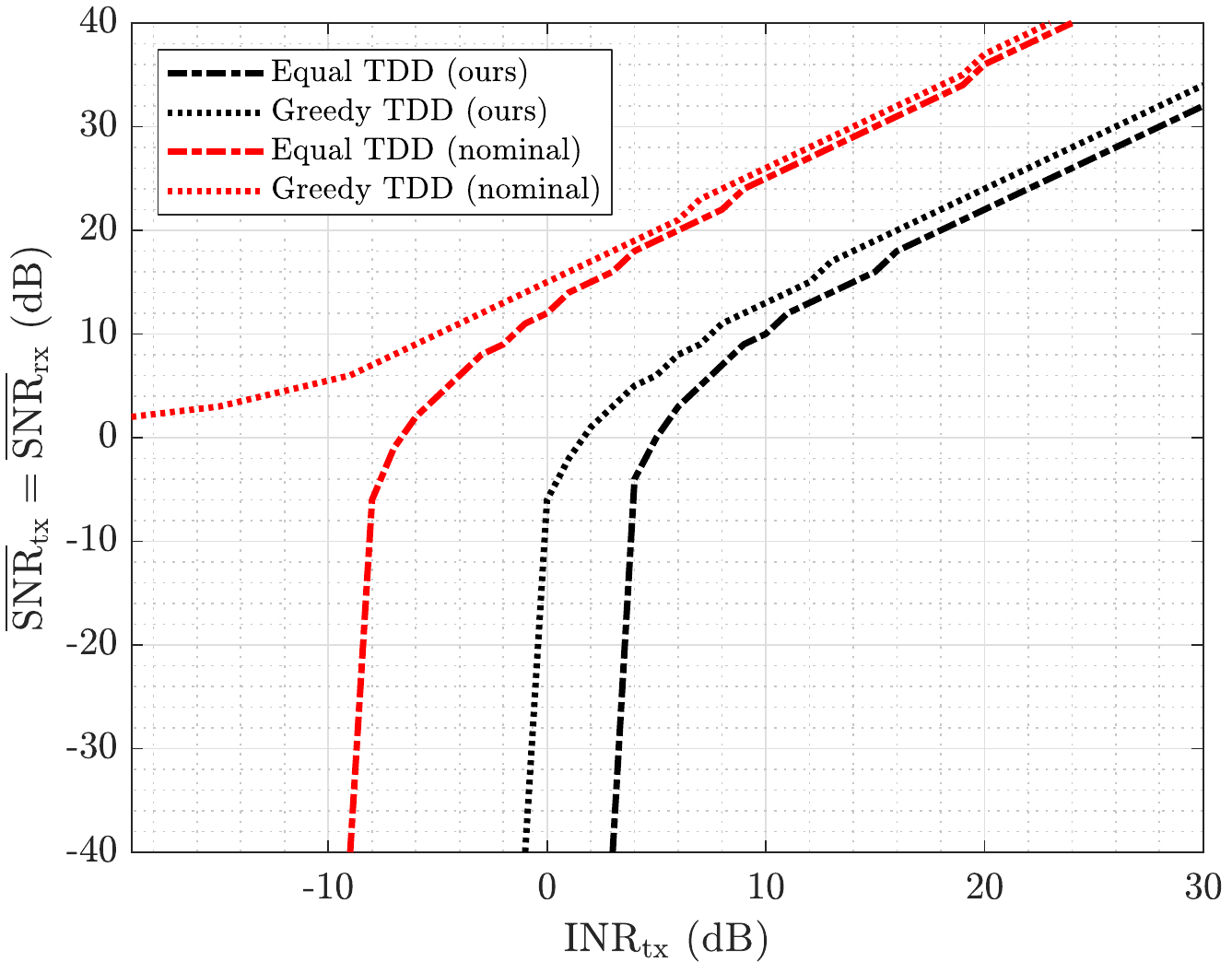}
    \caption{Caption.}
    \label{fig:}
\end{figure}
}

\comment{
---

\subsection{How Does Full-Duplexing with \steer Compare to that with \Naive Beam Selection?}

How to choose $\inrulthresh$

SNR comparable, SINR gain

Full-duplex capacity

Our codeboook offers greater robustness to self-interference while offering similar SNR performance.
As a result, we offer greater uplink SINR and comparable downlink SINR.
This allows us to tolerate more downlink INR in fact, and allows us to acheive 

\subsection{How Does Full-Duplexing with Our Codebook Compare to Other Multiplexing Strategies?}

Rate region.

Spectral efficiency curve?

TDMA, SDMA, Uplink and downlink of each

\subsection{When Should We Full-Duplex instead of Half-Duplex?}

Asymmetric SNRs, which users are trouble

---

\subsection{Choosing INR thresh.}

\subsection{Comparing our design versus TDD and GTDD.}

\subsection{Comparing our design versus nominal codebook.}

rate gain

SNR dist, INR dist, SINR dist

\subsection{INR/SNR region.}

\subsection{Vary SNRrx and SNRtx.}

\subsection{Problem children in azimuth.}

---

\subsection{How Does Full-Duplexing with Our Codebook Compare to Other Multiplexing Strategies?}

SNR CDF relative to SNR bar

How far do beams typically have to deviate (angles probabilities)

SE versus TDD and GTDD for various (Delta theta, Delta phi).

INR thresh?

\begin{figure}
    \centering
    \includegraphics[width=\linewidth,height=0.3\textheight,keepaspectratio]{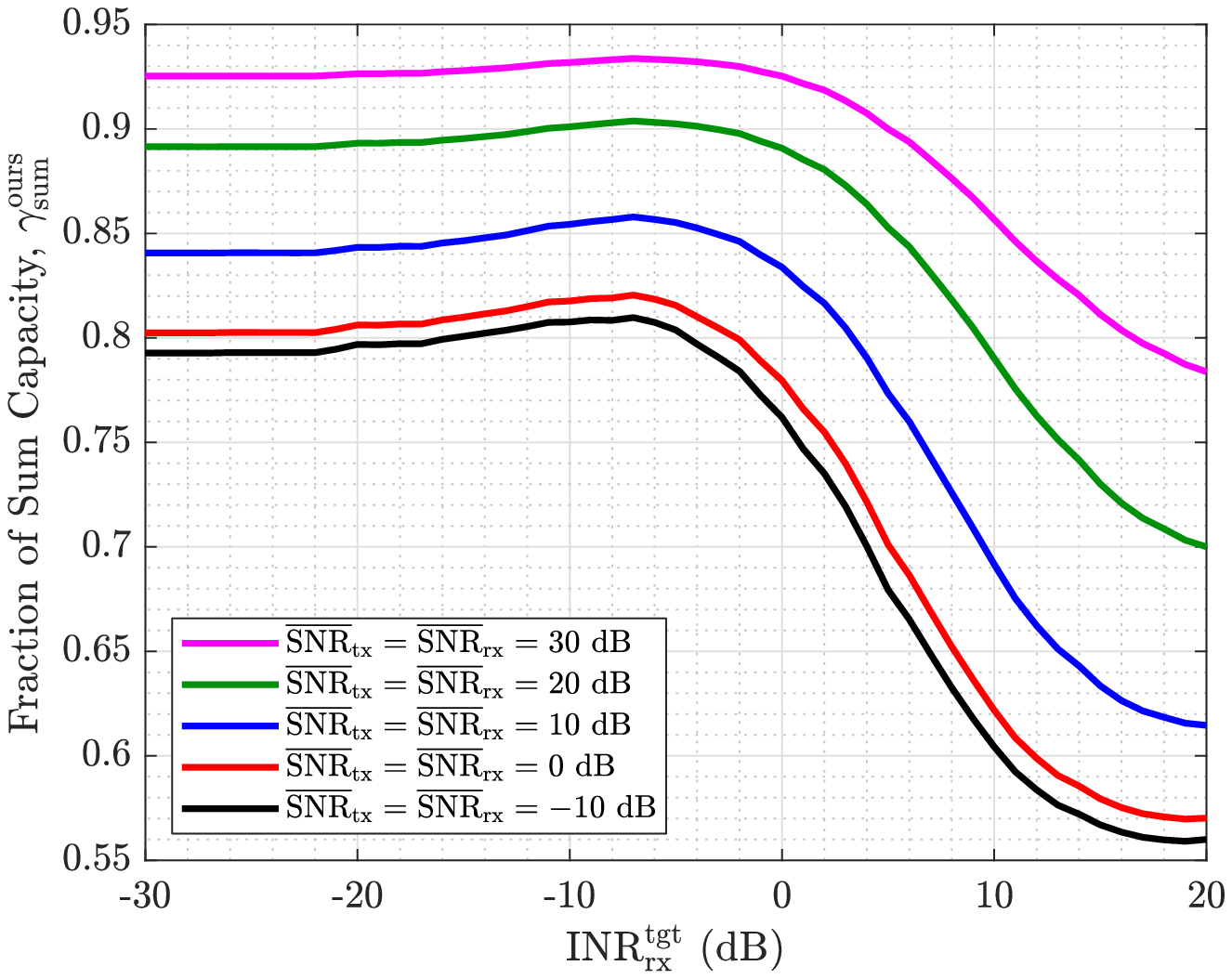}
    \caption{The fraction $\capfracsumours$ of the sum capacity $\capsumcb$ achieved by our design as a function of design parameter $\inrrxthresh$ for various $\snrtxbar = \snrrxbar$.}
    \label{fig:}
\end{figure}

\begin{figure*}
    \centering
    \subfloat[Caption a.]{\includegraphics[width=0.475\linewidth,height=\textheight,keepaspectratio]{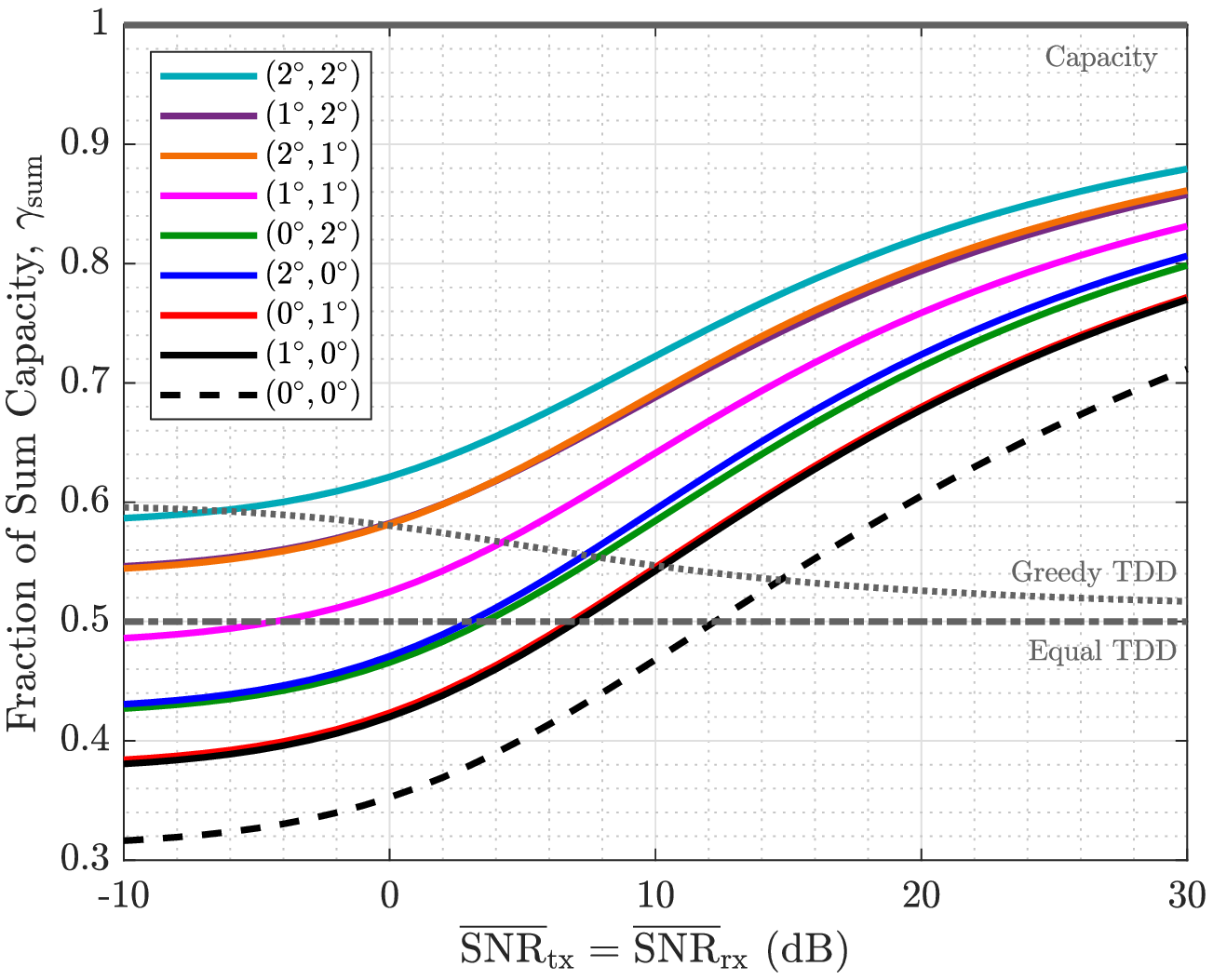}
        \label{fig:subfig-a}}
    \quad
    \subfloat[Caption b.]{\includegraphics[width=0.475\linewidth,height=\textheight,keepaspectratio]{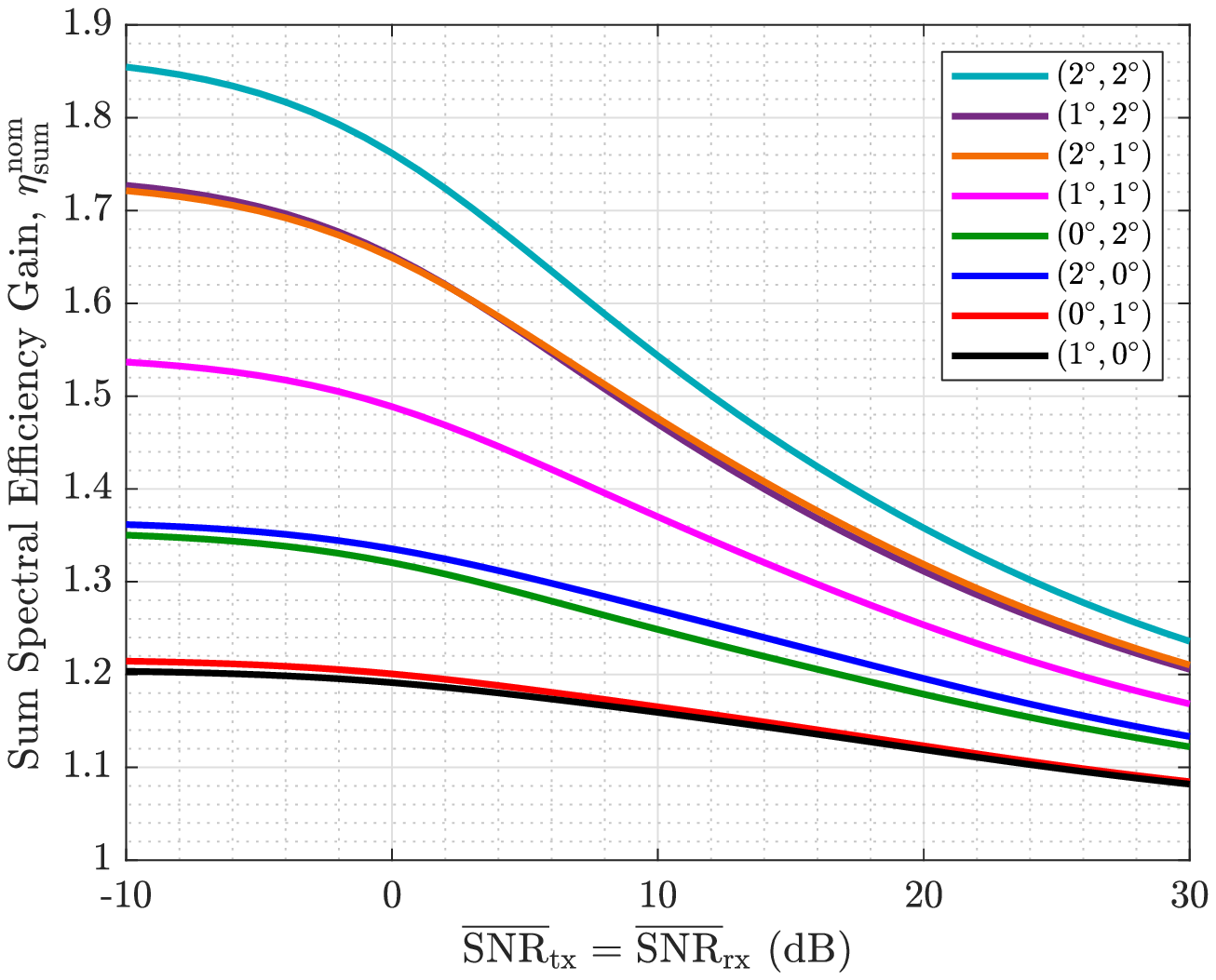}
        \label{fig:subfig-b}}
    \caption{Caption here.}
    \label{fig:subfigs}
\end{figure*}

\begin{figure*}
    \centering
    \subfloat[Caption a.]{\includegraphics[width=0.475\linewidth,height=\textheight,keepaspectratio]{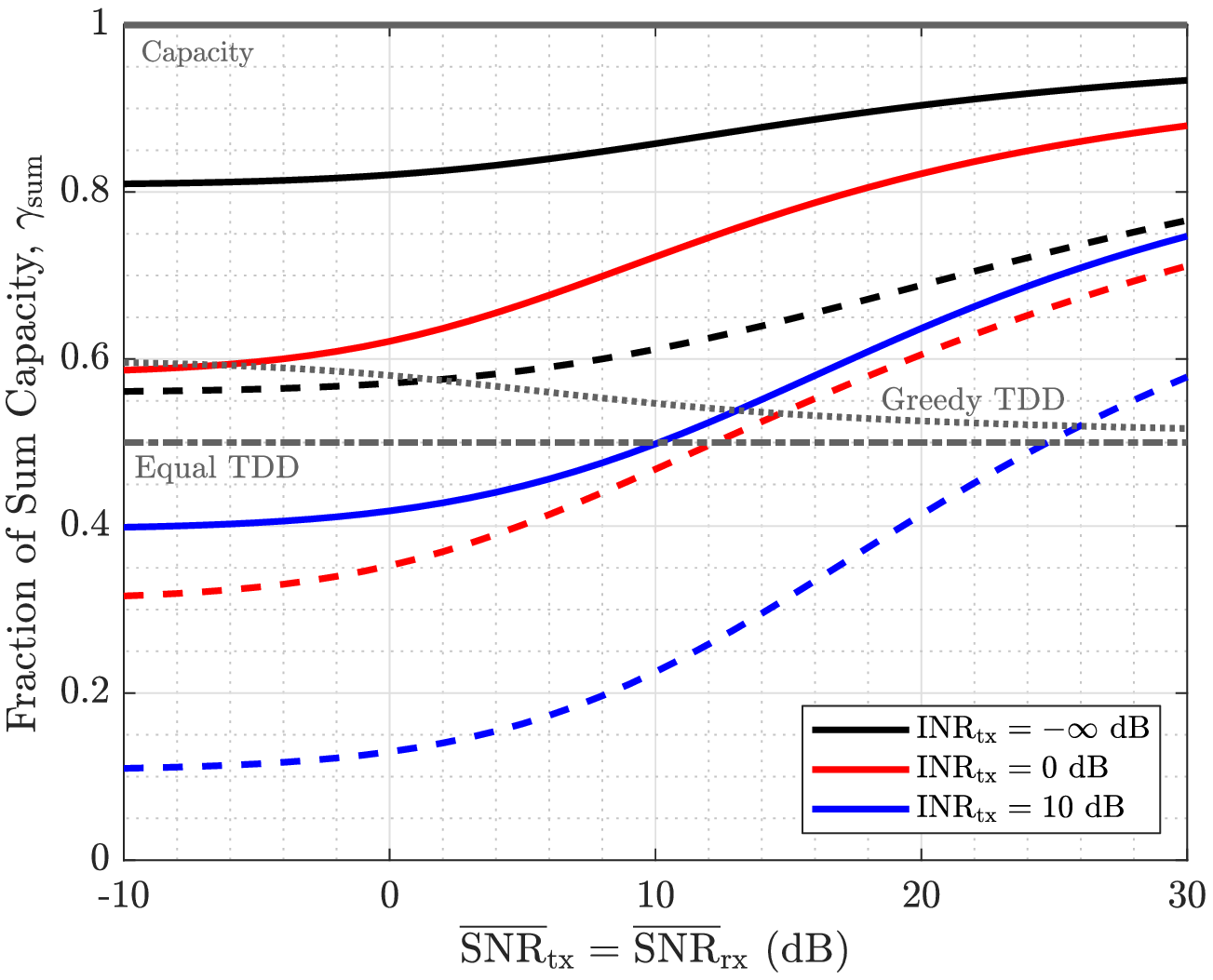}
        \label{fig:subfig-a}}
    \quad
    \subfloat[Caption b.]{\includegraphics[width=0.475\linewidth,height=\textheight,keepaspectratio]{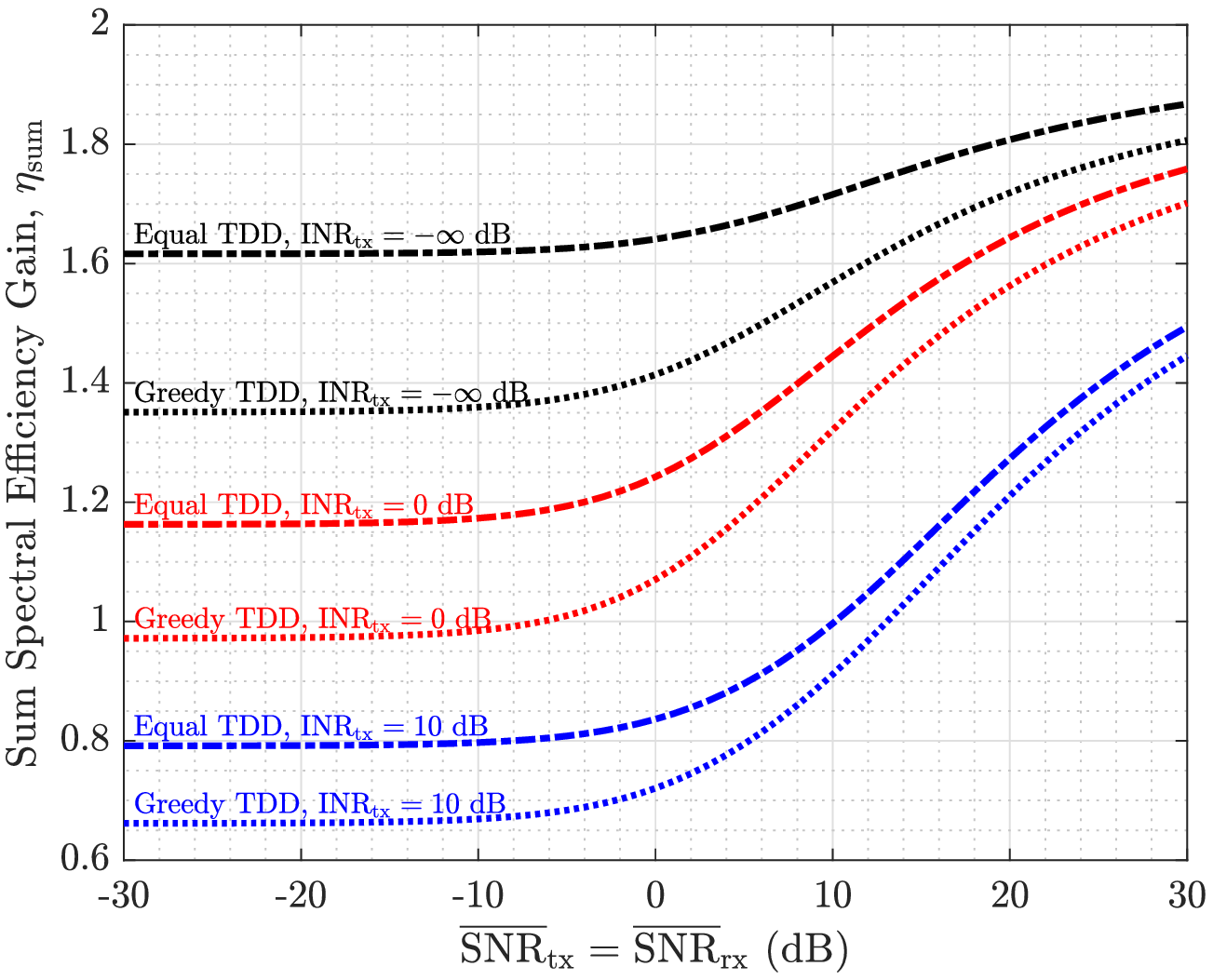}
        \label{fig:subfig-b}}
    \caption{Caption here.}
    \label{fig:subfigs}
\end{figure*}

\subsection{How Does Our Codebook Compare to a \Naive Beam Selection Approach for Full-Duplex?}

Rate gain,

\begin{figure*}
    \centering
    \subfloat[Caption a.]{\includegraphics[width=0.475\linewidth,height=\textheight,keepaspectratio]{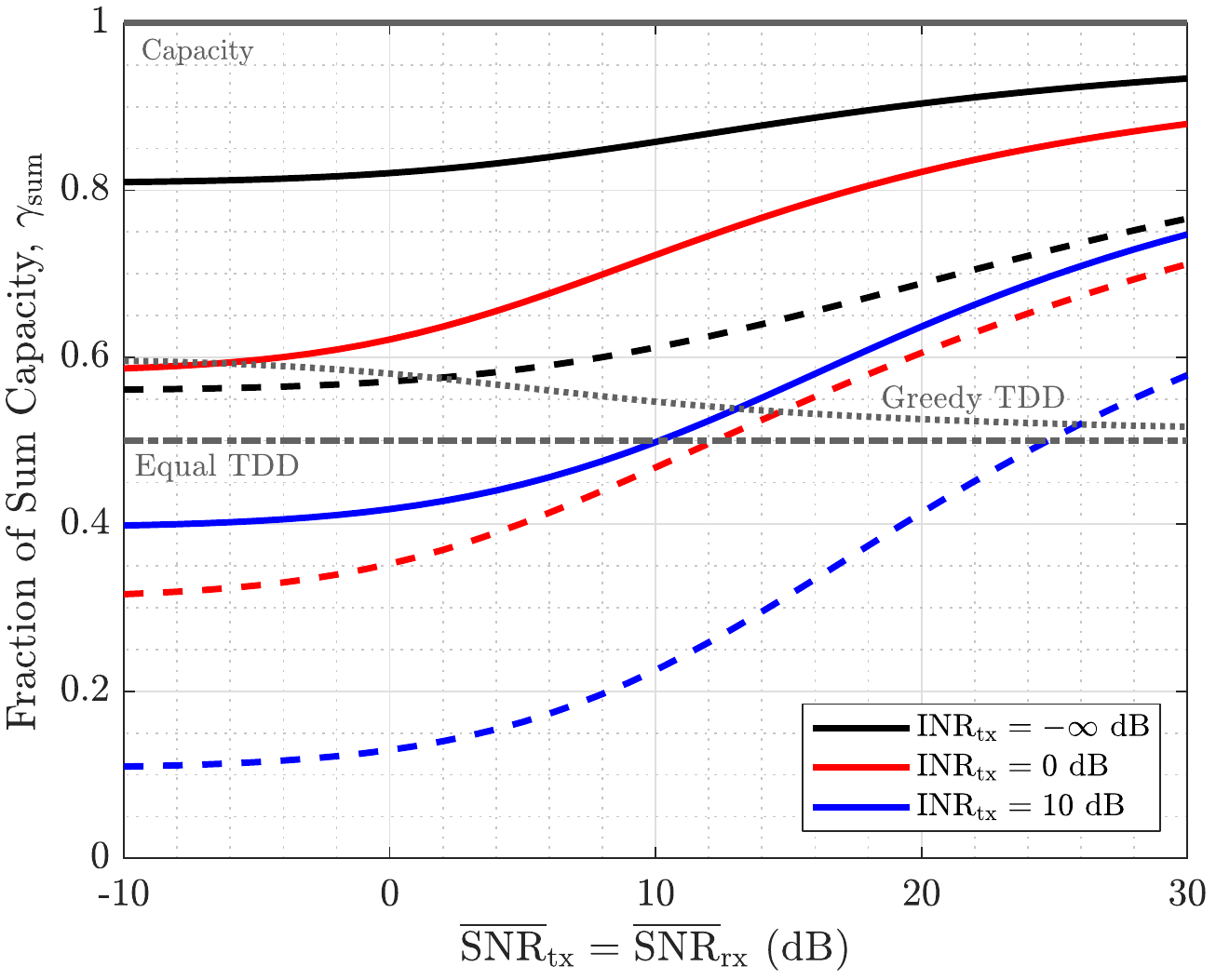}
        \label{fig:subfig-a}}
    \quad
    \subfloat[Caption b.]{\includegraphics[width=0.475\linewidth,height=\textheight,keepaspectratio]{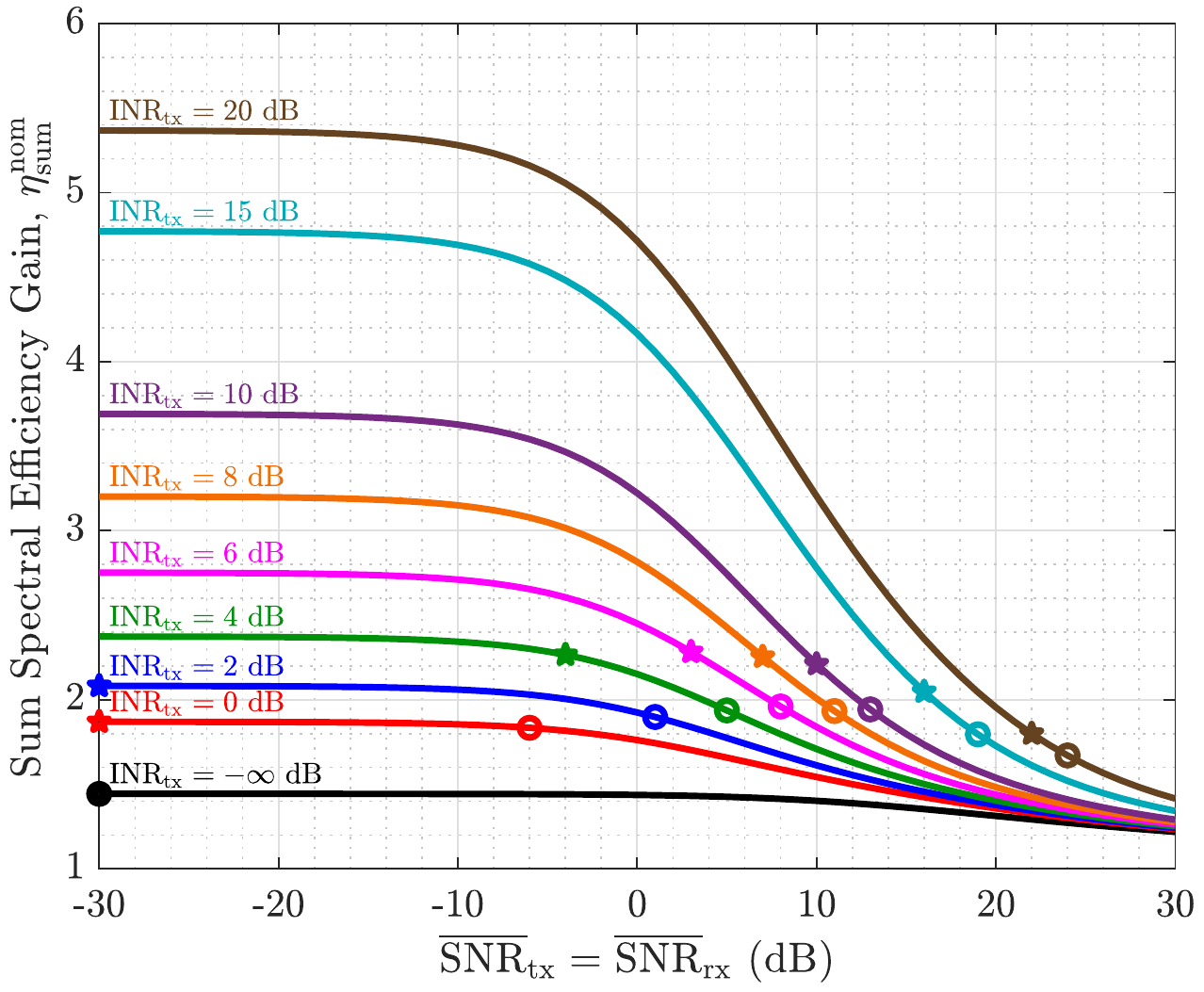}
        \label{fig:subfig-b}}
    \caption{Caption here.}
    \label{fig:subfigs}
\end{figure*}

Delta SNR CDF, Delta INR CDF, Delta SINR CDF.

\begin{figure*}
    \centering
    \subfloat[Caption a.]{\includegraphics[width=0.475\linewidth,height=\textheight,keepaspectratio]{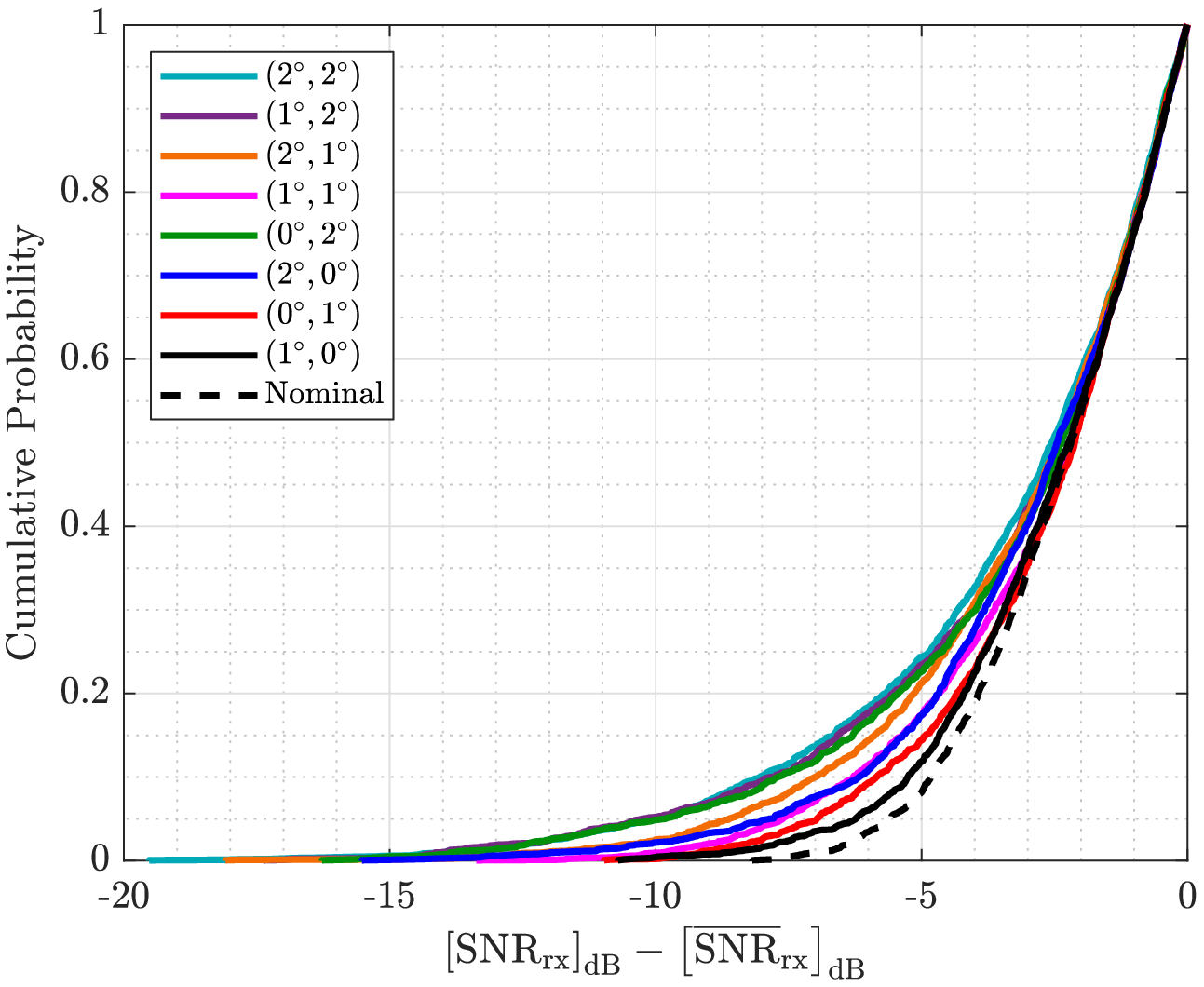}
        \label{fig:subfig-a}}
    \quad
    \subfloat[Caption b.]{\includegraphics[width=0.475\linewidth,height=\textheight,keepaspectratio]{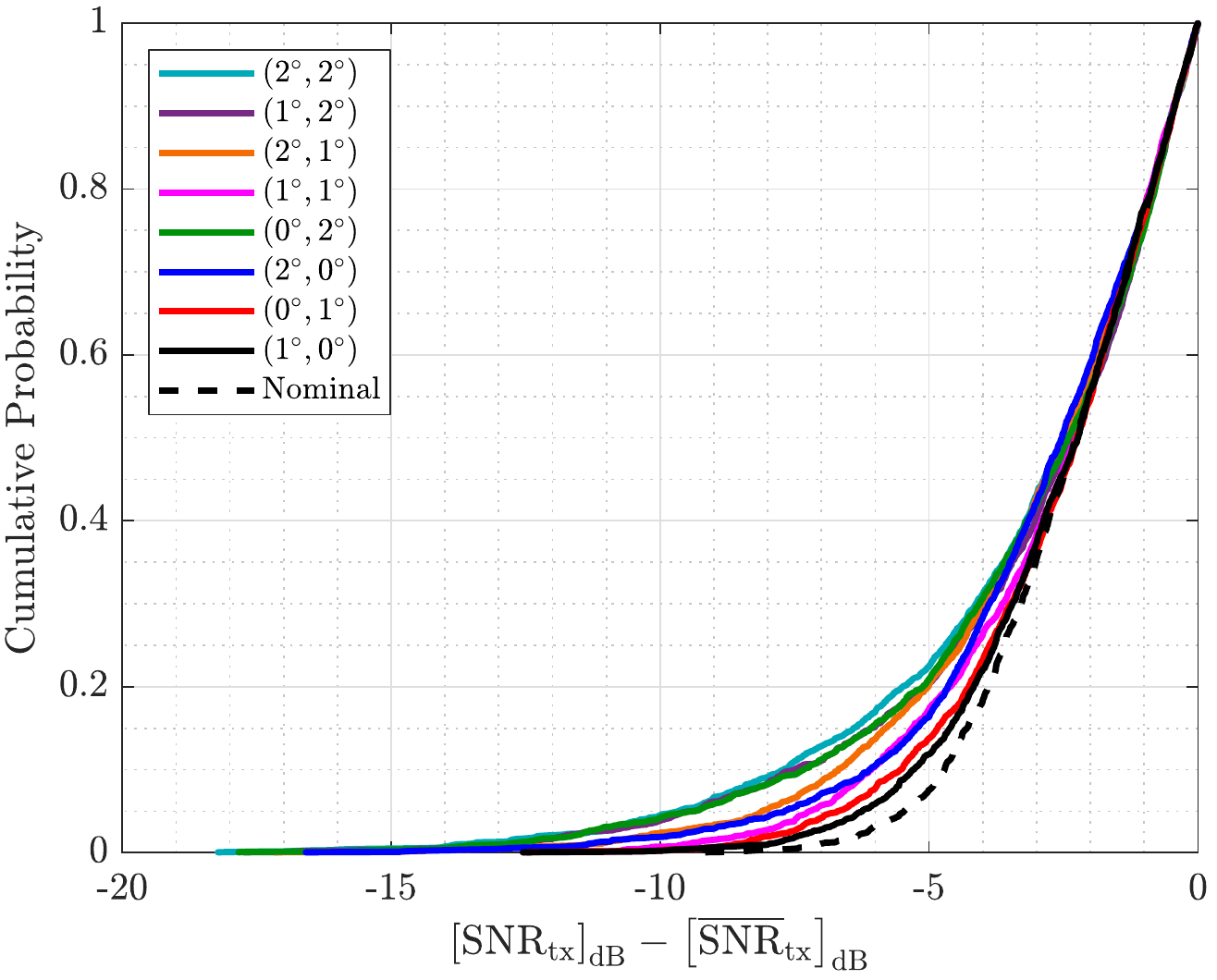}
        \label{fig:subfig-b}}
    \caption{Caption here.}
    \label{fig:subfigs}
\end{figure*}

\begin{figure*}
    \centering
    \subfloat[Caption a.]{\includegraphics[width=0.475\linewidth,height=\textheight,keepaspectratio]{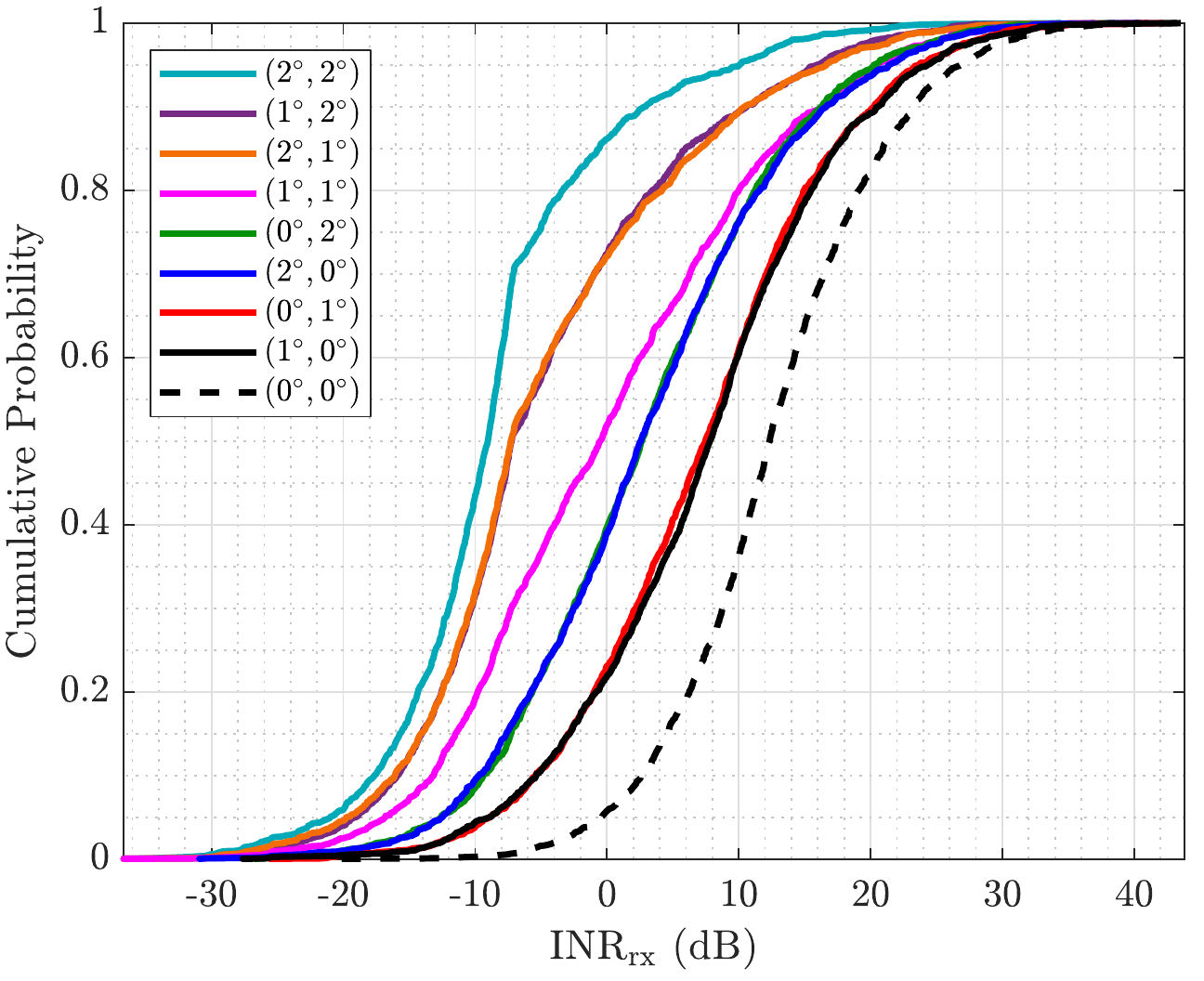}
        \label{fig:subfig-a}}
    \quad
    \subfloat[Caption b.]{\includegraphics[width=0.475\linewidth,height=\textheight,keepaspectratio]{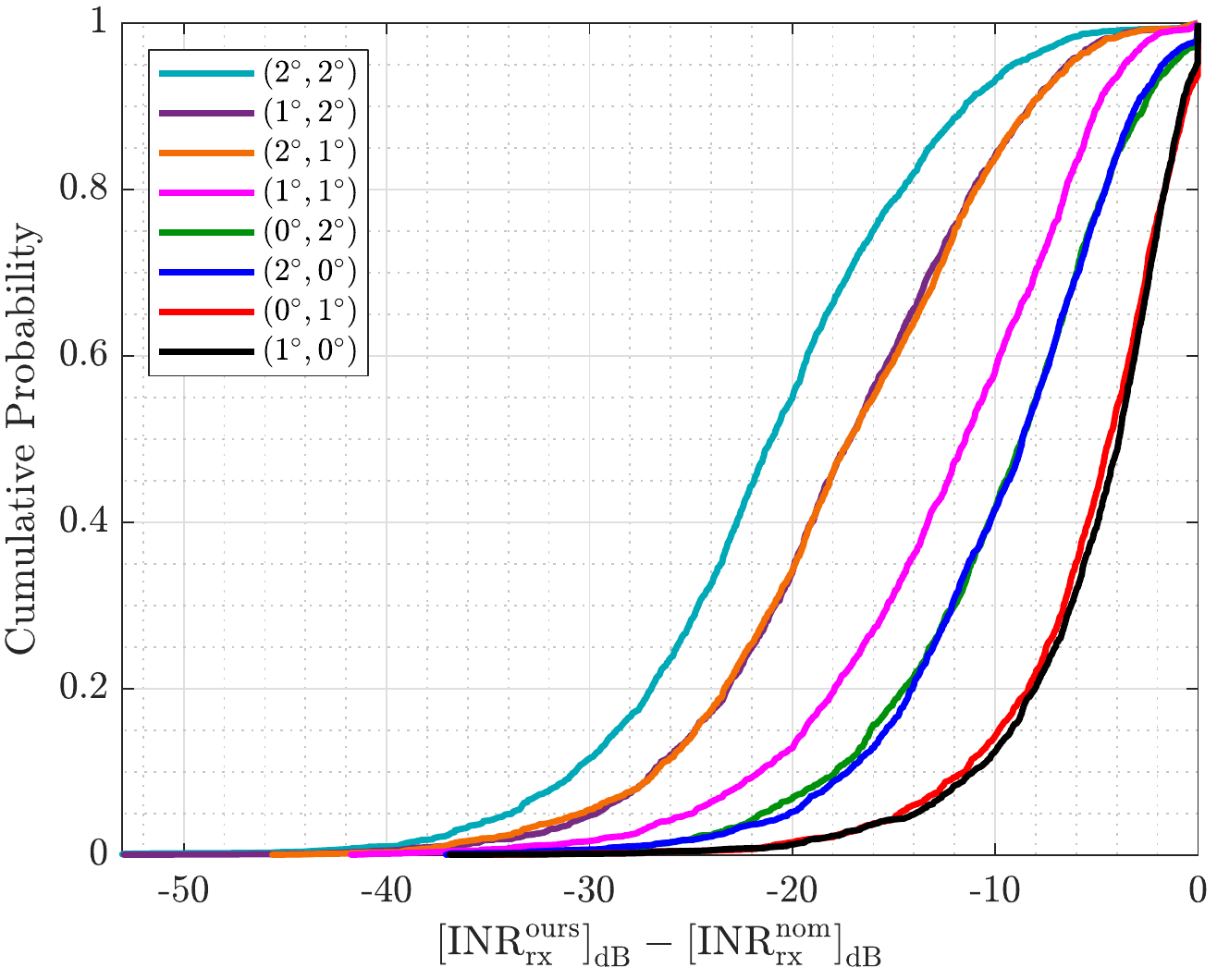}
        \label{fig:subfig-b}}
    \caption{Caption here.}
    \label{fig:subfigs}
\end{figure*}

\begin{figure*}
    \centering
    \subfloat[Caption a.]{\includegraphics[width=0.475\linewidth,height=\textheight,keepaspectratio]{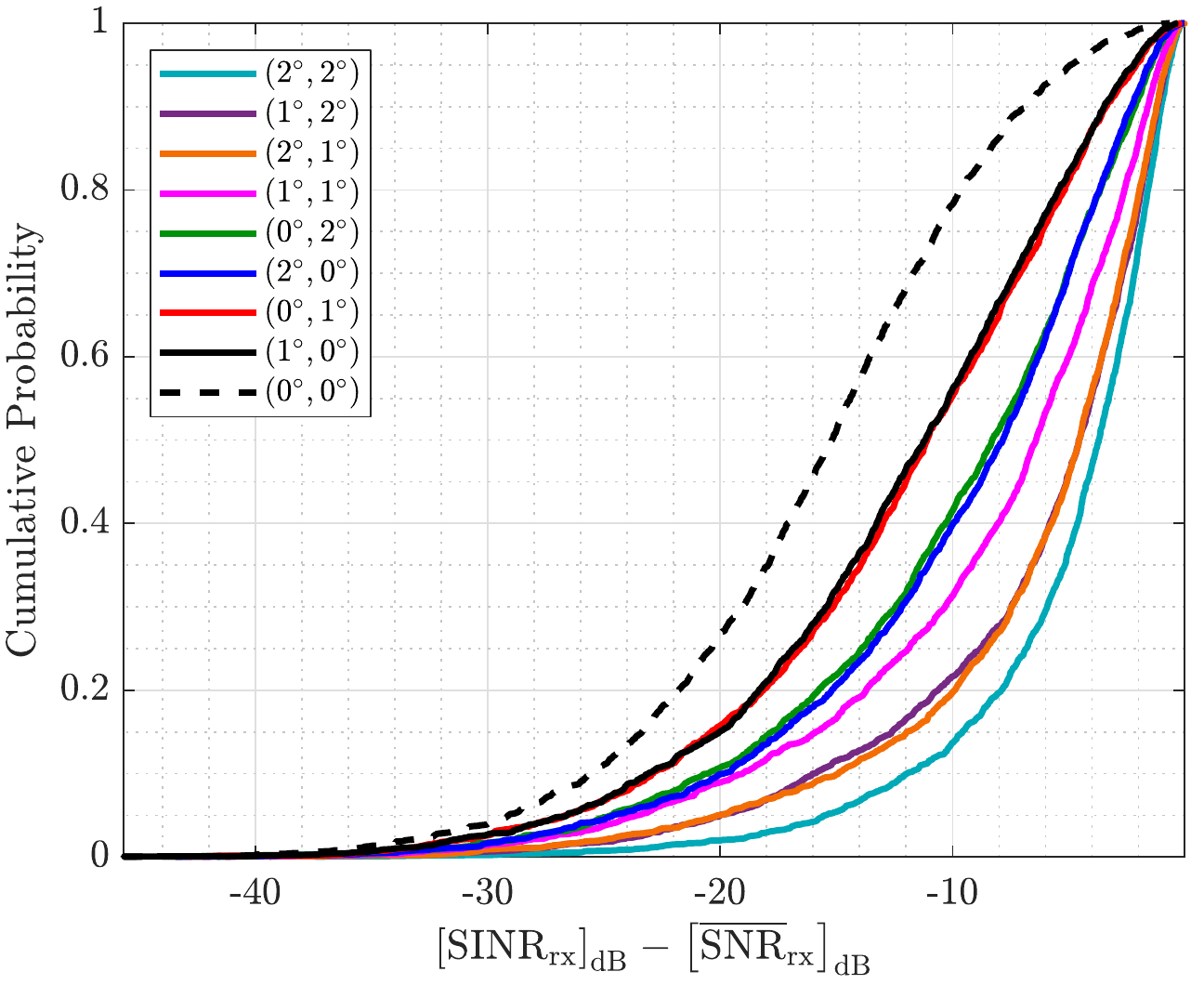}
        \label{fig:subfig-a}}
    \quad
    \subfloat[Caption b.]{\includegraphics[width=0.475\linewidth,height=\textheight,keepaspectratio]{plots/main_frisbee_v04_19}
        \label{fig:subfig-b}}
    \caption{Caption here.}
    \label{fig:subfigs}
\end{figure*}

\subsection{When Should We Full-Duplex instead of Half-Duplex?}

\begin{figure}
    \centering
    \includegraphics[width=\linewidth,height=0.28\textheight,keepaspectratio]{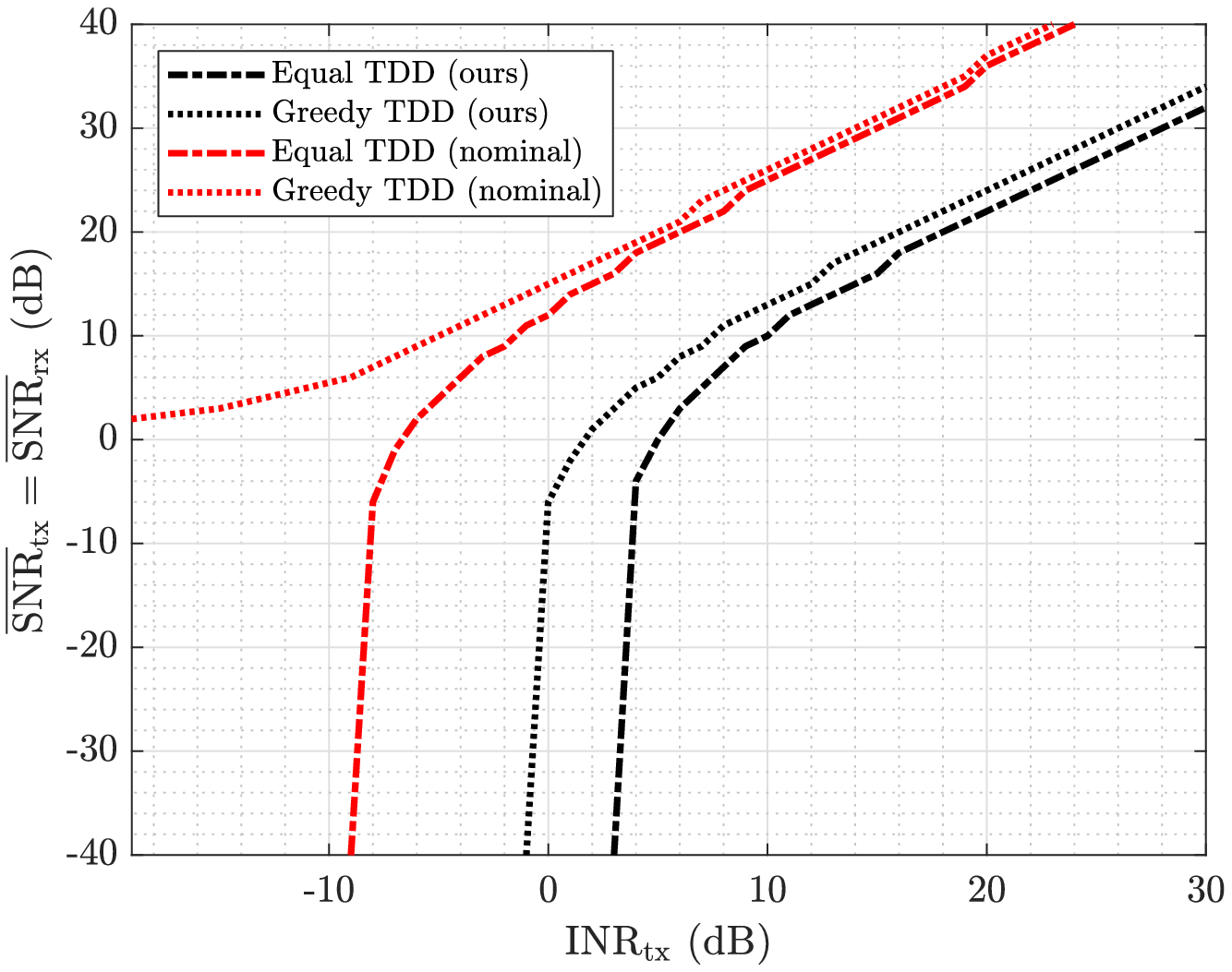}
    \caption{Caption.}
    \label{fig:}
\end{figure}

Vary INRtx, SNRtx, SNRrx.

\pagebreak

\subsection{Performance Metrics}
True capacities.
\begin{align}
\captx &= \logtwo{1 + \snrtxbar} \\
\caprx &= \logtwo{1 + \snrrxbar} \\
\capsum &= \captx + \caprx
\end{align}

SNR achieved by conventional codebooks.
\begin{align}
\snrtxnom &= \max_{\vf \in \precb} \ \snrtx \\
\snrrxnom &= \max_{\vw \in \comcb} \ \snrrx 
\end{align}

\begin{align}
\snrtxnom &\triangleq \snrtxbar \cdot \frac{\bars{\vhtx\ctrans \vf\thphtxbar}^2}{\Na} \\
\snrrxnom &\triangleq \snrrxbar \cdot \frac{\bars{\vw\thphrxbar\ctrans \vhrx}^2}{\Na} 
\end{align}

\begin{align}
\snrtxours &\triangleq \snrtxbar \cdot \frac{\bars{\vhtx\ctrans \vf\thphtxopt}^2}{\Na} \\
\snrrxours &\triangleq \snrrxbar \cdot \frac{\bars{\vw\thphrxopt\ctrans \vhrx}^2}{\Na} 
\end{align}

\begin{align}
\inrrxnom &\triangleq \inrrx\thphtxrxbar \\
\inrrxours &\triangleq \inrrx\thphtxrxopt
\end{align}

\begin{align}
\sinrrxnom &\triangleq \frac{\snrrxnom}{1 + \inrrxnom} \\
\sinrrxours &\triangleq \frac{\snrrxours}{1 + \inrrxours} 
\end{align}

Capacities achieved by conventional codebooks.
\begin{align}
\captxcb &= \logtwo{1 + \snrtxnom} \\
\caprxcb &= \logtwo{1 + \snrrxnom} \\
\capsumcb &= \captxcb + \caprxcb
\end{align}

Equal TDD rates.
\begin{align}
\setxtdd &= 0.5 \cdot \captxcb \\
\serxtdd &= 0.5 \cdot \caprxcb \\
\sesumtdd &= 0.5 \cdot \capsumcb
\end{align}

Greedy TDD rates.
\begin{align}
\setxgtdd &= \ind{\snrtxnom \geq \snrrxnom} \cdot \caprxcb \\
\serxgtdd &= \ind{\snrtxnom < \snrrxnom} \cdot \caprxcb \\
\sesumgtdd &= \maxop{\captxcb,\caprxcb}
\end{align}

Rates achieved by nominal codebook (with SI).
\begin{align}
\setxnom &= \logtwo{1 + \sinrtxnom} \\
\serxnom &= \logtwo{1 + \sinrrxnom} \\
\sesumnom &= \setxnom + \serxnom
\end{align}

Rates achieved by our codebook (with SI).
\begin{align}
\setxours &= \logtwo{1 + \sinrtxours} \\
\serxours &= \logtwo{1 + \sinrrxours} \\
\sesumours &= \setxours + \serxours
\end{align}

Improvement of our codebook over equal TDD.
\begin{align}
\segaintxtdd &= \frac{\setxours}{\setxtdd} \\
\segainrxtdd &= \frac{\serxours}{\serxtdd} \\
\segainsumtdd &= \frac{\setxours + \serxours}{\setxtdd + \serxtdd}
\end{align}

Improvement of our codebook over greedy TDD.
\begin{align}
\segaintxgtdd &= \frac{\setxours}{\setxgtdd} \\
\segainrxgtdd &= \frac{\serxours}{\serxgtdd} \\
\segainsumgtdd &= \frac{\setxours + \serxours}{\setxgtdd + \serxgtdd}
\end{align}

Improvement of our codebook over conventional codebook.
\begin{align}
\segaintxnom &= \frac{\setxours}{\setxnom} \\
\segainrxnom &= \frac{\serxours}{\serxnom} \\
\segainsumnom &= \frac{\setxours + \serxours}{\setxnom + \serxnom}
\end{align}

Fraction of the codebook capacity achieved by equal TDD.
\begin{align}
\capfractxtdd &= \frac{\setxtdd}{\captxcb} \\
\capfracrxtdd &= \frac{\serxtdd}{\caprxcb} \\
\capfracsumtdd &= \frac{\setxtdd + \serxtdd}{\captxcb + \caprxcb}
\end{align}

Fraction of the codebook capacity achieved by greedy TDD.
\begin{align}
\capfractxgtdd &= \frac{\setxgtdd}{\captxcb} \\
\capfracrxgtdd &= \frac{\serxgtdd}{\caprxcb} \\
\capfracsumgtdd &= \frac{\setxgtdd + \serxgtdd}{\captxcb + \caprxcb}
\end{align}

Fraction of the codebook capacity achieved by conventional codebook.
\begin{align}
\capfractxnom &= \frac{\setxnom}{\captxcb} \\
\capfracrxnom &= \frac{\serxnom}{\caprxcb} \\
\capfracsumnom &= \frac{\setxnom + \serxnom}{\captxcb + \caprxcb}
\end{align}

Fraction of the codebook capacity achieved by our codebook.
\begin{align}
\capfractxours &= \frac{\setxours}{\captxcb} \\
\capfracrxours &= \frac{\serxours}{\caprxcb} \\
\capfracsumours &= \frac{\setxours + \serxours}{\captxcb + \caprxcb}
\end{align}

Notice that
\begin{align}
\segainsumnom = \frac{\capfracsumours}{\capfracsumnom}
\end{align}
etc.
}

\section{Conclusion} \label{sec:conclusion}

In this work, we present \steer, a measurement-driven beam selection methodology for full-duplex \mmwave systems that leverages small shifts of the steering directions of the transmit and receive beams to significantly reduce self-interference and deliver high beamforming gain.
Evaluation of \steer through measurements with a 28 GHz phased array platform along with further simulation highlights its ability to reduce self-interference to levels near or below the noise floor, offering noteworthy spectral efficiency gains over half-duplex and full-duplex operation that uses conventional beam selection.
\steer can facilitate the deployment of full-duplex \mmwave systems to deliver high-throughput, low-latency wireless connectivity, while importantly supporting existing beam alignment schemes in 5G.
Valuable future work would include further measuring the time dynamics and small-scale spatial variability of \mmwave self-interference, which would ultimately drive design decisions of \steer at deployment.
% Also, the study and measurement of cross-link interference in full-duplex \mmwave systems would be a useful future contribution.
Also, the design of beamforming codebooks that are inherently robust to self-interference and the integration of full-duplex beam selection into \mmwave network standards would be useful future contributions.
\edit{Extending \steer to multi-user/multi-beam systems, along with evaluating \steer using \mmwave platforms at other carrier frequencies, in different configurations, and in a variety of settings, would be necessary future work.}

\section*{Acknowledgments}

I.~P.~Roberts is supported by the National Science Foundation Graduate Research Fellowship Program under Grant No.~DGE-1610403. 
Any opinions, findings, and conclusions or recommendations expressed in this material are those of the authors and do not necessarily reflect the views of the National Science Foundation.

% \section*{References} \label{sec:bibliography}
% \printbibliography[heading=none]
\bibliographystyle{bibtex/IEEEtran}
\bibliography{bibtex/IEEEabrv,refs}

\end{document}